\documentclass[longauth]{aa}
\usepackage[colorlinks,linkcolor=blue,citecolor=blue]{hyperref}

\usepackage{graphicx} 
\usepackage{lscape}
\usepackage{txfonts}
\usepackage{natbib}

\begin{document}

\title{The eROSITA Final Equatorial-Depth Survey (eFEDS): Catalog of galaxy clusters and groups\footnote{The full cluster catalog is available in electronic form at the CDS via anonymous ftp to \url{cdsarc.u-strasbg.fr} (130.79.128.5) or via \url{http://cdsweb.u-strasbg.fr/cgi-bin/qcat?J/A+A/}}}

\author{
A.~Liu\inst{1}\thanks{e-mail: \href{mailto:liuang@mpe.mpg.de}{\tt liuang@mpe.mpg.de}},
E.~Bulbul\inst{1},
V.~Ghirardini\inst{1},
T.~Liu\inst{1},
M.~Klein\inst{2},
N.~Clerc\inst{3}, 
Y.~\"Ozsoy\inst{1}, 
M.~E.~Ramos-Ceja\inst{1}, 
F.~Pacaud\inst{4},
J.~Comparat\inst{1}, 
N.~Okabe\inst{5,6,7}, 
Y.~E.~Bahar\inst{1}, 
V. Biffi\inst{2,8,9}, 
H.~Brunner\inst{1}, 
M.~Br\"uggen\inst{10}, 
J.~Buchner\inst{1},
J.~Ider Chitham\inst{1}, 
I.~Chiu\inst{11,12,13}, 
K. Dolag\inst{2},
E.~Gatuzz\inst{1},  
J.~Gonzalez\inst{1}, 
D.~N.~Hoang\inst{10}, 
G.~Lamer\inst{14}, 
A.~Merloni\inst{1}, 
K.~Nandra\inst{1}, 
M.~Oguri\inst{15,16,17}, 
N.~Ota\inst{4,18}, 
P.~Predehl\inst{1},
T.~H.~Reiprich\inst{4}, 
M.~Salvato\inst{1}, 
T.~Schrabback\inst{4},
J.~S.~Sanders\inst{1},
R.~Seppi\inst{1}, 
Q.~Thibaud\inst{3}}

\institute{
\inst{1}{Max Planck Institute for Extraterrestrial Physics, Giessenbachstrasse 1, 85748 Garching, Germany}\\
\inst{2}{Universitaets-Sternwarte Muenchen, Fakultaet fuer Physik, LMU Munich, Scheinerstr. 1, 81679 Munich, Germany}\\
\inst{3}{IRAP, Université de Toulouse, CNRS, UPS, CNES, Toulouse, France}\\
\inst{4}{Argelander-Institut f{\"{u}}r Astronomie (AIfA), Universit{\"{a}}t Bonn, Auf dem H{\"{u}}gel 71, 53121 Bonn, Germany}\\
\inst{5}{Physics Program, Graduate School of Advanced Science and Engineering, Hiroshima University, 1-3-1 Kagamiyama, Higashi-Hiroshima, Hiroshima 739-8526, Japan}\\
\inst{6}{Hiroshima Astrophysical Science Center, Hiroshima University, 1-3-1 Kagamiyama, Higashi-Hiroshima, Hiroshima 739-8526, Japan}\\
\inst{7}{Core Research for Energetic Universe, Hiroshima University, 1-3-1, Kagamiyama, Higashi-Hiroshima, Hiroshima 739-8526, Japan}\\
\inst{8}{INAF - Osservatorio Astronomico di Trieste, via Tiepolo 11, I-34143 Trieste, Italy}\\
\inst{9}{IFPU - Institute for Fundamental Physics of the Universe, Via Beirut 2, I-34014 Trieste, Italy}\\
\inst{10}{University of Hamburg, Hamburger Sternwarte, Gojenbergsweg 112, 21029 Hamburg, Germany}\\
\inst{11}{Tsung-Dao Lee Institute, and Key Laboratory for Particle Physics, Astrophysics and Cosmology, Ministry of Education, Shanghai Jiao Tong University, Shanghai 200240, China}\\
\inst{12}{Department of Astronomy, School of Physics and Astronomy, and Shanghai Key Laboratory for Particle Physics and Cosmology, Shanghai Jiao Tong University, Shanghai 200240, China}\\
\inst{13}{Academia Sinica Institute of Astronomy and Astrophysics (ASIAA), 11F of AS/NTU Astronomy-Mathematics Building, No.1, Sec. 4, Roosevelt Rd, Taipei 10617, Taiwan}\\
\inst{14}{Leibniz-Institut für Astrophysik Potsdam (AIP), An der Sternwarte 16, 14482 Potsdam, German}\\
\inst{15}{Research Center for the Early Universe, The University of Tokyo, 7-3-1 Hongo, Bunkyo-ku, Tokyo 113-0033, Japan}\\
\inst{16}{Department of Physics, The University of Tokyo, 7-3-1 Hongo, Bunkyo-ku, Tokyo 113-0033, Japan}\\
\inst{17}{Kavli Institute for the Physics and Mathematics of the Universe (Kavli IPMU, WPI), The University of Tokyo, 5-1-5 Kashiwanoha, Kashiwa, Chiba 277-8582, Japan}\\
\inst{18}{Department of Physics, Nara Women's University, Kitauoyanishi-machi, Nara, 630-8506, Japan}\\
}


\titlerunning{eFEDS cluster and group catalog}
\authorrunning{Liu et al.}

\abstract
 {}
{The eROSITA Final Equatorial-Depth Survey has been carried out during the performance verification phase of the Spectrum-Roentgen-Gamma/eROSITA telescope and was completed in November 2019. This survey is designed to provide the first eROSITA-selected sample of clusters and groups and to test the predictions for the all-sky survey in the context of cosmological studies with clusters of galaxies.    }
{In the area of $\sim$140 square degrees covered by eFEDS, 542 candidate clusters and groups of galaxies were detected as extended X-ray sources with the \texttt{eSASS} source detection algorithm. We performed imaging and spectral analysis of the 542 cluster candidates with eROSITA X-ray data and studied the properties of the sample. }
{We provide the catalog of candidate galaxy clusters and groups detected by eROSITA in the eFEDS field down to a flux of $\sim10^{-14}~{\rm erg}~{\rm s}^{-1}~{\rm cm}^{-2}$ in the soft band (0.5--2~keV) within 1\arcmin. The clusters are distributed in the redshift range $z=$[0.01, 1.3] with a median redshift $z_{\rm median}=0.35$. With eROSITA X-ray data, we measured the temperature of the intracluster medium within two radii, 300~kpc and 500~kpc, and constrained the temperature with $>2\sigma$ confidence level for $\sim 1/5$ (102 out of 542) of the sample. The average temperature of these clusters is $\sim$2~keV. Radial profiles of flux, luminosity, electron density, and gas mass were measured from the precise modeling of the imaging data. The selection function, the purity, and the completeness of the catalog are examined and discussed in detail. The contamination fraction is $\sim1/5$ in this sample and is dominated by misidentified point sources. The X-ray luminosity function of the clusters agrees well with the results obtained from other recent X-ray surveys. 
We also find 19 supercluster candidates in this field, most of which are located at redshifts between 0.1 and 0.5, including one cluster at $z\sim0.36$ that was presented previously. }
{The eFEDS cluster and group catalog at the final eRASS equatorial depth provides a benchmark proof of concept for the eROSITA All-Sky Survey extended source detection and characterization. We confirm the excellent performance of eROSITA for cluster science and expect no significant deviations from our pre-launch expectations for the final all-sky survey.  }


\maketitle

\begin{figure*}
\begin{center}
\includegraphics[width=0.99\textwidth, trim=55 5 105 50, clip]{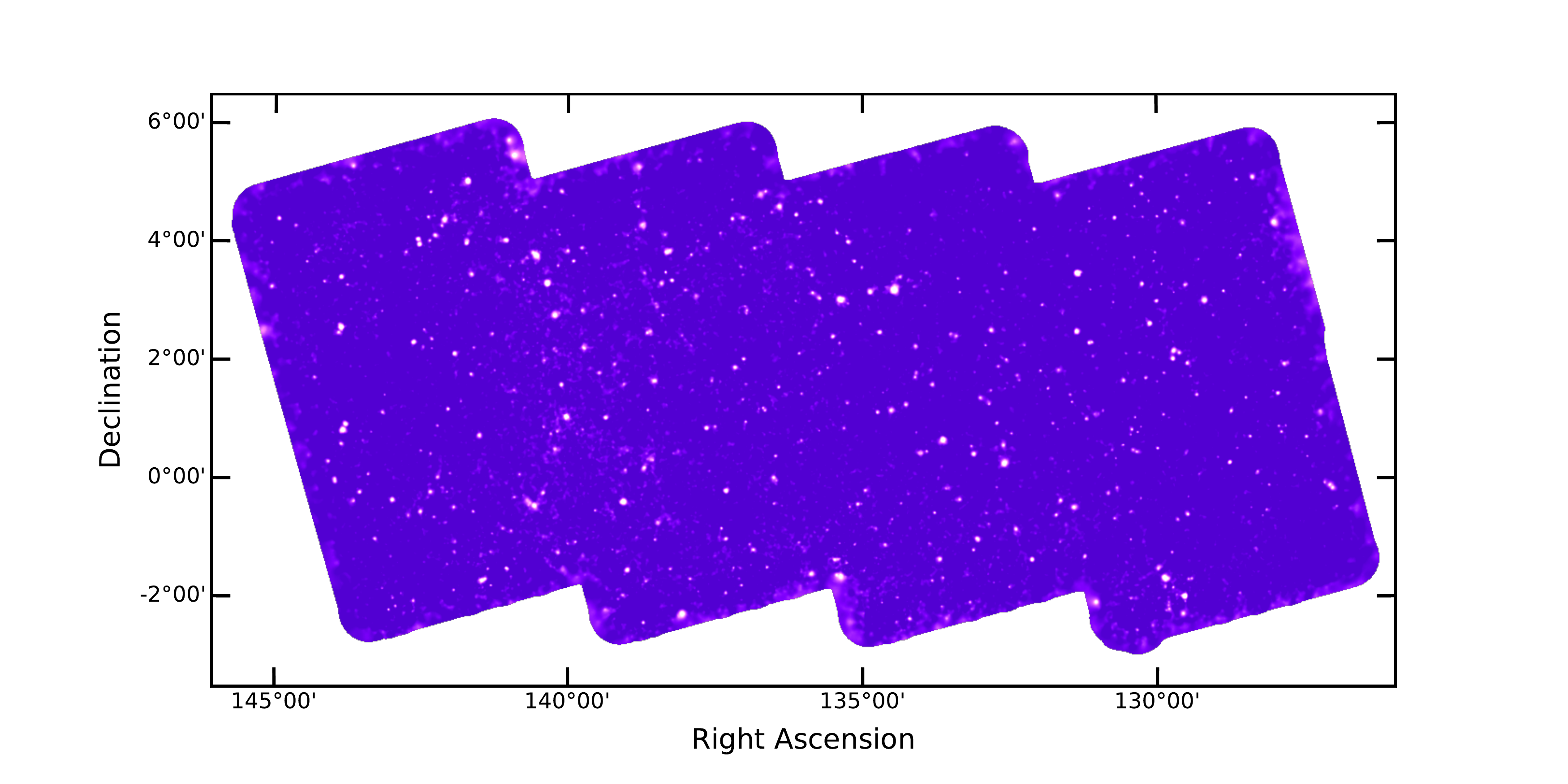}
\caption{Exposure-corrected point-source-free 0.2--2.3~keV adaptively smoothed eROSITA image of the eFEDS field. In detail, a point source mask is constructed by excluding regions around each point source in which the surface brightness is above 10\% of the background surface brightness. The \texttt{accumulate\_counts} program from the contour binning package \citep{2006Sanders} is used to calculate a map containing the radius around each pixel (including those in masked areas) that encloses at least 36 counts (excluding masked sources). This map is used to smooth the input image and exposure map with Gaussians, excluding masked areas, where $\sigma$ is given by the radius in each pixel. The smoothed image and exposure map were divided to create the exposure-corrected image shown here.}
\label{fig:image_efeds_large}
\end{center}
\end{figure*}
\section{Introduction}
The extended ROentgen Survey with an Imaging Telescope
Array \citep[eROSITA,][]{2021Predehl} on board the Spectrum-Roentgen-Gamma (SRG) mission is a German-Russian X-ray telescope launched on July 13, 2019. With the large collecting area (1365~cm$^2$ at 1~keV), moderate angular resolution (on-axis half-energy width, HEW, $\sim$18\arcsec at 1.49~keV), and wide energy band coverage (0.2--10~keV) \citep{2021Predehl}, eROSITA will provide an X-ray all-sky survey with unprecedented sensitivity. The complete eROSITA All-Sky Survey (eRASS) will be about 25 times more sensitive than the ROSAT All-Sky Survey \citep[RASS,][]{1999voges} in the soft X-ray band (0.2--2.3~keV), and will be the first ever true X-ray imaging all-sky survey in the hard band (2.3--10~keV) \citep{Merloni2012}. The final eROSITA All-Sky Survey will consist of eight complete scans of the X-ray sky by the end of 2023, each lasting for six months. The first and second scans have already been completed in June and December 2020. 

One of the main science goals of eROSITA is to study cosmology by detecting a large number of galaxy clusters and groups (for simplicity, we use the term ``clusters'' to refer to the assembly of X-ray emitting galaxy clusters and groups unless noted otherwise), which are the main extended sources in the X-ray sky. Clusters are the most massive gravitationally bound systems in the Universe. Located in the area of science where cosmology and astrophysics meet, clusters play a unique role in tracing the formation and evolution of the large-scale structure \citep[see][for a review]{allen2011} and the various astrophysical processes on smaller scales \citep[e.g.,][]{2002Rosati}. Studies on cluster-related cosmology and astrophysics require a large sample of clusters with a clean selection function and an accurate mass calibration \citep[see, e.g.,][]{2019Pratt}. 

\begin{figure*}
\begin{center}
\includegraphics[width=0.99\textwidth, trim=70 10 45 50, clip]{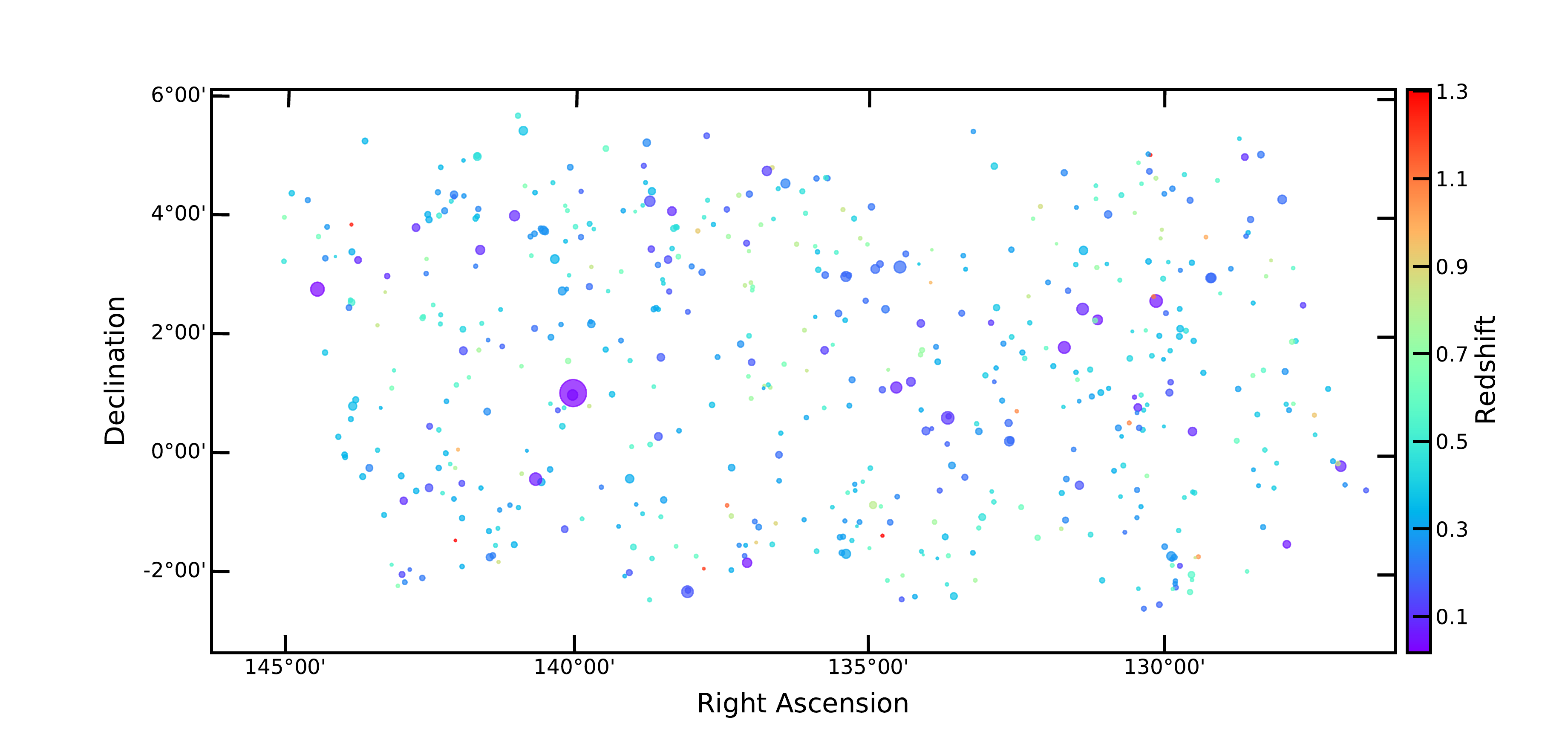}
\caption{Distribution of the 542 cluster candidates in eFEDS. The color code represents the redshift of the cluster, provided by MCMF \citep[][]{2018Klein,2019Klein}. The radius of the circle is equal to $R_{500}$, within which the average density is 500 times the critical density at the cluster redshift. } 
\label{fig:cat}
\end{center}
\end{figure*}

The first X-ray imaging all-sky survey, RASS, was performed by ROSAT in the 1990s \citep{1999voges}. It surveyed the X-ray sky in the soft (0.1--2.4~keV) band. RASS-based catalogs, such as ROSAT Brightest Cluster Sample \citep[BCS,][]{1998Ebeling}, Northern ROSAT All-Sky Galaxy Cluster
Survey \citep[NORAS,][]{2000bohringer}, ROSAT-ESO Flux Limited X-ray Galaxy Cluster Survey \citep[REFLEX,][]{2001boehringer}, Massive Cluster Survey \citep[MACS,][]{2001Ebeling}, Clusters In the Zone of Avoidance \citep[CIZA,][]{2002Ebeling}, and several other catalogs compiled in more recent years \citep[see, e.g,][]{Piffaretti2011,2014Boehringer,2017boehringer,2019Klein,2020Finoguenov}, contain a few thousand clusters in total, reaching a flux limit of $\sim 10^{-12}~{\rm erg}~{\rm s}^{-1}~{\rm cm}^{-2}$. However, the sensitivity of ROSAT limits its capability of detecting high-redshift and low-mass clusters. In the past ten years, the Sunyaev–Zeldovich (SZ) effect surveys, such as South Pole Telescope \citep[SPT,][]{2015Bleem}, Atacama Cosmology Telescope \citep[ACT,][]{2013Hasselfield}, and {\sl Planck} \citep{planck2014}, have also made a remarkable contribution to increase the ICM-based cluster sample. Nevertheless, the thermodynamical and chemical properties of the SZ selected clusters also need to be measured based on X-ray follow-up observations. Moreover, X-ray surveys are more sensitive in detecting low-mass nearby clusters compared to other bands such as SZ and optical, and they are less affected by projection effects. It is therefore necessary to establish a larger and deeper X-ray cluster sample that effectively extends to high-redshift and low-mass regimes. 

In the years after RASS, a number of small- and medium-area X-ray surveys have been conducted based on {\sl XMM-Newton} and {\sl Chandra} \citep[see, e.g.,][]{2007Hasinger,2007Cappelluti,2007Finoguenov,2007Pierre,2010Finoguenov,2012Clerc,2015Finoguenov,2016Pierre,2019Gozaliasl,2021Koulouridis}. Although the sky coverage of most of these surveys is smaller than 100 square degrees, hundreds of clusters have been detected, reaching fluxes of $10^{-15}~{\rm erg}~{\rm s}^{-1}~{\rm cm}^{-2}$, because of the high sensitivity of the instruments and the sufficient depth of the surveys. With well-understood selection functions and extensive follow-up observations, these samples play an important role in the study of cluster physics and the constraints of cosmology \citep[see, e.g.,][]{2016Pacaud,2017Ridl,2018Adami,2018Pacaud} and fill the gap between RASS and the next generation of X-ray all-sky survey. As the successor of ROSAT, one of the main design goals of eROSITA is to provide a larger X-ray selected sample of clusters. eROSITA is expected to detect $\sim 10^5$ clusters in the complete all-sky survey \citep{Merloni2012,Pillepich2012}, up to redshifts $z\sim1.5$, down to masses of $M_{500}\sim10^{13}M_{\odot}$, and reaching a flux limit of $f_{0.5-2~{\rm keV}}\sim10^{-14}~{\rm erg}~{\rm s}^{-1}~{\rm cm}^{-2}$. This cluster catalog will inevitably be a valuable resource for testing and constraining cosmological models with the aim to trace the evolution of the large-scale structure and to study cluster-related astrophysics. 

A proof-of-concept mini-survey, the eROSITA Final Equatorial-Depth Survey (eFEDS), was designed to demonstrate the survey science capabilities of eROSITA. The observations of the eFEDS field were performed between November 4 to 7, 2019, during the performance verification phase. The eFEDS field is located at 126$^{\circ}$ < R.A. < 146$^{\circ}$ and $-3^{\circ}$ < Dec. < $+6^{\circ}$, covering a solid angle of approximately 140 square degrees (137.9~deg$^2$ have a vignetting-corrected exposure time in the 0.5--2~keV band of 0.1~ks or more), with the similar depth as the full eROSITA All-Sky Survey in the equatorial regions. The average exposure times are $\sim2.2$ ks and $\sim1.2$ ks before and after correcting for vignetting effects. The eFEDS field has also been observed with a broad array of multiwavelength survey instruments from optical to radio bands. In particular, the photometric data from the Hyper Suprime-Cam (HSC) Subaru Strategic Program
\citep[HSC-SSP;][]{HSC1stDR,HSC1styrOverview,Miyazaki18HSC,Komiyama18HSC,Kawanomoto18HSC,Furusawa18HSC,Bosch18HSC,Haung18HSC,Coupon18HSC,CAMIRA-HSC,HSC2ndDR}, DECaLS \citep[Dark Energy Camera Legacy Survey,][]{2019Dey}, SDSS \citep[Sloan Digital Sky Survey,][]{SDSS}, 2MRS \citep[2MASS Redshift Survey,][]{2MRS}, and GAMA \citep[Galaxy And Mass Assembly,][]{GAMA} surveys are used for optical confirmation and redshift determination of the clusters, and HSC, in particular, will provide weak-lensing masses for eFEDS clusters \citep[][]{Klein2021,chiu2021}. A combination of these with the X-ray properties measured with eROSITA data will enable the calibration of scaling relations between X-ray observables and cluster halo mass \citep[][Bahar et al., in prep]{Ghirardini2021c}. Being the largest contiguous survey at the final Equatorial depth, it provides an ideal setup for testing the predictions for the cluster number density in the survey and offers rich cluster science through its multiwavelength coverage \citep[e.g.,][]{2021Ghirardini,Pasini2021}. 

In this paper, we present the catalog of candidate galaxy clusters and groups detected in eFEDS, and provide the first results of the X-ray analysis on these clusters based on eROSITA data. 
More detailed studies of the eFEDS clusters will also be presented in a series of accompanying and forthcoming papers. \citet{Klein2021} present the optical identification of the eFEDS cluster candidates. \citet{chiu2021}, Ramos et al. (in prep.), and Ota et al. (in prep.) perform an optical and weak-lensing analysis on the eFEDS clusters. Radio properties of the clusters are studied in \citet{Pasini2021}. ICM morphology and X-ray scaling relations based on eROSITA data are studied in \citet{Ghirardini2021c} and Bahar et al. (in prep.), respectively. Spectroscopic follow-up results will be provided in Ider Chitham et al. (in prep.). The catalog of cluster candidates that are misidentified as point sources will be presented in Bulbul et al. (in prep.). Moreover, the main eFEDS X-ray source catalog is provided in \citet{Brunner2021}. The eFEDS simulation results are introduced in detail in \citet{LiuT2021}.

The paper is organized as follows. In Sect.~\ref{sec:catalog} we describe the eROSITA observations of the eFEDS field and the source detection in detail and discuss the potential contamination in the sample. We also summarize in this section the optical confirmation and redshift determination of the clusters on the basis of optical photometric and spectroscopic survey data. In Sect.~\ref{sec:slctfnc} we examine the selection function of the cluster sample. In Sect.~\ref{sec:observables} we provide the details of the X-ray data analysis we performed in this work and of the X-ray observables. In Sect.~\ref{sec:xlf} we compute the X-ray luminosity function of the cluster sample. In Sect.~\ref{sec:sc} we perform a search for superclusters in the eFEDS field on the basis of the spatial distribution and redshifts of the clusters. Our conclusions are summarized in Sect.~\ref{sec:conclusions}. Throughout this paper, we adopt the concordance $\Lambda$CDM cosmology with $\Omega_{\Lambda} =0.7$, $\Omega_{\mathrm m} =0.3$, and $H_0 = 70$~km~s$^{-1}$~Mpc$^{-1}$. Quoted error bars correspond to a 1$\sigma$ confidence level unless noted otherwise.

\section{eFEDS extended source catalog}
\label{sec:catalog}

\subsection{eROSITA observation and data calibration}
\label{sec:data_detec}
The data are processed with the eROSITA Standard Analysis Software System \citep[{\tt eSASS},][]{Brunner2021}\footnote{version {\tt eSASSusers\_201009}.}. The details of the data reduction and calibration are described in \citet{Brunner2021} and \citet{2020Dennerl}. We here provide a short summary of the analysis steps. Pattern recognition and energy calibration are applied for all the seven eROSITA telescope modules (TMs) to produce calibrated event lists. The event lists are then filtered after the determination of good time intervals, dead times, corrupted events and frames, and bad pixels. Using star-tracker and gyro data, we assign celestial coordinates to the reconstructed X-ray photons, which can then be projected on the sky so that images and exposure maps can be produced. We select all valid pixel patterns here, that is, single, double, triple, and quadruple events, but use only photons that are detected at off-axis angles $\le$ 30$\arcmin$. By doing this, we remove the photons in the corners of the square CCDs where the vignetting and point spread function (PSF) calibrations are currently less accurate.

\subsection{Source detection strategy}
\label{sec:source_detection}
The details of the source detection for eFEDS are presented in \citet{Brunner2021}. We briefly summarize the compilation of the extended source catalog.
The source detection was performed using the tool {\tt erbox} in \texttt{eSASS} on the merged 0.2--2.3~keV image of all seven eROSITA TMs. {\tt erbox} is a modified sliding-box algorithm that searches for sources on the input image that are brighter than the expected background fluctuation at a given image position. The source detection procedure contains the following steps. As the first step, {\tt erbox} is applied to scan the X-ray image with a local sliding window and returns a list of candidate sources that are enhancements with respect to the  background above a certain threshold. The background is interpolated from a frame-shaped region around the detection window. The candidate sources identified in the first step are then excised from the original image. The resulting source-free image is then used to create a background map through adaptive filtering using the {\tt erbackmap} tool in \texttt{eSASS}. The source detection with {\tt erbox} is then repeated, but using the new background map created in the last step, producing a new list of candidate sources. The {\tt erbox+erbackmap} iteration is run three times to enhance the reliability of the background map and the sensitivity of the detection algorithm.

In the second step, the source parameters for each candidate, such as the detection likelihood, the extent likelihood, and the extent, are determined by fitting the image with the source model, which is a $\beta$-model convolved with the calibrated PSF, in which $r_c$ equals to the extent of the source, and is set free to vary between 8\arcsec and 60\arcsec for extended sources. This step is performed using the {\tt ermldet} tool in \texttt{eSASS}. 

We applied these source detection procedures on the 0.2--2.3~keV image of the eFEDS field. Setting the minimum detection likelihood, $\mathcal{L_{\rm det}} \equiv$ -ln($P$), at 5 and the minimum extent likelihood $\mathcal{L_{\rm ext}}$ at 6, we detect 542 candidate extended sources over the full eFEDS field. This corresponds to an extended source density of $\text{about four sources}$~ per square degree at the equatorial depth. The eROSITA image of the eFEDS field in which the extended sources are highlighted is shown in Fig.~\ref{fig:image_efeds_large}. The distribution of the sources in the field with redshift information (see Section~\ref{sec:optical}) is shown in Fig.~\ref{fig:cat}. 

\subsection{Optical confirmation and redshift determination }
\label{sec:optical}

\begin{figure}
\begin{center}
\includegraphics[width=0.49\textwidth, trim=0 0 35 30, clip]{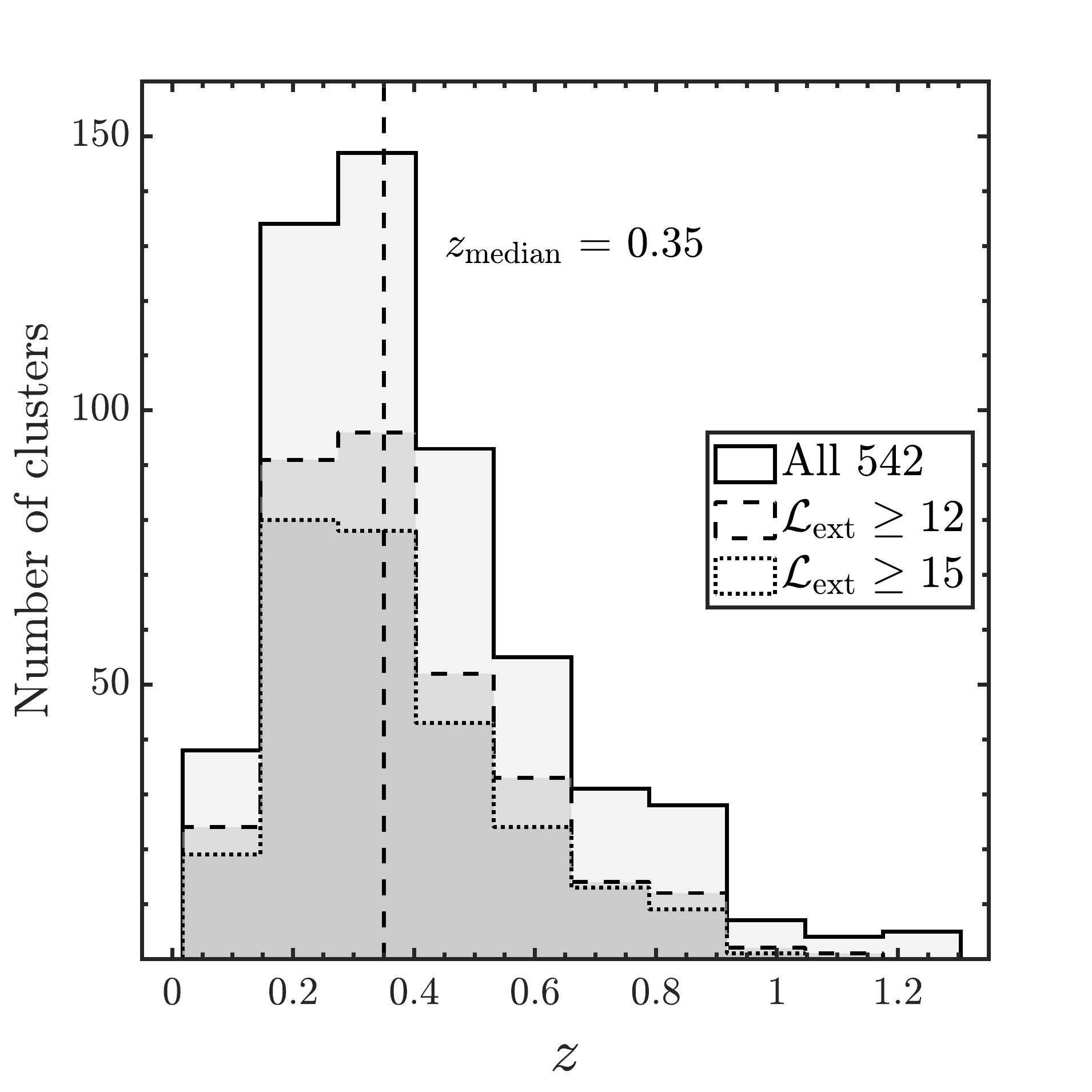}
\caption{Redshift histogram of the 542 cluster candidates in the redshift range of 0.01 to 1.3. The median redshift of the sample $z_{\rm median}$=0.35 is shown as the vertical dashed line. We also plot the results after cutting the sample with different thresholds on the extent likelihood $\mathcal{L}_{\rm ext}$.
}
\label{z_hist}
\end{center}
\end{figure}

The photometric redshifts of the clusters are determined based on the multicomponent matched filter (MCMF) cluster confirmation tool \citep[see][for more details]{2018Klein,2019Klein}. The MCMF tool takes optical photometric datasets and searches for galaxy overdensities along the line of sight at the position of a cluster candidate. To do this, the galaxy cluster richness is measured as a function of redshift using a red-sequence technique and within an aperture defined from the X-ray count rate. Peaks in richness versus redshift space are fit by specific peak profiles, and redshift and richnesses are recorded for multiple possible counterparts along the line of sight. MCMF is then run on random lines of sight excluding regions around X-ray candidates. 
MCMF on eFEDS was run on two photometric datasets with different optical filters and depth: DECaLS $g, r, z$ and unWISE W1 bands, and HSC $g, r, i, z$ bands. DECaLS is free from strong calibration issues and provides full coverage of the eFEDS field, but is $\sim2$ magnitudes shallower than HSC-SSP data. The deep HSC-SSP data in the $g, r, i, z$ bands provide good photometric redshifts out to high redshifts ($z\sim1.3$), where DECaLS has shallower data and misses the $i$ band. The results of both MCMF runs were then combined to the final catalog. 
Spectroscopic redshifts, on the other hand, were then derived by cross-matching the clusters with public spectroscopic redshifts including SDSS up to DR16 \citep{SDSS}, GAMA \citep{GAMA}, and 2MRS \citep{2MRS} and requiring either three or more redshifts consistent with the photo-$z$ estimate and within $R_{500}$ \citep[estimated using the relation between X-ray count rate and mass, see][]{Klein2021} or a spectroscopic redshift of the brightest cluster galaxy (BCG). We note that the searching radius of $R_{500}$ is large enough because it is well beyond the typical offsets between the X-ray and optical center or the BCG of galaxy clusters \citep[e.g.,][]{2020Seppi}. In total, we provide spectroscopic redshifts for 297 clusters. 

In Fig.~\ref{z_hist} we plot the histogram of the 542 redshifts. The redshift of the clusters in the sample ranges from 0.01 to 1.3, and the median value is $z_{\rm median} = 0.35$. The redshift distribution is very close to the prediction of \citet{Pillepich2012} when a mass cut at $M_{500}=5\times10^{13}~M_{\odot}$ is assumed (see their Fig.~3). 

In our companion paper on the optical confirmation and redshifts \citep{Klein2021}, we provide a detailed
description of the follow-up and analysis of its results. We refer to that paper for more details.

\subsection{Crossmatch with published X-ray and SZ cluster catalogs}
The eFEDS field is also covered by other X-ray and SZ surveys. We therefore matched our catalog to the published cluster catalogs. For X-ray clusters, we used the MCXC \citep[Meta-Catalogue of X-ray detected Clusters of galaxies,][]{Piffaretti2011} and the CODEX \citep[COnstrain Dark Energy with X-ray clusters,][]{2020Finoguenov}, both of which are mainly based on the ROSAT all-sky survey (MCXC also contains clusters from ROSAT pointed observations and {\sl XMM-Newton} observations). For SZ clusters, we used the most recent ACT-DR5 cluster catalog \citep{2021Hilton} and the PLANCKSZ2 catalog \citep{2016planckb}. The matching distance was determined as 2\arcmin\ for X-ray catalogs and the ACT catalog and 3\arcmin\ for the PLANCKSZ2 catalog. We do not present the crossmatch results with optical cluster catalogs here because the selection of clusters in optical surveys is quite different from that of ICM-based surveys \citep[see, e.g.,][]{2012Wen,2016Rykoff}. The results of the match are listed in Table~\ref{match}. In summary, we find 1 and 43 matches with the MCXC and CODEX catalogs and 10 and 57 matches with the PLANCKSZ2 and ACT catalogs, also including multiple-to-one matches. After removing the cases that were matched in multiple catalogs, we find 86 matches in total between the 542 eFEDS clusters and these published catalogs, corresponding to $\sim16\%$ of the whole sample. This indicates that the majority of the clusters are detected for the first time in ICM-based surveys. The redshift comparisons of the common clusters are given in \citet{Klein2021}.

We briefly compare the luminosity ($L_{500}$ in the 0.1--2.4~keV band) provided in the CODEX catalog and measured in this work (see Sect.~\ref{sec:observables} for the measurement of luminosity) for the clusters in common. Out of the 43 common clusters, we find 13 clusters with a luminosity difference larger than $2\sigma$.  CODEX measured a higher luminosity than our results for all these 13 clusters. Clearly, this disagreement can be due to many possible reasons. First, the redshifts are different: in 4 out of the 13 clusters, the redshifts we adopt in this work are different by $>15\%$ from those used in the CODEX catalog. Second, the centers of the clusters and the radii on which the luminosity was measured differ strongly in the two catalogs: the offsets in the center of the 13 clusters range from $\sim$30\arcsec to 2\arcmin.  eROSITA has a better angular resolution than ROSAT, therefore the luminosity measured with ROSAT for these 13 clusters might be biased by unresolved active galactic nuclei (AGN). However, we did not further compare the measurements of X-ray observables in the two catalogs because this would require revisiting the details in the analysis of the CODEX clusters, which is beyond the goal of this paper.

\begin{table}
\caption{Results of crossmatching with published cluster catalogs. }
\label{match}
\begin{center}
\begin{tabular}[width=0.5\textwidth]{lcccc}
\hline\hline
Catalog & Survey type & Ref. & $d_{\rm match}$ & $N_{\rm match}$ \\
\hline
MCXC  & X-ray & (1) & 2\arcmin & 1 \\
CODEX  & X-ray & (2) & 2\arcmin & 43 \\
PLANCKSZ2  & SZ & (3) & 3\arcmin & 10 \\
ACT  & SZ & (4) & 2\arcmin & 57 \\
\hline
\end{tabular}
\tablefoot{References: (1) \citet{Piffaretti2011}, (2) \citet{2020Finoguenov}, (3) \citet{2016planckb}, (4) \citet{2021Hilton}. }
\end{center}
\end{table}
\subsection{Contamination in the catalog}
\label{sec:contami}
\begin{figure}
\begin{center}
\includegraphics[width=0.499\textwidth, trim=0 90 20 95, clip]{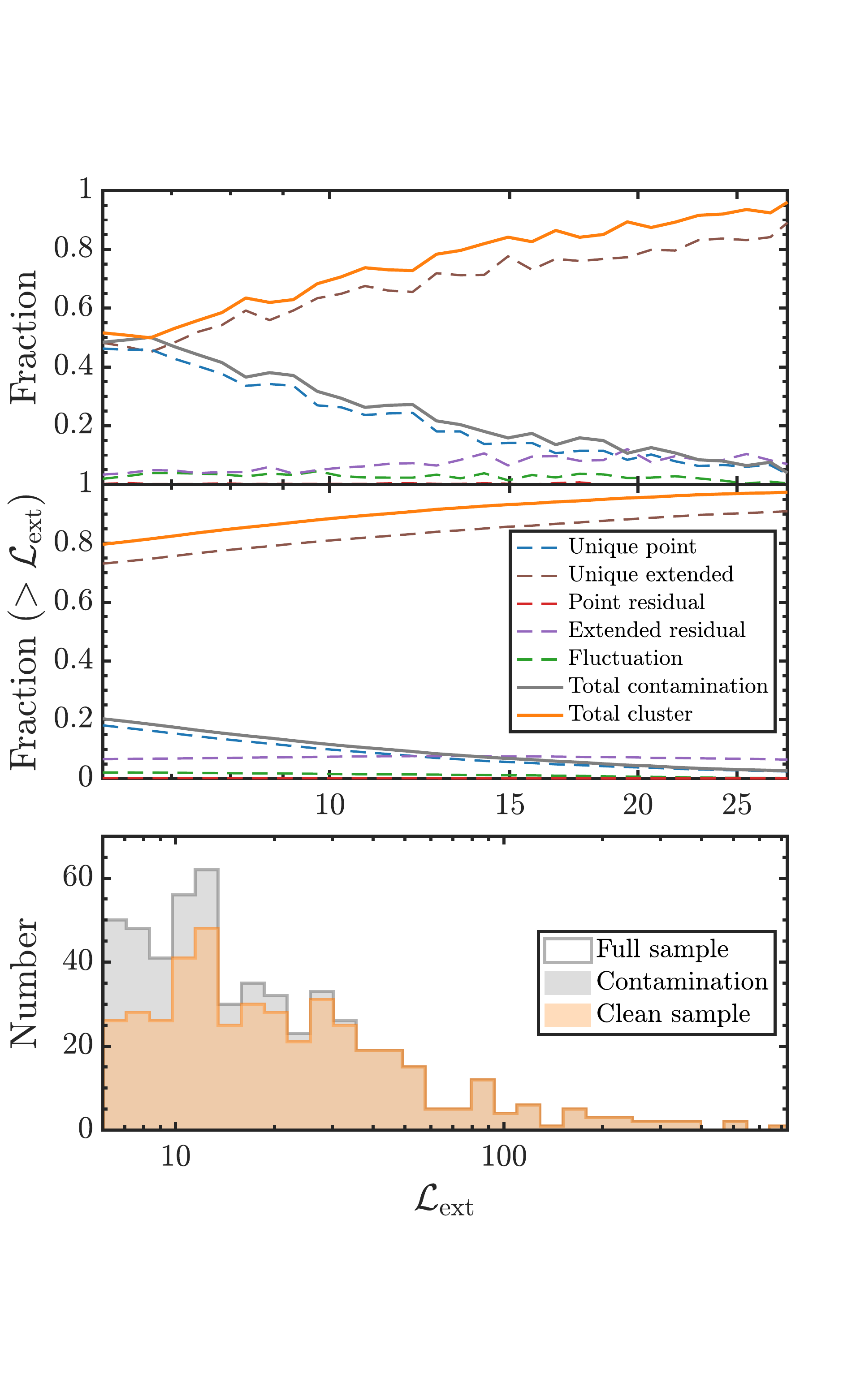}
\caption{Contamination fraction assessment in our sample based on simulations. {\sl Upper panel}: Cumulative detection fraction of sources in the five classes as a function of extent likelihood. The orange solid line shows the total fraction of sources classified as real clusters, namely, the sum of unique extended source and extended source residual. The total contamination is plotted with the solid gray line based on summing all the other three classes. {\sl Middle panel}: Cumulative detecting fraction of sources in the five classes. {Lower panel}: Distribution of extent likelihood of the clusters in our catalog. The clean sample after removing the fraction of contamination is shown in orange. The gray bar represents the contamination in each bin. }
\label{fig:simu}
\end{center}
\end{figure}

We used a set of realistic simulations of the eFEDS field using {\tt sixte-2.6.2}\footnote{https://www.sternwarte.uni-erlangen.de/research/sixte/} \citep{2019Dauser} and {\tt simput-2.4.10} to assess the contamination fraction in our catalog. The details of these simulations and their results are provided in \citet{LiuT2021} and \citet{Brunner2021}. The source detection and matching were performed on the simulated field in the same way as described in detail in  Sect.~\ref{sec:source_detection}. Here we use the simulations to estimate the total contamination in the catalog.

We divided the detected sources into the following five classes.
Class 1 includes unique point sources, which are classified as counterparts of input AGN or stars. Class 2 includes unique extended sources, which are classified as counterparts of input extended sources.
Class 3 consists of point source residuals. In this case, the input point source associated with the detected source has already been detected as a point source or an extended source. However, a part of its signal, likely in the outer wing of the PSF that cannot be perfectly fit by the source detection algorithm, is detected as another point source. 
Class 4 sources are extended source residuals, which are similar to class 3, but the input source is an extended source. The input extended source has already been detected, but a part of its photons, which probably represents the signal from substructures or fluctuations in its outskirts, is detected as another extended source. Finally, class 5 includes sources without input counterparts, and classified as spurious sources due to background fluctuation. 

We plot the detection fraction of the sources in each class as a function of the extent likelihood in Fig.~\ref{fig:simu}. A cut at $\mathcal{L_{\rm ext}} > 6$ is applied to be consistent with our catalog. In Fig.~\ref{fig:simu} we consider classes 1, 3, and 5, the unique point source, point source residual, and background fluctuation, as contamination sources (i.e., noncluster sources) in the catalog. Classes 2 and 4, the unique extended source and extended source residual, are classified as real clusters. In particular, we remark that we considered class 4 as real extended sources instead of  spurious detections because substructures in cluster outskirts are commonly observed and it is reasonable to identify them as separated sources. 
The total fractions of contamination sources and clusters are also plotted in Fig.~\ref{fig:simu} as solid lines. The top panel of Fig.~\ref{fig:simu} shows that the contamination fraction can reach $\sim50\%$ at $\mathcal{L}_{\rm ext} = 6$ and decreases to $<20\%$ at $\mathcal{L}_{\rm ext} = 15$. We also found that the contamination in the catalog is dominated by class 1, namely, the unique point sources, which are mostly AGN that are misidentified as extended sources. This is consistent with our previous simulation results \citep[see][Fig.~9]{2018Clerc}, which show that sources with a low extent and a low extent likelihood have a higher probability of being point sources that are misidentified as extended. Combining the result with the distribution of $\mathcal{L}_{\rm ext}$ of our sample (see the middle and lower panels of Fig.~\ref{fig:simu}), we find that the total contamination fraction in our sample is $\sim1/5$, corresponding to a total number of $\sim110$ noncluster sources. Because most of these noncluster sources have low $\mathcal{L}_{\rm ext}$, they can be excluded by setting an $\mathcal{L}_{\rm ext}$ threshold much higher than 6. For example, a simple cut at $\mathcal{L}_{\rm ext}\ge 15$ delivers a subsample of $\sim270$ clusters with a purity $>$90\%, even though the sample volume is unavoidably reduced. We list in Table~\ref{tab:table2} the properties of the subsamples corresponding to different thresholds of $\mathcal{L}_{\rm ext}$. The redshift distributions of the subsamples are plotted in Fig.~\ref{z_hist}. We remark that with the current eFEDS X-ray data alone, we are not able to perfectly clean the sample without significantly decreasing the sample volume. Deeper X-ray observations and multiwavelength follow-ups are needed to further clean the sample. 
For example, \citet{Klein2021} presented an approach to clean the sample that uses the optical information. This approach is found to remove a significant part of the contaminants. We refer to this paper for more details about the results and discussion of the optical cleaning of the sample. 

We stress that our selection function is built based on the simulations of the X-ray sky coupled with the standard extended source detection procedure. Currently, the selection function does not include any information about the optical cleaning. The modeling of the total selection function including both X-ray and optical information and cross-talk between them is being developed and subject to future work (Clerc et al. in prep.). To be consistent with our selection function, we include all the 542 cluster candidates in our further analysis in this paper unless noted otherwise. We recommend using the X-ray extent and detection likelihood selections where the provided selection function is used for sample studies.

\begin{table*}
    \centering
    \caption{Properties of the subsamples corresponding to different thresholds of $\mathcal{L}_{\rm ext}$. }
    \label{tab:table2}
    \begin{tabular}{lccccc}
        \hline\hline
        Selection  & Number of clusters & $z_{\rm median}$ & Flux limit (0.5--2~keV, 1~arcmin) & Completeness & Purity \\ \hline
        Full sample & 542 & 0.35 & 10$^{-14}$~erg~s$^{-1}$~cm$^{-2}$ & 40\% & 80\% \\
        $\mathcal{L}_{\rm ext}\ge 12$ & 325 & 0.34 & $1.7\times10^{-14}$~erg~s$^{-1}$~cm$^{-2}$ & 44\% & $>$85\%  \\
        $\mathcal{L}_{\rm ext}\ge 15$ & 267 & 0.33 & $2\times10^{-14}$~erg~s$^{-1}$~cm$^{-2}$ & 47\% & $>$90\%  \\
        \hline\hline
    \end{tabular}
    \tablefoot{The completeness corresponds to the flux limit for each subsample. Purity is estimated from the simulation results in \citet{LiuT2021}. }
\end{table*}

\section{Selection function}
\label{sec:slctfnc}
This section describes how the cluster sample completeness and the parameters it depends on were inferred. Simulations of the eFEDS field accounting for the instrument response function and the scanning strategy are described in \citet{LiuT2021}. This relies on realistic methods to simulate the emission of galaxy clusters and AGN \citep{2019Comparat, 2020Comparat}.

\subsection{eFEDS field simulations}
\label{sec:sims}
We recall here the main steps for producing the simulations. Eighteen independent realizations of the field were simulated in order to increase the statistics in the selection function derivation. A full-sky light cone of dark matter halos was created based on the numerical N-body simulation \citep[UNIT 1 inverse,][]{2019Chuang}, which assumes a Planck-CMB cosmology \citep{2016Planck}. The halos were associated with X-ray emitting sources, namely AGN and galaxy clusters. The model of the X-ray emitting black hole population is empirical and inherits from the halo abundance matching technique, which is particularly efficient at reproducing the observed stellar mass, the luminosity and specific accretion rates distributions \citep{2019Comparat}. AGN fluxes were derived from a set of template spectra folded with a redshift- and luminosity-dependent obscuration model. For halos and subhalos with masses $M_{500c}$ above $5\times 10^{13}~M_{\odot}$, cluster images were drawn from a library of emission measure profiles inferred from actual datasets, leading to a reproduction of the observed scaling relations between mass, luminosity, and temperature \citep{2020Comparat}. The assignment of halos and profiles depends on their mass, redshift, and dynamical state; a large positional offset between the dark matter center of mass and its highest density peak will preferentially lead to a low central value of the emission measure profile. An ellipsoidal shape is given to cluster images that follows the halo triaxial shape. Halos of lower masses are simulated with the same method, but we decreased their flux to obtain realistic fluxes for groups. The method suffers from the existing Malmquist bias in the library of profiles we used and creates groups that are too bright.
The source spectra are absorbed by an amount depending on the local Galactic absorption column density. Stellar sources as well as cosmic X-ray background and emission from the Galaxy are accounted for using measurements acquired during the calibration and performance verification phases. The SIXTE simulation software \citep{2019Dauser} is fed with the actual spacecraft attitude file during the eFEDS mapping, hence the simulations faithfully reproduce the exposure variations in the field. The resulting event lists are processed similarly to the true eFEDS event lists, see Sect.~\ref{sec:data_detec}. In particular, extended sources are identified using identical thresholds in detection and extent likelihood as in the real catalog.
\subsection{Finding source counterparts}
Matching the input galaxy cluster catalogs and the extended source lists poses a challenge because the sky density of the input galaxy cluster catalogs is higher than that of the extended source lists. Fixing a matching radius or performing a nearest neighbor search might lead to unrealistic matches because of projection effects. In order to mitigate this issue, we applied a Bayesian matching procedure that is based on the NWAY software and formalism \citep{2018Salvato}. Cross-correlation between the twocatalogs avoids specifying an explicit scale length, except for a maximum radial search of $3\arcmin$ around each source, which is set to save computational cost. We accounted for angular extents by artificially increasing the positional uncertainties of input and detected sources. These uncertainties were set to 10\% of a cluster virial radius and to the best-fit source extent. The NWAY algorithm was run twice, according to whether the input cluster list or the detection list was considered as the parent catalog. During the two independent runs, a probability value $p_i$ for an association to be valid was assigned to each pair composed of an input cluster and a detected source. An additional value $p_{\rm any}$ representing the probability for a source from the parent catalog to have at least one counterpart in the other catalog was also issued. These probabilities account for chance associations and provide a ranking of the most likely counterparts. Their values depend on the position of the sources, their positional uncertainties, the solid angle of the survey field, and the source density in both catalogs. In order to distinguish complex cases involving multiple substructures, heavy projection effects, or substantial source splitting by the detection algorithms, we introduced a prior on the true flux distribution of detected sources. This prior updates the values of $p_i$ and $p_{\rm any}$. It was obtained by multiplying a loose version of the selection functions presented in \citet{2018Clerc} by the input flux distribution. When this prior was applied, NWAY favored brighter sources when it searched for counterparts to a detected source in the input cluster catalog. Finally, thresholds on $p_{\rm any}$ and $p_{i}$ (two sets of thresholds for both runs) enabled selecting valid matches. The final list of matches is found to be rather insensitive to the exact value of these thresholds: many of the $p_{\rm any}$ values are distributed either around zero or around one, and choosing $p_{i} > 0.1$ proved to be efficient in selecting the correct counterpart.
Valid matches with the highest $p_{i}$ among all possible associations were denoted as primary matches; the matches that were primary in both runs were considered solid matches between the input and detection lists.
\subsection{Training classifiers}
The main outcome of the matching procedure is a list of simulated clusters with associated properties and a flag indicating whether a source is reliably matched to a detection. These properties are either intrinsic to a cluster (e.g., luminosity, redshift, and central emission measure) or extrinsic (e.g., local exposure time and absorbing column density). We reformulated the problem of cluster selection into a binary classification problem and explored two methods to predict the detection probability of a cluster as a function of its properties.

The first set of selection functions relies on a random forest algorithm \citep[e.g.,][]{2001Breiman}. We used the implementation present in the scikit-learn package\footnote{Version 0.23.2} \citep{2011Pedregosa}. Properties attached to clusters or features were normalized, and their one-dimensional distribution was rendered as flat as possible through histogram equalization. A subsample from the initial list was created by randomly selecting two-thirds of the detected and undetected clusters. The remaining sources were kept for testing and evaluation purposes. The random forest classifier randomly drew 1001 decision trees and averaged their results in order to provide a low-variance estimator of the probability of detection, $p_{\rm det}$. Each tree was built by sampling the initial list of clusters with replacements. The thresholds at each node of a tree were chosen by minimizing the Gini impurity. The uncertainty on the value of $p_{\rm det}$ was obtained with a custom implementation of the unbiased infinitesimal jackknife variance estimator proposed by \citet{2014Wager}.
Various algorithm parameters were tested and optimized by balancing execution time and classifier scores. These scores measure the fractional number of objects in the validation subsample that are correctly classified (with $p_{\rm det}=0.5$ selected as classification threshold). Scores of our classifiers typically reach 80--90\%, depending on the nature and the number of selected features. In order to prevent the classifier from extrapolating out of the training zone, we used a simple convex-hull algorithm \citep{1996Barber}.

The selection of features relevant for cluster selection functions was based on practical concerns as well as on careful examination of the classifier outcome. Of the variables we tested, the cluster (true) 0.5--2~keV X-ray counts, its flux in the same band, the redshift, mass, and luminosity were found to be the most discriminating, as expected from earlier sensitivity studies. We re-formed our choice of parameters by examining the feature importance of the random forest classifier, and we found that the central emission measure as well as the local exposure time play additional roles in the selection.

\begin{figure}
\begin{center}
\includegraphics[width=0.49\textwidth, trim=0 10 0 10, clip]{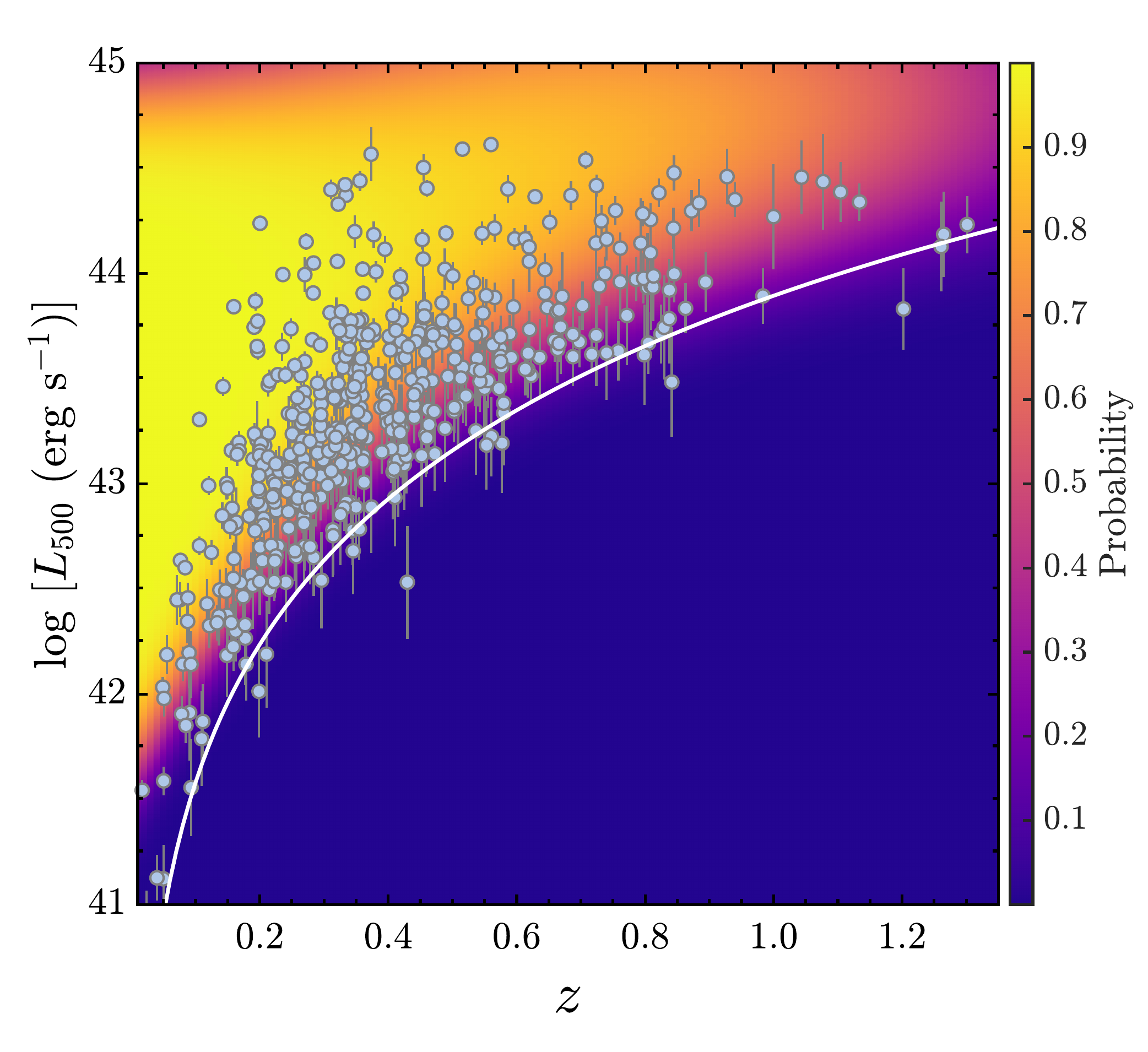}
\caption{Detection probability as a function of luminosity $L_{500}$ in the 0.5--2~keV band and redshift $z$. The data points with error bars are the $L_{500}$ for the clusters with $>2\sigma$ measurements. Details of the computation of $L_{500}$ are presented in Sect.~\ref{sec:xlf}. The white curve shows the flux limit at $1.5\times10^{-14}~{\rm erg}~{\rm s}^{-1}~{\rm cm}^{-2}$, corresponding to an average completeness level of $\sim40\%$. }
\label{fig:sel}
\end{center}
\end{figure}

The second set of selection functions relied on Gaussian processes \citep[e.g.,][]{2006Rasmussen}. This classifier constructs a latent variable, modeled as a Gaussian process whose covariance function (kernel) is a squared exponential function. It is transformed into a probability by means of a sigmoid (inverse probit) function, and the algorithm uses the Bernoulli likelihood of the training sample in order to maximize the marginal likelihood and find the best hyperparameters for the kernel, namely its amplitude and scale lengths along each feature dimension. In order to increase speed in execution time, we used the stochastic variational Gaussian process (SVGP) algorithm \citep{2015Hensman} implemented in the GPy library\footnote{GPy, since 2012: A Gaussian process framework in Python; Version 1.9.9; \url{http://github.com/SheffieldML/GPy}}. Among the additional modeling hypotheses, this algorithm imposes ten inducing points to create an auxiliary random process summarizing the latent function. The selection probability $p_{\rm det}$ was obtained by integrating out the latent variable distribution against the link function (the sigmoid); uncertainties on $p_{\rm det}$ were approximated by folding the 1$\sigma$ envelope of the latent function into the link function.
The resulting selection functions appear smoother than those trained with random forests and their uncertainty range is better controlled; however, they cannot capturing small-scale variations in the feature space as well as selection functions trained with random forests. Nevertheless, both flavors of the selection function provide equivalent performances. In Fig.~\ref{fig:sel} we show the selection probability as a function of cluster luminosity and redshift using Gaussian processes. 

\section{X-ray analysis and cluster X-ray properties}
\label{sec:observables}
\begin{figure*}[h!]
\begin{center}
\includegraphics[width=0.495\textwidth, trim=-6 20 40 40, clip]{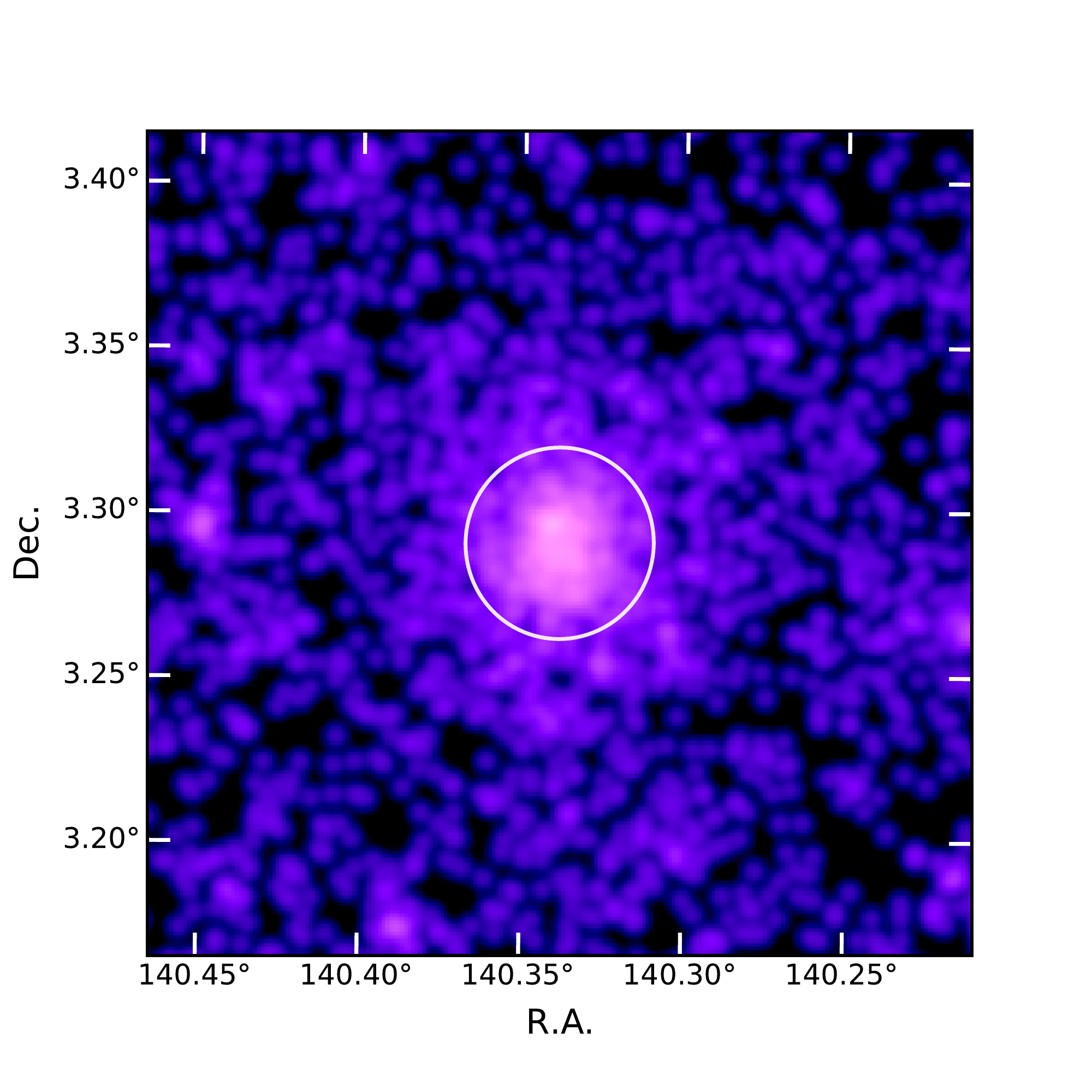}
\includegraphics[width=0.495\textwidth, trim=-6 20 40 40, clip]{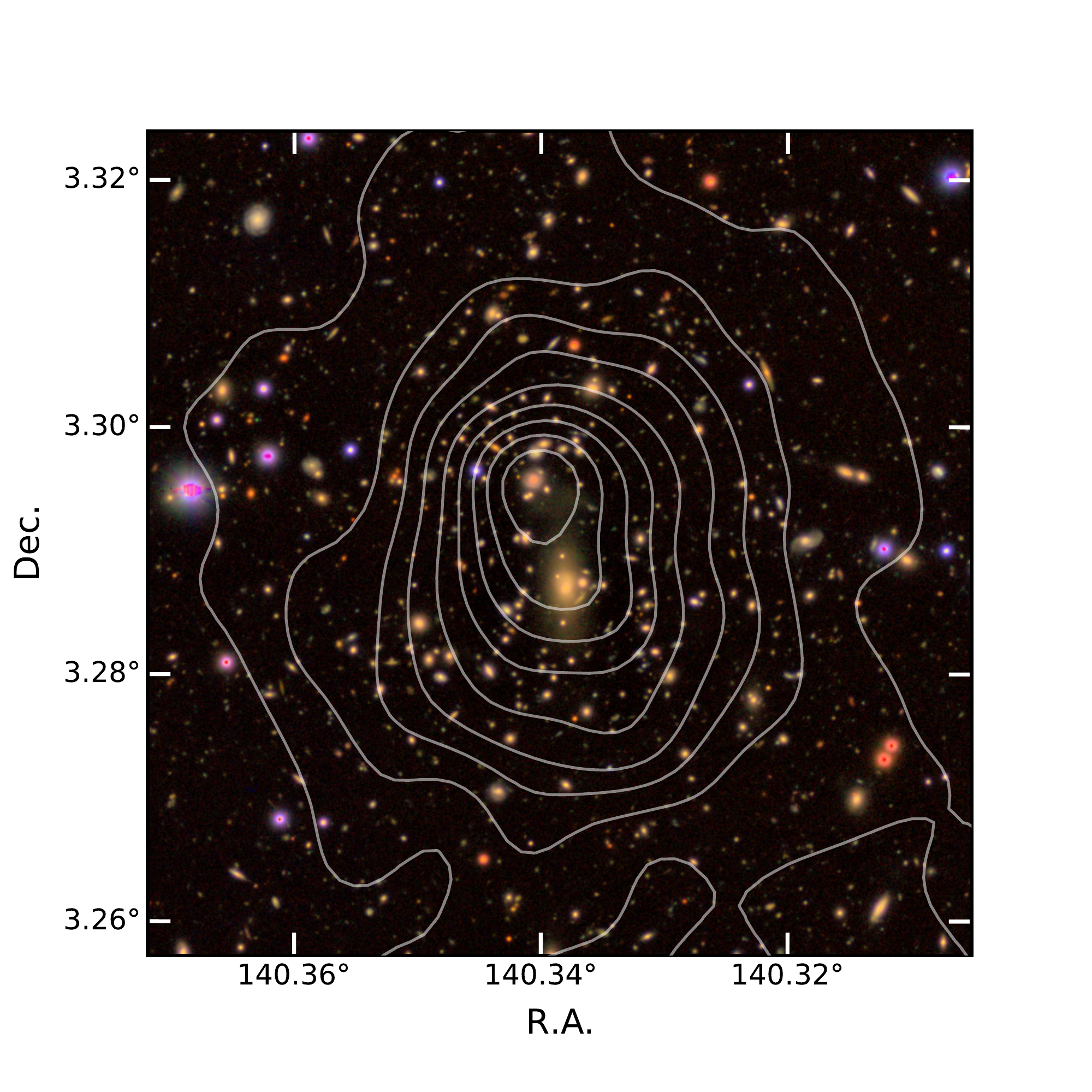}
\includegraphics[width=0.99\textwidth, trim=10 0 30 10, clip]{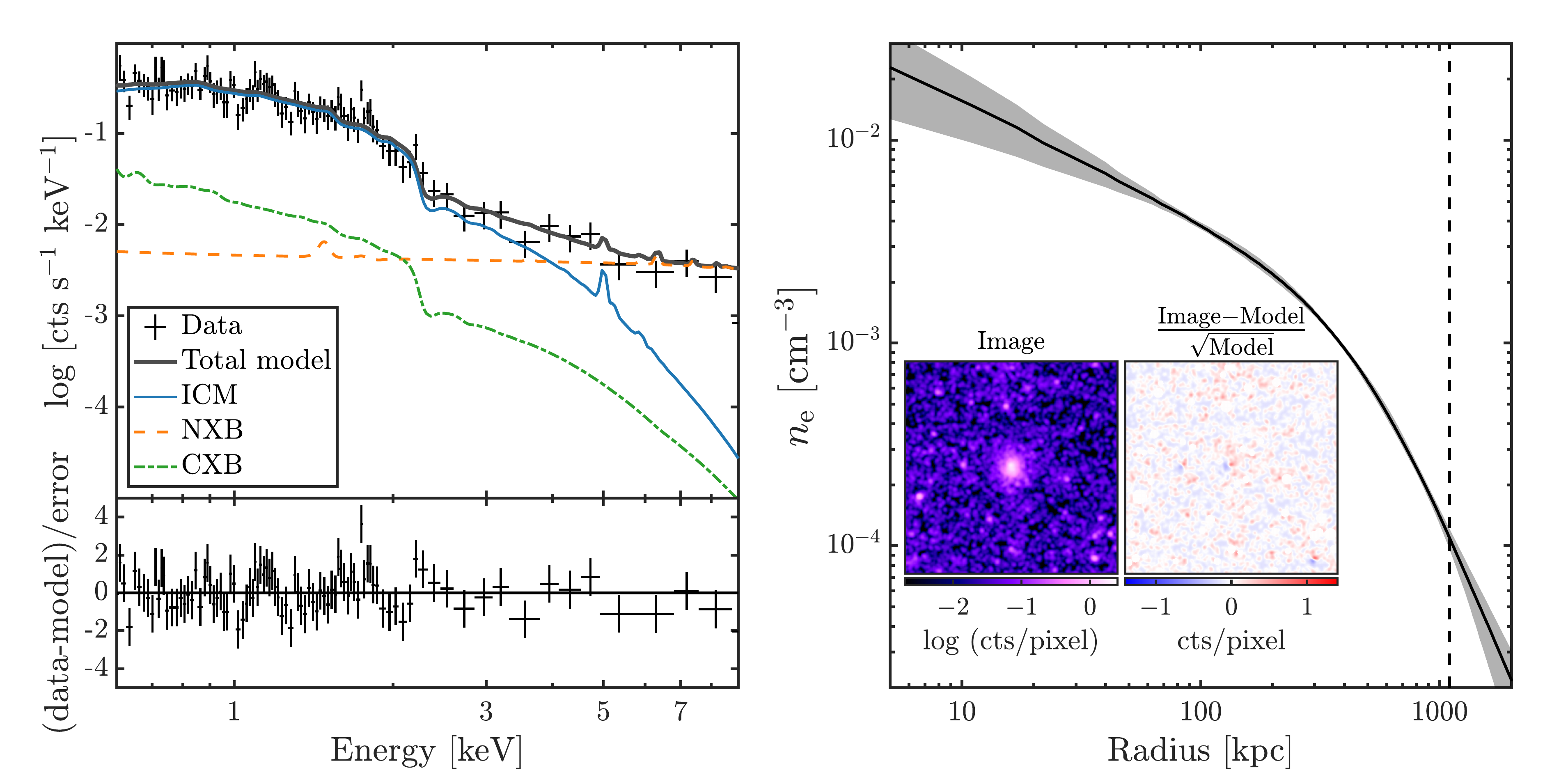}
\caption{Results of imaging and spectral analysis for cluster eFEDS~J092121.2+031726 at redshift 0.333 (spectroscopic) as an example. This cluster is detected with $\mathcal{L}_{\rm ext}=478.6$ and $\mathcal{L}_{\rm det}=1729.8$ and has one of the highest S/Ns in our sample. The temperature and soft-band luminosity within 500~kpc are $5.2_{-0.8}^{+1.3}~{\rm keV}$ and $2.14_{-0.08}^{+0.08}\times 10^{44}~{\rm erg}~{\rm s}^{-1}$. {\sl Upper left}: Soft-band (0.5--2~keV) eROSITA image. The white circle marks the region of $r=$500~kpc. The image is smoothed with a Gaussian with FWHM=12$\arcsec$. {\sl Upper right}: {\sl Subaru} HSC image of the central region in the ($z, i, r$) bands. Superimposed in white are the X-ray contours. {\sl Lower left}: Spectrum within the 500~kpc region. The spectra and the corresponding responses of the seven TMs are merged for clarity. The total model, ICM model, non X-ray background, and cosmic X-ray background are plotted separately.  {\sl Lower right}: Electron density profile obtained from the spectral and imaging analysis. The dashed vertical line shows the $R_{500}$ of this cluster, 1.1~Mpc, estimated from the $L-M$ scaling relation. The inset images show the result of 2D image fitting. The original soft-band image is on the left, and the residual image is on the right.  }
\label{fig:example}
\end{center}
\end{figure*}
In this section we describe the X-ray analysis of the 542 cluster candidates we performed on the basis of the performance verification phase eROSITA observations. The analysis mainly consists of two parts, imaging analysis and spectral analysis. These parts aim at providing a measurement of the average cluster temperature and the radial profiles of other observables such as flux, luminosity, electron density, and gas mass. As an example, the image, spectrum, and the measured electron density profile of one of the most massive clusters in the field, eFEDS~J092121.2+031726, are shown in Fig.~\ref{fig:example}.

\subsection{Imaging analysis, luminosity, and gas mass}
\label{sec:imaging}

The X-ray imaging analysis for the eFEDS field has been extensively described in \citet{2021Ghirardini}. We only highlight the main steps here. We note that in the imaging analysis of this work, we do not conduct further analyses of the morphology and dynamical status of the clusters. An in-depth study of the morphological parameters of the eFEDS clusters will be performed in \citet{Ghirardini2021c}.

The imaging analysis in this work is based on a direct image fitting, building a cluster model, projecting it onto the plane of the sky, and adding to it the model images of instrumental and sky backgrounds. 
Images and exposure maps (vignetted and unvignetted) were extracted in the soft 0.5--2~keV energy band using the \texttt{eSASS} tools {\tt evtool} and {\tt expmap}.

To model the image for each cluster, we adopted a forward modeling approach. We started from the \citet{vikhlinin2006a} electron number density model, 

\begin{equation}
n_{\mathrm e}^2(r) = n_0^2 \cdot \left( \frac{r}{r_{\mathrm c}} \right)^{-\alpha} \cdot \left( 1 + \left( \frac{r}{r_{\mathrm c}} \right)^2 \right)^{-3\beta+\alpha/2} \cdot \left(  1 + \left( \frac{r}{r_{\mathrm s}} \right)^3 \right)^{-\epsilon/3},
\end{equation}

\noindent where $n_0$ is the normalization factor, $r_{\mathrm c}$ and $r_{\mathrm s}$ are core and scale radii, $\alpha$ controls the slope of the density profile in core and intermediate radii, and $\epsilon$ controls the change of slope at large radii. The priors on our parameters were $\epsilon < 5$ \citep[as suggested by][]{vikhlinin2006a}, $\beta > 1/3$, and $\alpha > 0$, and we froze $r_{\mathrm s} = r_{\mathrm c}$. The number density model was then projected onto the 2D image plane, convolved with the eROSITA PSF and multiplied by the vignetted exposure map. The resulting cluster model image was finally matched to the count image to obtain the best fit,

\begin{equation}
S=\frac{1}{4 \pi(1+z)^{4}} \int n_{\mathrm{e}} n_{\mathrm{p}} \Lambda(kT, Z) {\rm d}l,
\end{equation}

\noindent where $n_{\rm e}$ and $n_{\rm p}$ are the number density of electron and proton, respectively, and we assumed $n_{\rm e}=1.2n_{\rm p}$. $\Lambda(kT, Z)$ is the band-averaged cooling function, dependent on temperature and metallicity, and d$l$ is the integral along the line of sight \citep{2010bulbul}. 

To compute the ICM mass of a cluster from the electron density profile, we used the enclosed ICM mass within a given aperture obtained by integrating the best-fit density model,
\begin{equation}
M_{\mathrm{ICM}} = 4\pi \mu_\mathrm{e} m_\mathrm{p}\int_{0}^{R} n_\mathrm{e}(r)\ r^{2}\ {\rm d}r,
\end{equation}

\noindent where we assumed 0.3 solar abundance of the ICM, adopting the solar abundance table, including the He abundance, from \citet{asplund2009}. The average nuclear charge and mass are $A\sim1.4$ and $Z\sim1.2$, and $\mu_{\rm e}=A/Z\sim1.17$.

The particle background map was obtained by folding the instrumental background parameter to the unvignetted exposure map. The sky-background component, including contribution from the cosmic X-ray background, Galactic halo, and the Local Bubble, was added to the particle background after being folded by the vignetted exposure map to create the total background model. 

The faint point sources within the images were excised, while the bright ones were modeled as delta functions convolved with the PSF to eliminate residual emission due to the wings of the PSF. The resultant model image was fit with the soft-band image of eFEDS observations using the Monte Carlo Markov chain (MCMC) code \textit{emcee} \citep{2013emcee} to find the best-fit parameters, consisting of the parameters of the number density model. 
We integrated the surface brightness profile and converted it into luminosity and flux by constructing an {\tt apec} model in {\tt Xspec}, adopting the ICM temperature measured from the spectral analysis. Clearly, the energy conversion factor (ECF) depends on the temperature, which itself varies with radial distance from the cluster center. 
Taking advantage of the weak dependence of this conversion on temperature \citep{2016Pacaud}, we simply adopted the temperature measured within 500~kpc to compute the ECF for any radius. The MCMC chains for spectral and imaging analysis were used to compute the ECF. By integrating the full probability distributions of temperature and surface brightness, we self-consistently estimated the uncertainties on the luminosity and flux measurements. 

\begin{figure}
\begin{center}
\includegraphics[width=0.49\textwidth, trim=0 20 15 40, clip]{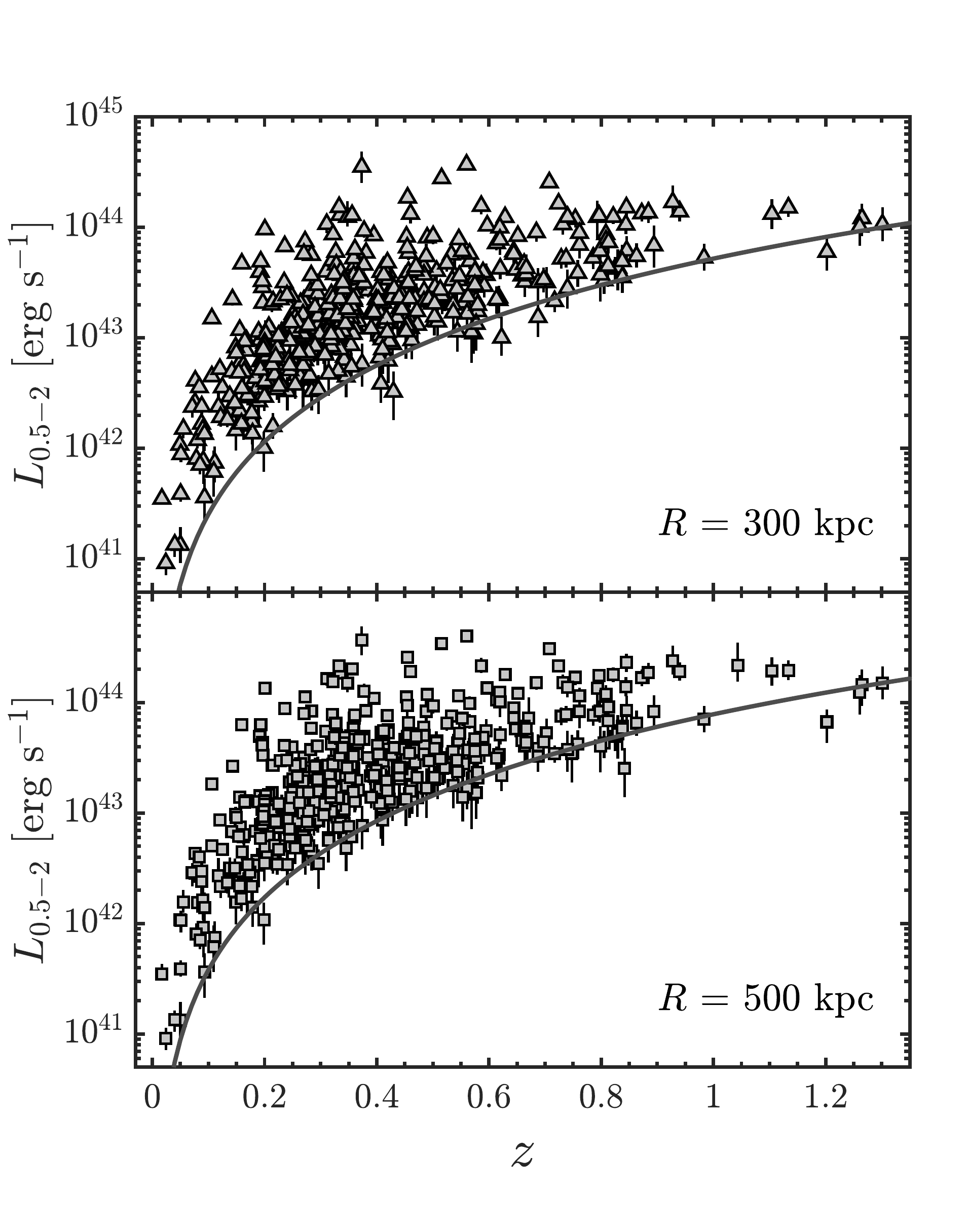}
\caption{Luminosity redshift distribution of the clusters with significant luminosity measurements ($>2\sigma$) for 300~kpc (upper panel) and 500~kpc (lower panel). The black curve shows the flux limit: $10^{-14}~{\rm erg}~{\rm s}^{-1}~{\rm cm}^{-2}$ for 300~kpc and $1.5\times10^{-14}~{\rm erg}~{\rm s}^{-1}~{\rm cm}^{-2}$ for 500~kpc (without a $K$ correction). }
\label{fig:lum_dist}
\end{center}
\end{figure}

We obtained significant ($>2\sigma$) luminosity measurements for $\sim90$\% of the clusters although we only obtained temperature measurements with the same significance level for a much lower fraction (see Sect.~\ref{sec:spectroscopy}) because the imaging data were used to measure the luminosity, and because conversion factor from surface brightness to luminosity depends only weakly on temperature. The luminosity-redshift distribution of the eFEDS cluster sample is shown in Fig.~\ref{fig:lum_dist}. For clarity, we only plot the data points with $>2\sigma$ significance for each measurement. 
The figure clearly shows the flux-limited nature of the sample and the absence of high-luminosity clusters at low redshift due to the small volume, both of which are common for a flux-limited survey such as eFEDS. These biases should be taken into account for further analyses of cosmological parameters and scaling relations.

\subsection{Spectral analysis}
\label{sec:spectroscopy}
\begin{figure*}
\begin{center}
\includegraphics[width=0.99\textwidth, trim=0 0 10 0, clip]{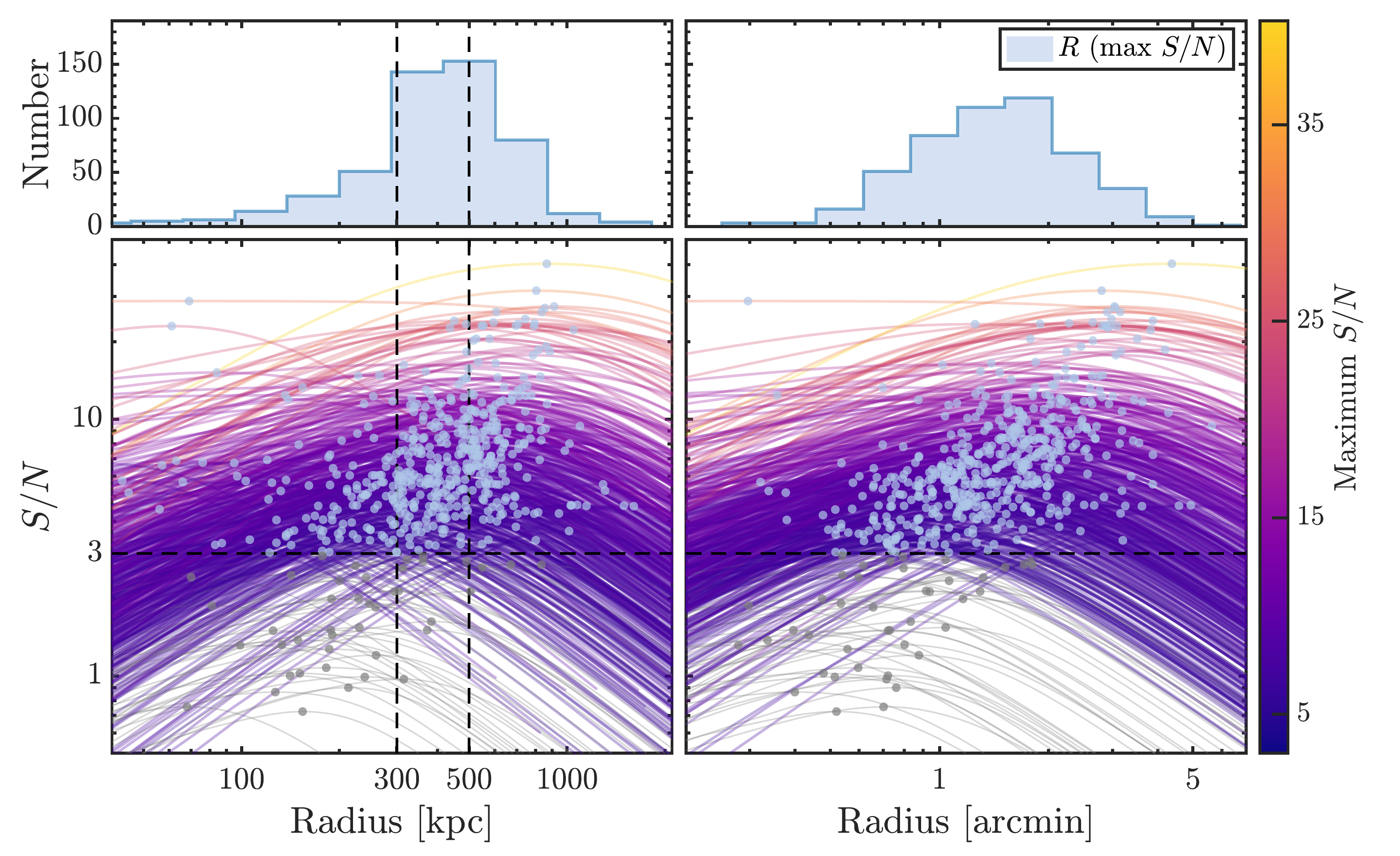}
\caption{Signal-to-noise ratio profiles for all the clusters. The net count rates are computed from the best-fit model of the soft-band image (see Sect. \ref{sec:imaging} for more details). For each cluster, the S/N as a function of radius is shown as a curve color-coded by its maximum value. The maximum S/N and the corresponding radii for all clusters are marked with blue dots. Clusters with maximum S/N lower than 3 are plotted in dark gray. The histogram in the upper left panel shows the distribution of the maximum S/N, only considering clusters with maximum S/N higher than 3, in order to remove the contamination from low-significance clusters. }
\label{snprof}
\end{center}
\end{figure*}

We extracted spectra and computed ancillary response files (ARFs) and redistribution matrix files (RMFs) from the seven TMs using the \texttt{eSASS} algorithm {\tt srctool} with the latest version of the calibration database. 
For the extraction radius that we used for the spectral analysis, we opted for two fixed physical apertures, 300~kpc and 500~kpc. The first was chosen for a fair comparison with similar flux-limited surveys, for instance, the XXL survey \citep[][]{2016Pacaud}. The second is a compromise between the aim to include more photons and the rapidly decreasing signal-to-noise ratio (S/N) in the cluster outskirts. In Fig. \ref{snprof} we verify the selection of the two extraction radii by plotting the S/N in the soft band as a function of extraction radius. We remark that while the radius of 300~kpc was determined regardless of the S/N, most of the clusters reach maximum S/N around 500~kpc. We note that due to the wide PSF of the eROSITA mirrors, photons originating from particular regions in the sky may become detected in the adjacent regions on the detector. This is known as the PSF spilling effect. This effect is particularly strong for high-redshift clusters ($z>1.5$), where the 300~kpc regions are comparable to the 26\arcsec  FOV average PSF HEW of eROSITA. Because we do not have clusters at $z>1.5$ in this sample, we omitted the systematic uncertainties due to PSF spilling in this work.

The background spectra were extracted within a [2500--4000]~kpc annulus centered at each cluster centroid after masking the emission from point sources and other clusters. The inner radius of the background region corresponds to $\sim3R_{500}$ for a cluster with $M_{500}=2\times10^{14}~M_{\odot}$ at $z=0.3$. We verified this choice of background radius on the eFEDS image and confirm that the inner radius of 2500~kpc extended well beyond the ICM emission in our data for all the clusters. The background spectra include the contribution from unvignetted instrumental background due to galactic cosmic rays \citep{Freyberg2020} and vignetted cosmic X-ray background due to unresolved point sources, local hot bubble and Galactic halo emission. Our final total background model then includes an unabsorbed {\tt apec} model for the local hot bubble and two absorbed {\tt apec} models for the Galactic halo and the Local Group at $\sim0.25$~keV and at a slightly higher temperature at $\sim0.75$~keV \citep{2000Kuntz,2008Snowden,2012bBulbul}. The cosmic X-ray background due to unresolved sources w s modeled by an absorbed power-law, where the index was frozen to 1.46 \citep[see, e.g.,][]{2017Luo} and the normalization was determined by fitting the local background spectra. The fitting band was restricted to 0.8--9~keV for the TMs affected by the light leak \citep[TM5 and TM7, see][]{2021Predehl}, while the 0.6--9~keV spectrum was used for all the other TMs. A more conservative way to remove the light leak is to entirely ignore the spectra of TMs 5 and 7. However, this will lose a significant fraction of good photons and further reduces the S/N. We compared our fitting results with those obtained by ignoring TMs 5 and 7 for a sample of 50 clusters with the highest S/N. We find that the temperatures measured with and without TMs 5 and 7 are consistent within 1$\sigma$  statistical uncertainty for all the cases. The normalization of spectra of TMs 5 and 7 is more affected by the calibration issues and light leak. However, we find that the difference in normalization with and without TMs 5 and 7 is also within 2$\sigma$. Because the electron number density and luminosity are not directly determined from the spectroscopy (and the normalization) but from the imaging analysis, we do not expect that our measurements are significantly affected by calibration issues related to TMs 5 and 7. We therefore decided to keep TMs 5 and 7 in the spectral fitting and only ignored the energy range below 0.8~keV, as already mentioned.

Because most of our detected clusters lie in the low-count regime, we strictly used C-statistics in our fits \citep{cash1979} and modeled our background instead of subtracting it from the total spectrum. The ICM emission was fit with the {\tt apec} thermal plasma emission model \citep{smith2001,2012Foster}. The solar abundance table from \citet{asplund2009} was adopted. The Galactic hydrogen absorption was modeled using {\tt tbabs} \citep{2000Wilms}, where the column density $n_{\rm H}$ was fixed to $n_{\rm H,tot}$ provided by \citet{2013Willingale} at the cluster position. This value takes neutral hydrogen and also molecular hydrogen into account. The current version of the {\tt tbabs} model includes the most accurate atomic data available for neutral species (e.g., they consider the smearing of the K-photoabsorption edge through Auger decay)\footnote{https://pulsar.sternwarte.uni-erlangen.de/wilms/research/tbabs}. The spectral fitting was done using {\tt Xspec} version 12.11.1 \citep{arnaud1996}.

\begin{figure}
\begin{center}
\includegraphics[width=0.49\textwidth, trim=0 20 15 40, clip]{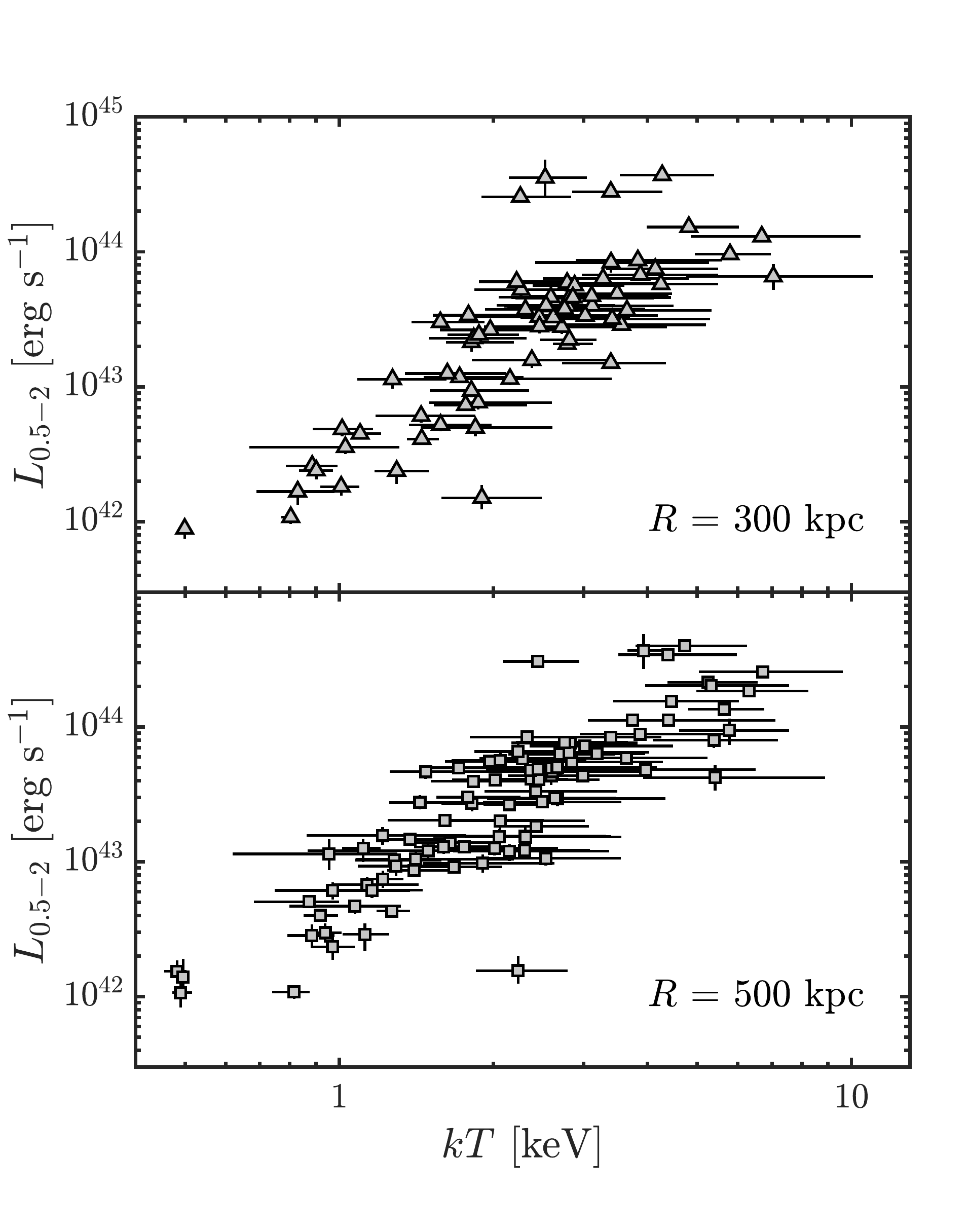}
\caption{Luminosity plotted as a function of temperature for the clusters with $>2\sigma$ temperature measurements in either of the two radii. }
\label{fig:lt}
\end{center}
\end{figure}

Due to the shallow depth of the eFEDS survey, we do not have significant metallicity measurements for all clusters in the sample (see Fig.~\ref{fig:example} for an example). Therefore we fixed the metallicity to 0.3~$Z_\odot$ \citep[see, e.g.,][]{liu2020} for all clusters for a uniform treatment of the data. The redshift of the cluster was set at the value provided by MCMF, as described in Sect.~\ref{sec:optical}. We used spectroscopic redshifts where available and adopted photometric redshifts for the rest of the clusters. We note that the redshifts of the ICM and member galaxies may be slightly different, especially in disturbed clusters \citep[e.g.,][]{2015Liu}, therefore an ideal approach is to allow the redshift to vary within a small range. However, given the low number of photons in our spectra (only a few clusters in our sample have $\sim1000$ net counts in the full band within 500~kpc), we were unable to obtain a significant constraint on the redshift from the X-ray spectral analysis for the vast majority of the sample. Thawing the redshift in the model does not improve the spectral fitting. We therefore fixed the redshift parameter to the MCMF values mentioned above for all the clusters. Moreover, the measurement of redshift from X-ray spectra requires well-understood gain calibration of the CCD \citep[see, e.g.,][]{2020Sanders}. Therefore we will explore the X-ray redshift determination in future studies. We also note that we ignored the multiple phase nature of the ICM within our extraction radii by using a single {\tt apec} model in the spectral fitting. This would unavoidably result in larger residuals particularly in the soft band (see Fig.~\ref{fig:example}, lower left panel). However, considering the low S/N in our data, we argue that an averaged spectroscopic-like temperature is the only quantity we can measure, and that constraining the multitemperature structures in the ICM is well beyond the data quality and the goal of this paper.

Due to the shallow survey data, we were unable to obtain a robust constraint on temperature for most of the clusters for which we only detected fewer than 100 counts in the 0.5--2~keV band within 500~kpc. Another problem in temperature measurement is the contamination of misidentified AGN in the sample. Finally, we were able to measure accurate temperatures of only 102 clusters ($\sim1/5$ of the full sample) at $>2\sigma$ significance level. All these clusters have $>100$ counts in the 0.5--2~keV band within 500~kpc. Most of them (100/102) have $L_{\rm ext}>10$, thus the contamination level of this subsample is much lower than the average of the full sample. We plot the $L-T$ relations for 300~kpc and 500~kpc in Fig.~\ref{fig:lt}. 
The ICM temperature of these 102 clusters ranges from $\sim$0.5~keV to $\sim$7~keV, and the average temperature is $\sim$2~keV, implying that they are dominated by low-mass clusters and groups \citep[see, e.g.,][Fig.~3]{2014Borm}. As expected owing to the large effective area in the soft band, we are more sensitive to low-mass cluster and group populations than other surveys.
Moreover, our ability to measure hot clusters at $>5$~keV is limited due to the reduced sensitivity of eROSITA in the energy band $>3$~keV. This is clearly reflected in Fig.~\ref{fig:lt}:  the 
ICM temperatures of hot clusters are poorly constrained. Fig.~\ref{fig:lt} also shows that several outliers appear at high luminosity but relatively low temperature. For example, eFEDS~J083811.8-015934, eFEDS~J092339.0+052654, and eFEDS~J093520.9+023234 have $L_{\rm 0.5-2~keV,~300~kpc} = 3.70_{-0.31}^{+0.34}\times10^{44}, 3.55_{-1.02}^{+1.30}\times10^{44}$, and $2.79_{-0.19}^{+0.17}\times10^{44}~{\rm erg~s^{-1}}$, and $kT_{\rm 300~kpc} = 4.27_{-0.74}^{+1.12}, 2.52_{-0.38}^{+0.53}$, and $3.39_{-0.54}^{+0.89}~{\rm keV}$. We verified the X-ray and optical data for these cases and found that they are all massive clusters, with estimated $R_{500}$ (see Sect.~\ref{sec:xlf}) of $>1.2$~Mpc, much larger than the 300~kpc and 500~kpc extraction radii we chose. The low temperatures we measured are probably due to the cool core that dominates the emission within our extraction radii. This also reminds us that a meaningful research of the relations between the X-ray observables can only be made after an accurate measurement of the $R_{500}$ of the clusters, from which both the core-included and core-excluded X-ray observables can be measured.
A detailed study of the scaling relations between different X-ray observables, for example, $L-T$ and $L-M_{\rm gas}$ measured within $R_{500}$,
will be performed in a future work by Bahar et al. (in prep.).

We provide in Table~\ref{tab:main} the main X-ray observables of the 102 clusters with $>2\sigma$ temperature measurements within either 300~kpc or 500~kpc. The full table containing the X-ray analysis results of all the 542 clusters is available at \url{https://erosita.mpe.mpg.de/edr/eROSITAObservations/Catalogues}, and at the CDS via anonymous ftp to \url{cdsarc.u-strasbg.fr} (130.79.128.5) or via \url{http://cdsweb.u-strasbg.fr/cgi-bin/qcat?J/A+A/}.

\section{X-ray luminosity function}
\label{sec:xlf}
\begin{table}
\caption{\label{tab:xlf} X-ray luminosity function of the eFEDS clusters for the full sample and in two redshift bins. }
\centering
\begin{tabular}[width=0.5\textwidth]{llll}
\hline\hline
$L_{500}$ & d$n$/d$L$ & -1$\sigma$ & +1$\sigma$ \\
\hline
$0.01<z<1.3$ \\ 
\hline
0.96$_{-0.46}^{+0.58}$ & 3.29$\times 10^{-3 }$ & 1.76$\times 10^{-3 }$ & 5.06$\times 10^{-3}$ \\ 
1.93$_{-0.39}^{+0.33}$ & 8.10$\times 10^{-4 }$ & 3.08$\times 10^{-4 }$ & 1.35$\times 10^{-3}$ \\ 
2.83$_{-0.58}^{+0.45}$ & 5.35$\times 10^{-4 }$ & 2.67$\times 10^{-4 }$ & 8.19$\times 10^{-4}$ \\ 
4.07$_{-0.79}^{+0.71}$ & 2.76$\times 10^{-4 }$ & 1.54$\times 10^{-4 }$ & 4.11$\times 10^{-4}$ \\ 
5.55$_{-0.77}^{+1.42}$ & 1.38$\times 10^{-4 }$ & 8.64$\times 10^{-5 }$ & 1.96$\times 10^{-4}$ \\ 
8.40$_{-1.44}^{+1.75}$ & 6.79$\times 10^{-5 }$ & 4.52$\times 10^{-5 }$ & 9.22$\times 10^{-5}$ \\ 
12.74$_{-2.59}^{+2.04}$ & 3.09$\times 10^{-5 }$ & 2.21$\times 10^{-5 }$ & 4.07$\times 10^{-5}$ \\ 
17.75$_{-2.97}^{+3.79}$ & 1.22$\times 10^{-5 }$ & 8.93$\times 10^{-6 }$ & 1.62$\times 10^{-5}$ \\ 
26.45$_{-4.91}^{+4.93}$ & 5.47$\times 10^{-6 }$ & 3.95$\times 10^{-6 }$ & 7.29$\times 10^{-6}$ \\ 
38.11$_{-6.73}^{+7.60}$ & 2.30$\times 10^{-6 }$ & 1.73$\times 10^{-6 }$ & 2.98$\times 10^{-6}$ \\ 
54.67$_{-8.96}^{+11.93}$ & 8.61$\times 10^{-7 }$ & 6.68$\times 10^{-7 }$ & 1.10$\times 10^{-6}$ \\ 
79.67$_{-13.07}^{+17.35}$ & 2.52$\times 10^{-7 }$ & 1.70$\times 10^{-7 }$ & 3.53$\times 10^{-7}$ \\ 
108.37$_{-11.35}^{+32.98}$ & 7.02$\times 10^{-8 }$ & 4.59$\times 10^{-8 }$ & 9.98$\times 10^{-8}$ \\ 
163.34$_{-22.00}^{+42.58}$ & 2.73$\times 10^{-8 }$ & 1.91$\times 10^{-8 }$ & 3.62$\times 10^{-8}$ \\ 
239.08$_{-33.16}^{+60.92}$ & 1.04$\times 10^{-8 }$ & 7.31$\times 10^{-9 }$ & 1.39$\times 10^{-8}$ \\ 
\hline
$0.01<z<0.35$ \\ 
\hline
0.99$_{-0.49}^{+0.63}$ & 3.03$\times 10^{-3 }$ & 1.72$\times 10^{-3 }$ & 4.60$\times 10^{-3}$ \\ 
2.30$_{-0.68}^{+0.63}$ & 6.03$\times 10^{-4 }$ & 3.44$\times 10^{-4 }$ & 9.02$\times 10^{-4}$ \\ 
4.10$_{-1.18}^{+1.16}$ & 2.11$\times 10^{-4 }$ & 1.38$\times 10^{-4 }$ & 2.95$\times 10^{-4}$ \\ 
7.39$_{-2.12}^{+2.10}$ & 6.68$\times 10^{-5 }$ & 4.93$\times 10^{-5 }$ & 8.49$\times 10^{-5}$ \\ 
12.42$_{-2.92}^{+4.68}$ & 2.64$\times 10^{-5 }$ & 2.21$\times 10^{-5 }$ & 3.07$\times 10^{-5}$ \\ 
21.37$_{-4.27}^{+9.44}$ & 8.30$\times 10^{-6 }$ & 6.74$\times 10^{-6 }$ & 9.85$\times 10^{-6}$ \\ 
40.26$_{-9.45}^{+15.24}$ & 2.81$\times 10^{-6 }$ & 2.16$\times 10^{-6 }$ & 3.44$\times 10^{-6}$ \\ 
62.31$_{-6.80}^{+37.69}$ & 6.13$\times 10^{-7 }$ & 3.97$\times 10^{-7 }$ & 8.13$\times 10^{-7}$ \\ 
\hline
$0.35<z<1.3$ \\ 
\hline
12.91$_{-2.91}^{+1.68}$ & 3.38$\times 10^{-5 }$ & 1.86$\times 10^{-5 }$ & 4.97$\times 10^{-5}$ \\ 
18.19$_{-3.60}^{+3.10}$ & 1.25$\times 10^{-5 }$ & 8.15$\times 10^{-6 }$ & 1.75$\times 10^{-5}$ \\ 
25.53$_{-4.24}^{+5.54}$ & 5.36$\times 10^{-6 }$ & 3.48$\times 10^{-6 }$ & 7.28$\times 10^{-6}$ \\ 
37.01$_{-5.94}^{+8.33}$ & 2.29$\times 10^{-6 }$ & 1.63$\times 10^{-6 }$ & 3.07$\times 10^{-6}$ \\ 
53.23$_{-7.89}^{+12.93}$ & 8.40$\times 10^{-7 }$ & 6.14$\times 10^{-7 }$ & 1.09$\times 10^{-6}$ \\ 
78.67$_{-12.51}^{+17.88}$ & 2.49$\times 10^{-7 }$ & 1.64$\times 10^{-7 }$ & 3.56$\times 10^{-7}$ \\ 
107.43$_{-10.89}^{+33.45}$ & 6.69$\times 10^{-8 }$ & 4.24$\times 10^{-8 }$ & 9.61$\times 10^{-8}$ \\ 
159.54$_{-18.65}^{+46.05}$ & 2.63$\times 10^{-8 }$ & 1.76$\times 10^{-8 }$ & 3.55$\times 10^{-8}$ \\ 
245.82$_{-40.24}^{+54.18}$ & 8.60$\times 10^{-9 }$ & 5.49$\times 10^{-9 }$ & 1.18$\times 10^{-8}$ \\ 
\hline
\end{tabular}

\tablefoot{$L_{500}$ are in units of $10^{42}~{\rm erg}~{\rm s}^{-1}$. Luminosity function values are in units of [${\rm Mpc}^{-3}~(10^{44}~{\rm erg}~{\rm s}^{-1})^{-1}$].  }
\end{table}

\begin{figure*}
\begin{center}
\includegraphics[width=0.49\textwidth, trim=0 0 0 10, clip]{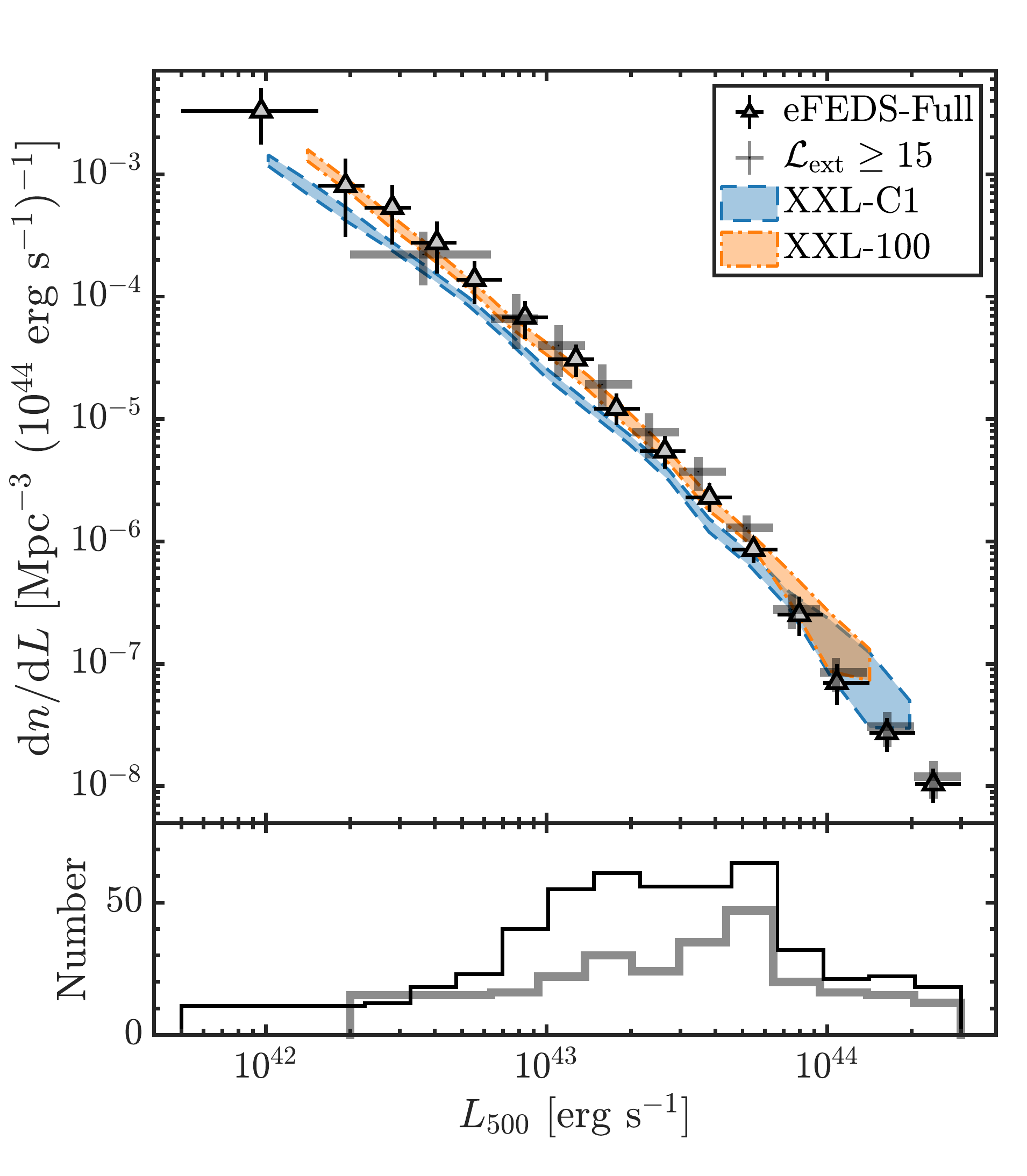}
\includegraphics[width=0.49\textwidth, trim=0 0 0 10, clip]{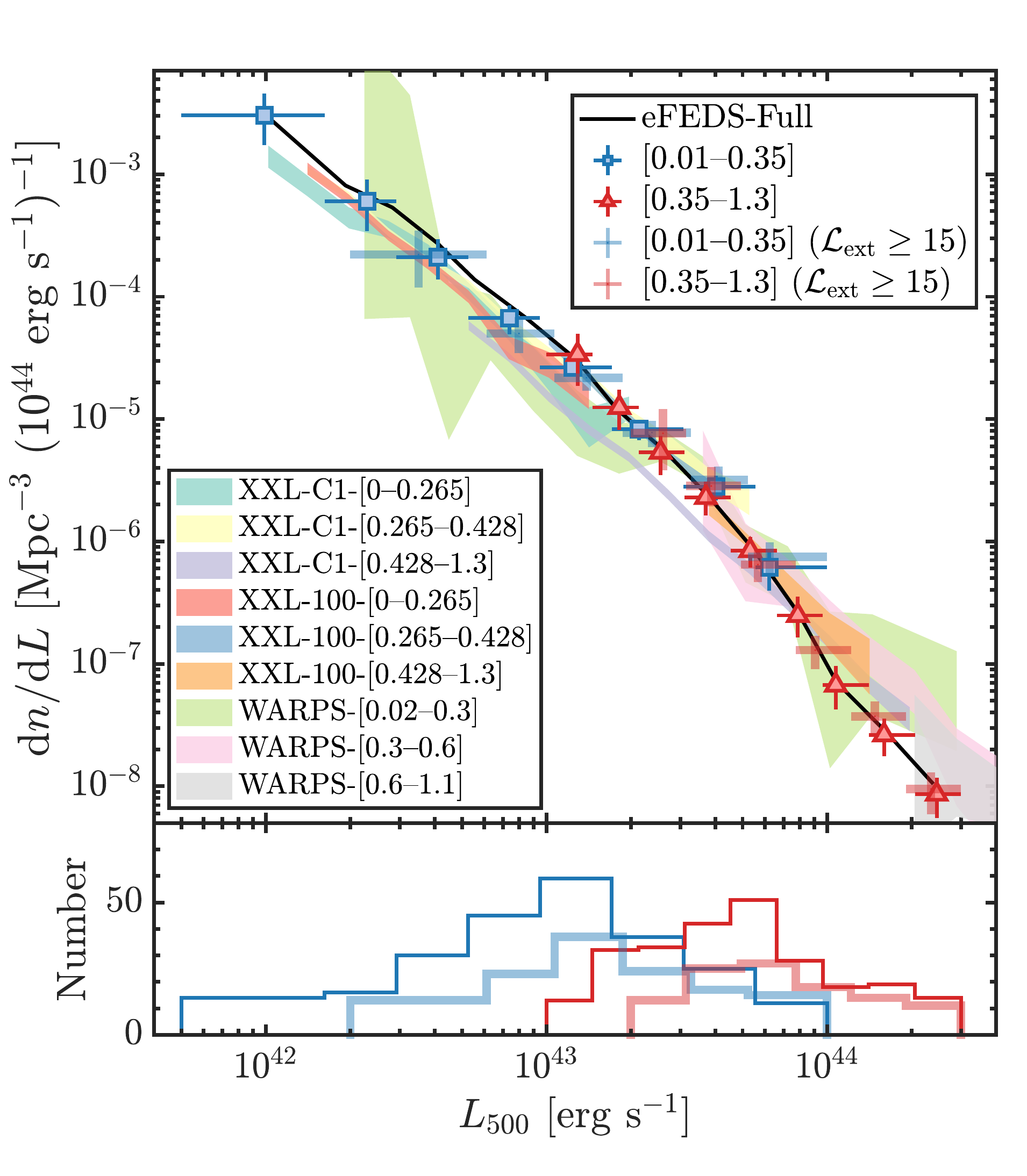}
\caption{X-ray luminosity function of the eFEDS cluster sample. {\sl Left panel:} Results for the full sample. {\sl Right panel:} Results in two redshift bins. The number of clusters in each bin is indicated in the lower panel. Some literature results are plotted as the shaded area for comparison: XXL-C1 \citep{2018Adami}, XXL-100 \citep{2016Pacaud}, and WARPS \citep{2013Koens}. Thick error bars show the results of the $\mathcal{L}_{\rm ext}\ge 15$ subsample. }
\label{fig:xlf}
\end{center}
\end{figure*}

The X-ray luminosity function (XLF) of a galaxy cluster survey is defined as the count of clusters per effective survey volume as a function of cluster luminosity. A widely adopted strategy is to measure XLF by dividing the cluster sample into luminosity bins \citep[see][for example]{2014Boehringer}, the XLF can be written as
\begin{equation}
\frac{{\rm d} n}{{\rm d} L}(\langle L_i\rangle)=\frac{1}{\Delta L_i} \sum_{j} \frac{1}{V_{\rm eff }[L_j, F_{\rm lim}, A(F_j)]/P(L_j, z_j)},
\end{equation}
\noindent where $\langle L_i\rangle$ and $\Delta L_i$ are the center luminosity (we used $L_{500}$ in 0.5--2~keV) and the width of the $i^{th}$ luminosity bin, $L_j$ is the luminosity of the $j^{th}$ cluster in the $i^{th}$ bin. $P(L_j, z_j)$ is the detection probability of a cluster with luminosity $L_j$ at redshift $z_j$, obtained from the selection function (see Sect.~\ref{sec:slctfnc}). $V_{\rm eff }[L_{j}, F_{\rm lim}, A(F_j)]$ is the survey-effective volume as a function of $L_j$ and the flux limit and sky coverage of the survey. The flux limit was set as $F_{\rm lim}=1.5\times 10^{-14}~{\rm erg}~{\rm s}^{-1}~{\rm cm}^{-2}$ (see Fig.~\ref{fig:sel}).
Because the exposure is nearly uniform across the whole eFEDS field above the flux limit, we used a constant sky coverage $A=126~{\rm deg}^2$, determined as the sky area with vignetted exposure higher than 0.5~ks, which is the lowest exposure at which a cluster with flux close to the flux limit is detected. $V_{\rm eff}$ was then computed as the comoving shell volume between redshift 0 and the redshift at which the cluster can be detected at the flux limit, scaled by the sky coverage. Because our sample has an average contamination fraction of $\sim1/5$, we finally scaled the ${\rm d}n/{\rm d}L$ by a factor of 0.8.

To be able to calculate the XLF of the eFEDS sample, the luminosity $L_{500}$ measured within $R_{500}$ must be known. 
There are several ways to measure the mass, thus computing the $R_{500}$, for a cluster, such as a dedicated weak-lensing analysis \citep[see][for a review]{2020Umetsu}, and a measurement of hydrostatic mass on the basis of deeper X-ray observation \citep[see, e.g.,][]{2019Ettori}. We estimated the $R_{500}$ in another way: the $R_{500}$ of each cluster was determined as the radius within which the mass $M_{500}$ and luminosity $L_{500}$ are consistent with cluster $L-M$ scaling relation.
While the majority of the published $L-M$ scaling relations in the literature focus on massive clusters \citep[e.g.,][]{2009Pratt,2009Vikhlinin, bulbul2019,2021Andrade-Santos},
a few studies also involved the lower end of the mass scale \citep[see, e.g.,][]{2011Eckmiller,2015Lovisari,2020Sereno}. Because the eFEDS sample consists of mostly low-mass clusters, we adopted the \citet{2015Lovisari} $L-M$ scaling relation, which is based on an X-ray selected bias-corrected sample of 20 galaxy groups and 62 massive HIFLUGCS clusters. We note that \citet{2015Lovisari} calculated the masses based on X-ray data and hydrostatic equilibrium assumption. The value of $R_{500}$ obtained from this approach is just a rough estimation. We discuss the impact of the choice of scaling relations on our results in Sect.~\ref{sec:scal}. The \citet{2015Lovisari} $L-M$ relation is

\begin{equation}
{\rm log}\left(\frac{L_{0.1-2.4}}{10^{43}~h^{-2}_{70}~{\rm erg~s}^{-1}}\right) = 1.39\times {\rm log}\left(\frac{M_{500}}{5\times10^{13}~h^{-1}_{70}~M_{\odot}}\right) - 0.12,
\end{equation}

\noindent where $L_{0.1-2.4}$ is the luminosity in 0.1--2.4~keV energy range within $R_{500}$, and is computed for each cluster in our sample using the same method as described in Sect.~\ref{sec:imaging}. We note that the 0.1--2.4~keV band luminosity is only used to estimate $R_{500}$ with the above scaling relation. We use the 0.5--2~keV band luminosity to compute the luminosity function.

The luminosity range [$5\times10^{41}$--$3\times 10^{44}$]~erg~s$^{-1}$~cm$^{-2}$ was divided into several bins with equal logarithmic width. The bins at low-$L$ and high-$L$ edges were then slightly adjusted in order to ensure that each bin included at least 10 clusters. The center of each bin was determined as the weighted-average luminosity of the clusters within it. The errors in ${\rm d}n/{\rm d}L$ were computed from 1000 bootstrapped samples for which the luminosity of each cluster was randomized considering the statistical uncertainty. This error was then added in quadrature with the Poisson error for the number of clusters in each bin. We investigated the evolution of the XLF by splitting the sample into two redshift bins, 0.01--0.35, and 0.35--1.3, each including $\sim250$ clusters. The XLFs of the full sample and the two redshift bins are listed in Table~\ref{tab:xlf} and plotted in Fig.~\ref{fig:xlf}. 

The XLF of galaxy clusters are available in the literature for a variety of samples \citep[see, e.g.,][]{1998Rosati,1998Vikhlinin,1999DeGrandi,2003Allen,2004Mullis,2007Boehringer,2013Koens,2014Boehringer,2016Pacaud,2018Adami,2020Finoguenov}. Here we compare our XLF with the results in several recent works based on different cluster samples, namely the WARPS \citep{2013Koens}, the XXL-100 \citep{2016Pacaud}, and the XXL-C1 sample \citep{2018Adami}. The comparison is shown in Fig.~\ref{fig:xlf}. 
We find good agreement in the XLFs of the full eFEDS sample and of the XXL-100 and XXL-C1 clusters. In particular, the eFEDS XLF is relatively closer to the result based on the XXL-100 sample, especially for low luminosities. When we divide the eFEDS cluster sample into two redshift bins, we do not observe any significant evolution of the XLF with redshift, consistent with the literature, as shown in the right panel of Fig.~\ref{fig:xlf}). However, we note that the contamination in the full eFEDS sample at different luminosity and redshift bins may induce a bias in the XLF measurements. Therefore we also computed the XLF with a purer subsample that we obtained by selecting the eFEDS clusters with extent likelihood $\mathcal{L}_{\rm ext}\ge 15$. This subsample contains $\sim$250 clusters in the luminosity range in which the XLF is computed, with an estimated purity $>90$\%. The selection function was also recomputed for this subsample according to the corresponding thresholds. As a result, the XLF of the subsample is fully consistent with that of the full sample, as shown in Fig.~\ref{fig:xlf}). 

\subsection{Effect of the scaling relation on the luminosity function}
\label{sec:scal}
Because $R_{500}$ and $L_{500}$ were determined using $L-M$ scaling relation, the XLF is dependent on the choice of $L-M$ relation we adopt. In this section we assess this effect by comparing the current XLF with the result we obtain with other $L-M$ relations. The general form of cluster $L-M$ relation reads ${\rm log}~(L)=a\cdot{\rm log}~(M)+b$, where the slope $a$ varies in the range [1, 2] in different works, depending on the cluster population that were studied \citep{2009Pratt,2010Arnaud,2013Giodini,2015Lovisari, bulbul2019,2020Sereno}. Several works have found that the $L-M$ relation for galaxy groups is steeper than for massive clusters \citep[see, e.g.,][]{2015Lovisari}. This adds complexity to our selection of the $L-M$ relation for our sample, which includes both low-mass groups and clusters. Even though we finally chose the result reported by \citet{2015Lovisari} based on a combined sample of groups and clusters, with $a=1.39$, it is necessary to examine the result when we adopt a steeper relation. We therefore recomputed $R_{500}$, $L_{500}$, and the corresponding XLF using the scaling relation in \citet{bulbul2019},

\begin{equation}
\frac{L_{\mathrm{500}}}{10^{44}~{\rm erg~s}^{-1}}= 4.12 \times\left(\frac{M_{500}}{M_{\mathrm{piv}}}\right)^{1.89}\cdot \left(\frac{E(z)}{E(z_{\mathrm{piv}})}\right)^{2}\cdot\left(\frac{1+z}{1+z_{\mathrm{piv}}}\right)^{-0.2},
\end{equation}
\noindent where $L_{500}$ is the 0.5--2~keV band luminosity within $R_{500}$, $M_{\rm piv}=6.35\times10^{14}~M_{\odot}$ and $z_{\rm piv}=0.45$. We found that while the large difference in the slopes of the two scaling relations (1.39 and 1.89) changed $R_{500}$ by about 10\%\  on average, the value of $L_{500}$ was only affected by 
a few percent. The XLFs also agree well. Clearly, the cluster sample based on which the scaling relation is obtained in either \citet{2015Lovisari} or \citet{bulbul2019} does not perfectly fit the cluster population in our sample, thus preventing us from having a more precise assessment of this bias. The accurate measurements of the quantities $R_{500}$, $L_{500}$, and $M_{500}$, of these clusters will have to rely on a weak-lensing analysis with high quality or on hydrostatic mass measurement on basis of deeper X-ray observations. Before submitting this paper, we obtained the mass $M_{500}$ derived from the scaling relation of count rate, mass, and redshift with a weak-lensing calibration, which will be presented in detail in \citet{chiu2021}. The $L_{500}$ are different by less than a few percent with respect to our results, and the change in the XLF is negligible.

\begin{table*}
\caption{Supercluster candidates detected in the eFEDS field. }
\label{sc}
\begin{center}
\begin{tabular}[width=0.9\textwidth]{lccccccc}
\hline\hline
ID & RA & Dec & $N$ & $z$ & $l(z)$ & $M_{\rm tot}$ & $C$ \\
   & [deg] & [deg] & & & [Mpc] & [$10^{15}~M_{\odot}$] &  \\
\hline
eFEDS-SC1 & 133.622 & +1.201 & 4 & 0.107 & 19.43 & 0.4 & 1 \\
eFEDS-SC2 & 132.914 & +0.729 & 6 & 0.195 & 21.39 & 1.2 & 0.77 \\
eFEDS-SC3 & 135.091 & +3.009 & 10 & 0.196 & 21.43 & 3.1 & 0.58 \\
eFEDS-SC4 & 140.531 & +3.925 & 6 & 0.269 & 24.33 & 2.2 & 0.87 \\
eFEDS-SC5 & 129.959 & $-$1.700 & 7 & 0.269 & 24.34 & 1.9 & 0.83 \\
eFEDS-SC6 & 139.996 & +2.435 & 5 & 0.282 & 24.86 & 1.8 & 1 \\
eFEDS-SC7 & 135.390 & $-$1.209 & 4 & 0.295 & 25.38 & 0.9 & 1 \\
eFEDS-SC8 & 133.289 & +1.397 & 4 & 0.324 & 26.17 & 1.0 & 0.57 \\
eFEDS-SC9 & 141.682 & $-$1.053 & 6 & 0.337 & 26.69 & 1.8 & 0.44 \\
eFEDS-SC10(*) & 130.690 & +1.045 & 4 & 0.342 & 27.02 & 0.8 & 0.86 \\
eFEDS-SC11 & 143.707 & +0.076 & 4 & 0.344 & 27.13 & 1.3 & 0.72 \\
eFEDS-SC12 & 143.862 & +0.641 & 4 & 0.358 & 28.23 & 3.0 & 1 \\
eFEDS-SC13 & 129.769 & +2.121 & 5 & 0.359 & 28.35 & 2.3 & 1 \\
eFEDS-SC14 & 130.666 & +1.226 & 8 & 0.414 & 33.45 & 2.6 & 0.49 \\
eFEDS-SC15 & 138.423 & +3.922 & 4 & 0.456 & 37.49 & 2.3 & 1 \\
eFEDS-SC16(*) & 129.502 & $-$2.060 & 6 & 0.565 & 47.02 & 4.4 & 0.77 \\
eFEDS-SC17(*) & 140.134 & +4.198 & 4 & 0.579 & 49.24 & 2.5 & 0.68 \\
eFEDS-SC18(*) & 137.574 & +3.182 & 4 & 0.619 & 58.18 & 2.1 & 0.65 \\
eFEDS-SC19(*) & 136.418 & +2.489 & 7 & 0.803 & 65.70 & 5.2 & 0.49 \\

\hline

\hline
\end{tabular}
\tablefoot{ We mark with an asterisk in brackets the superclusters in which for more than half of the members no spectroscopic redshifts are available. The coordinates and redshifts are the mean values of the member clusters. $C$ is the compactness; see text for more details.}
\end{center}
\end{table*}
\begin{figure}
\begin{center}
\includegraphics[width=0.49\textwidth, trim=5 0 40 30, clip]{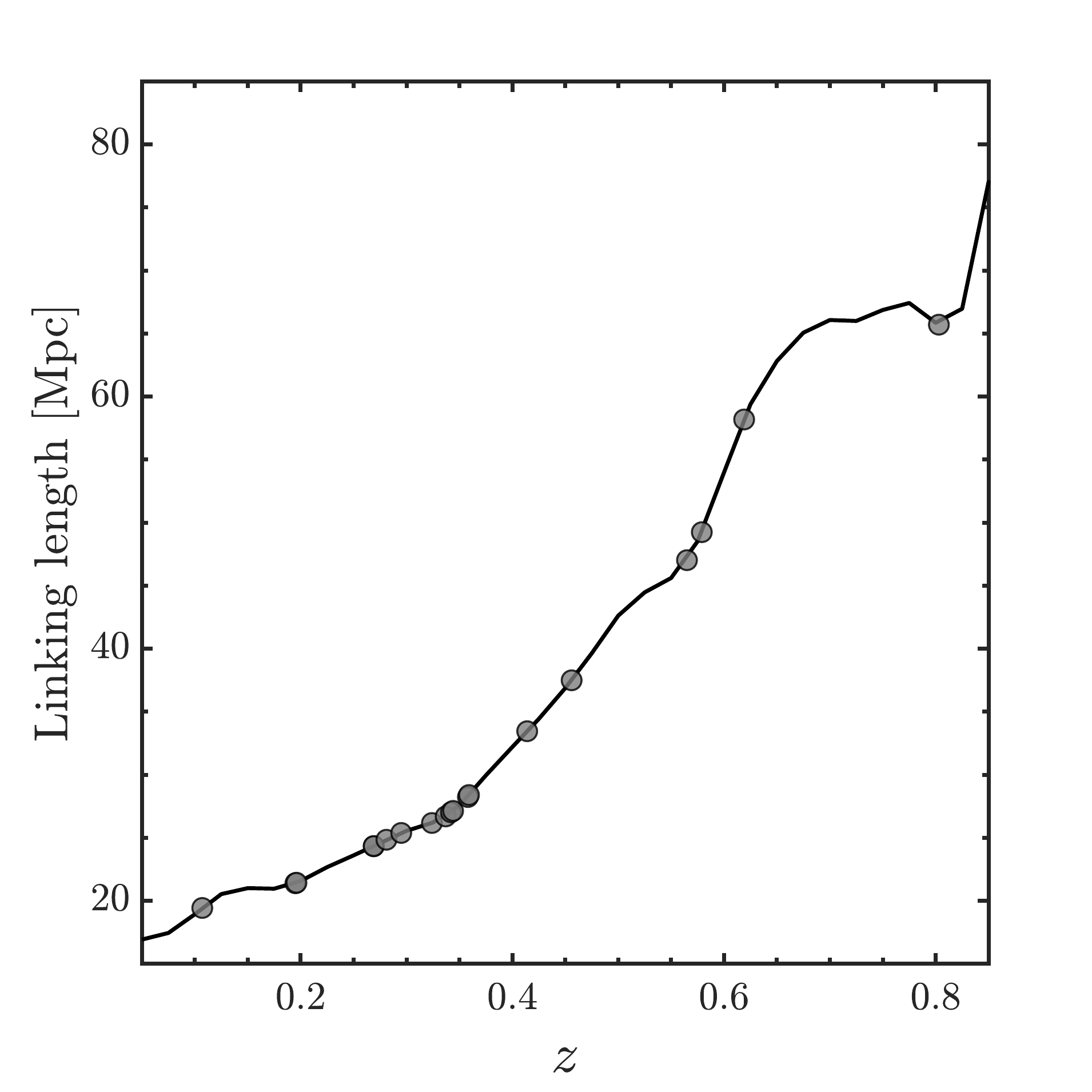}
\caption{Linking length in comoving distances as a function of cluster redshift in our sample. The circles mark the positions of the 19 supercluster candidates we detected.  }
\label{fig:lk}
\end{center}
\end{figure}
\section{Superclusters in the eFEDS field}
\label{sec:sc}

We further analyzed the spatial distribution of our sample of 542 galaxy clusters to search for superclusters. Early supercluster searches used optically detected catalogs of galaxy clusters such as Abell/ACO clusters \citep[see, e.g.,][]{1993Zucca,1994Einasto}. The first supercluster catalog based on X-ray selected clusters was presented in \cite{2001Einasto}, where 19 superclusters with multiplicity (the {\sl \textup{multiplicity}} ($N_{\rm cl}$) of a supercluster is defined as the number of its member clusters) $\ge 2$ were detected based on the early RASS catalog. In recent years, more superclusters have been discovered thanks to the increase in the volume and depth of X-ray cluster samples. For instance, \citet{2013Chon} detected 164 superclusters with $N_{\rm cl} \ge 2$ at $z\le 0.4$ based on the ROSAT-ESO FluxLimited X-ray (REFLEX II) cluster catalog. \citet{2018Adami} detected 35 superclusters with $N_{\rm cl} \ge 3$ and 39 cluster pairs out to $z\sim0.8$ based on the 365-cluster catalog from the XXL survey. 
    
We employed a friends-of-friends (FoF) algorithm to identify superclusters with an evolving linking length. This method that has been widely used in previous works \citep{1993Zucca,2001Einasto,2013Chon,2014Chow,2018Adami}.  This method is well suited for detecting superclusters because they are not virialized objects and thus often have irregular shapes. Our FoF algorithm starts by searching for neighboring clusters (the ``friends'') with a distance smaller than the linking length for a cluster in the sample. Then for each of the friends, the algorithm continues to search for friends of friends until no new friends can be found. Because the linking length varies with cluster redshift, we use the average linking length of the two clusters when we compute their relation to ensure that the search result is stable. 
    
The linking length of the FoF algorithm in comoving distances can be computed as follows \citep[see, e.g.,][Eqs.~1--3]{2013Chon}:
\begin{equation}
       l = \left(\frac{N(z_{\rm min},z_{\rm max})}{V(z_{\rm min},z_{\rm max},A)}\cdot f\right)^{-1/3},
\end{equation}

\noindent where $N$ is the number of clusters in the redshift bin, $V$ is the comoving volume of the shell as a function of the lower and upper bounds of the bin and the eFEDS survey area $A$. $f$ is the over-density factor. Therefore the linking length at a specific redshift corresponds to the maximum distance between two clusters defining a region that is  $f$ times overdense with respect to the average cluster number density at that redshift. The resulting linking length evolves with the redshift distribution of clusters and allows us to detect similar overdensities throughout the redshift range. We set $\Delta z=0.05$ and $f=10$, as was commonly used in previous works. In Fig.~\ref{fig:lk} we plot the linking length as a function of cluster redshift in our sample. The linking length increases from $\sim$20~Mpc at low-$z$ to $>$50~Mpc at high-$z$ due to the low number density of high-$z$ clusters in our catalog (see Fig.~\ref{z_hist}).
    
We classify the cluster systems found by the FoF algorithm with at least four members as supercluster candidates. Clearly, systems with two or three member clusters are also possible objects of interest, and the thresholds of supercluster multiplicity are usually lower than 4 \citep[e.g.,][]{2013Chon,2018Adami}. However, our cluster sample has a contamination fraction of $\sim~1/5$, which decreases the reliability of the superclusters with low multiplicity. We therefore determined $N_{\rm cl} \ge 4$ as our threshold. Fig.~\ref{fig:lk} also shows that the linking length starts to increase abnormally beyond redshift $z\sim0.8$. Clearly, the density of clusters at this high redshift in our sample is too low for superclusters to be detected. We therefore set an upper limit on the linking length: $l_{\rm max}$=70~Mpc, to avoid incorrect detections at high redshifts. 

With this approach, we detected 19 supercluster candidates in the eFEDS field. We list them in Table~\ref{sc}. 
The multiplicities of these 19 superclusters range from 4 to 10. The cut out X-ray images of these superclusters and the properties of their member clusters are provided in the appendix (see Table~\ref{tab:sc_mem} and Fig.~\ref{fig:sc_image}). The galaxy density maps from the HSC-SSP survey data at the corresponding redshifts for all the superclusters are presented in Fig.~\ref{fig:sc_image_hsc}. The association of the galaxy density maps and the X-ray images of the superclusters is good. Moreover, we also identified 46 and 14 cluster systems with two and three member clusters, respectively. As a comparison, \citet{2018Adami} detected 39 cluster pairs and 35 superclusters with $N_{\rm cl}>3$ in the XXL 365-cluster catalog. Considering the difference between eFEDS and XXL in sky coverage (140~deg$^2$ vs. 50~deg$^2$), sample size (542 vs. 365), and flux limit ($10^{-14}~{\rm erg}~{\rm s}^{-1}~{\rm cm}^{-2}$ vs. a few $10^{-15}~{\rm erg}~{\rm s}^{-1}~{\rm cm}^{-2}$), we do not find a significant inconsistency between the number of superclusters in eFEDS and XXL.

We estimated the mass for each supercluster by summing the virial mass of its member clusters. The virial masses of the clusters in our sample were estimated using the $M_{500}$ obtained from the \citet{2015Lovisari} $L-M$ scaling relation (see Sect.~\ref{sec:xlf}). We then converted $M_{500}$ into $M_{\rm vir}$ by assuming a Navarro-Ferenk-White (NFW) profile \citep{1997Navarro} with concentration $c\equiv r_{200}/r_{\rm s}=4$, which approximately gives $M_{\rm vir}\approx 2M_{500}$ \citep{2013Reiprich}.
    
We also adopted a new parameter, the compactness, $C$, to describe the spatial distribution of member clusters within the supercluster. $C$ is defined in the following way. We considered a sphere whose diameter is equal to the linking length. The position of this sphere was adjusted until the total mass of the enclosed member clusters reached its maximum value. $C$ is then defined as the ratio of the maximum enclosed mass to the total mass of the supercluster. If all the member clusters are distributed within the linking length, the compactness is accordingly 1. Clearly, $C$ is very sensitive to the reliability of the distance between member clusters and thus is affected by the uncertainties in photometric redshift measurements. The precision of $C$ is also dependent on the multiplicity of a supercluster.
Among the 19 superclusters candidates, we measure a compactness $C=1$ for 6 candidates. Most of them have a compactness higher than 0.5. Limited by the sample size and the narrow redshift span (most of the supercluster candidates are below $z=0.5$), we are not able to further analyze the compactness.
On the other hand, an increasing supercluster compactness from high-$z$ to low-$z$ can be anticipated and can be investigated in future works, provided that a larger and more complete supercluster sample is available, especially at high redshift.

With the 3D distribution of member clusters and their total mass, we further explored the collapsing probability for each supercluster candidate. When a $\Lambda$CDM cosmology is assumed, an object with an overdensity $\delta_{\rm c}\equiv \rho/\rho_{\rm c}$ higher than $\sim$1.7 will finally collapse. Clearly, because superclusters do not have regular shapes, their volume cannot be measured directly. Moreover, we only know the mass in the most massive member clusters that are detectable, while the fraction of mass in filaments and low-mass subhalos is unavailable.
Therefore the density within a supercluster cannot be measured accurately. Nevertheless, we can obtain a very conservative lower limit of the density by dividing the total virial masses of the member clusters by the volume of the smallest sphere that encloses all the member clusters. Interestingly, we still found two superclusters with this very conservative estimation: eFEDS-SC12 and eFEDS-SC13, with $\delta_{\rm c}=$ 2.2 and 4.7, implying that they will finally collapse with a high probability. The radii of the smallest sphere are 11.8~Mpc and 8.4~Mpc, respectively. This is consistent with their compactness: the two superclusters have a compactness equal to 1. Table~\ref{tab:sc_mem} and Fig.~\ref{fig:sc_image} show that the cluster members of these two superclusters are distributed very closely to each other. 
On the other hand, we are not able to constrain the collapsing time of the two superclusters, which obviously should take into account not only gravity, but also the effect of dark energy and the 3D distribution of mass within the supercluster. We therefore defer this study to a future work.
    
We also confirm the detection of the supercluster discovered by \citet{2021Ghirardini}. However, this supercluster is found to be fragmented into two parts, eFEDS-SC11 and eFEDS-SC12, with four members each. 
This is due to the change in the member cluster redshift. \citet{2021Ghirardini} adopted the photometric redshifts from HSC survey data that were available at the time of submission, while here in this work we use the latest spectroscopic redshifts (see Sect.~\ref{sec:optical}). This difference also highlights the importance of using the most accurate spectroscopic redshifts in the search for superclusters.
We also note that although the supercluster discovered in \citet{2021Ghirardini} is fragmented into eFEDS-SC11 and eFEDS-SC12 in this work, they are still very close to each other. The closest members of eFEDS-SC11 and eFEDS-SC12 are separated by $\sim$29~Mpc, which only slightly exceeds the linking length at their redshift, $\sim$28~Mpc. This $\sim1$~Mpc difference is within the virial radius of a cluster, therefore eFEDS-SC11 and eFEDS-SC12 probably belong to the same structure. We therefore confirm that our results are fully consistent with \citet{2021Ghirardini}.
    
Most of the supercluster candidates are located in the redshift range [0.1--0.5]. Four high-$z$ candidates lie beyond redshift 0.5: eFEDS-SC16 at 0.565, eFEDS-SC17 at 0.579, eFEDS-SC18 at 0.619, and eFEDS-SC19 at 0.803. However, we note that we only have photometric redshifts for most members of these superclusters (see Table~\ref{tab:sc_mem}). Moreover, the linking length rapidly increases to $>$50~Mpc at high redshifts due to the low number density of high-redshift clusters in our catalog. Considering the relatively large uncertainty in the photometric redshift and the low statistics at high-$z$, we note that these candidates need to be verified by future spectroscopic follow-up observations.

\section{Conclusions}
\label{sec:conclusions}
We have presented the first catalog of eROSITA-selected galaxy cluster and group candidates and their physical properties as detected in the sky area covered by the eROSITA Final Equatorial-Depth Survey (eFEDS). The optical follow-up for redshift and confirmation was performed using the MCMF tool on the basis of data from the HSC-SSP survey and the DECaLS survey. Our conclusions are summarized below.

In the area of 140 square degrees that is covered by eFEDS, we detected 542 candidate clusters and groups of galaxies down to the flux limit of $\sim10^{-14}~{\rm erg}~{\rm s}^{-1}~{\rm cm}^{-2}$ in the soft band (0.5--2~keV) within 1\arcmin. The clusters are distributed in the redshift range $z=$[0.01, 1.3], and the median redshift  is $z_{\rm median}=0.35$. The contamination fraction of noncluster sources in the sample, dominated by AGN misidentified as extended sources, is $\sim~1/5$, as obtained from simulations. After accounting for the contamination, we find a cluster density of $\sim3.2$ per square degree at this depth, in agreement with the expectations in \citet{Merloni2012}. Our simulations suggest that it is possible to obtain a sample with a purity of $>$90\% if a simple cut at $\mathcal{L}_{\rm ext}\ge 15$ is applied. This further selection delivers a subsample of $\sim270$ clusters. However, we note that the sample volume is unavoidably reduced as the high-redshift tail of the sample will have a smaller extent and will therefore be more severely cut by a higher $\mathcal{L}_{\rm ext}$ threshold. Most of these high-redshift clusters are real, however, and the simulations predict that a few more of them will be found among the point-source-detected population. In a following paper (Bulbul et al., in prep.), we will present a detailed assessment of the contribution of unresolved clusters to the eFEDS point-sources catalog. Further improvement of the characterization of the high-redshift clusters will be obtained through the recently completed SDSS-IV and SDSS-V spectroscopic follow-up campaign, which will be presented elsewhere (Ider Chitham et al., in prep.). We also found that $\sim16\%$ of the clusters can be matched with a counterpart in the published X-ray or SZ cluster catalogs, indicating that the majority of the clusters in the sample are detected through their ICM emission for the first time. 

We measured the ICM temperature within two parametric radii, 300~kpc and 500~kpc. Radial profiles of flux, luminosity, electron density, and gas mass were measured from the soft-band surface brightness image. For $\sim~1/5$ of the clusters in the sample (102/542), we obtain $>2\sigma$ constraints on temperature. The average temperature of these clusters is $\sim$2~keV. The clusters span the luminosity range [$10^{41}~{\rm erg}~{\rm s}^{-1}$, $\sim10^{44}~{\rm erg}~{\rm s}^{-1}$]. According to the $L-M$ scaling relation in \citet{2015Lovisari}, $\sim40\%$ of the clusters have $M_{500}<10^{14}~M_{\odot}$, indicating that our sample is dominated by galaxy groups and low-mass clusters, as expected. 

The selection function, the purity, and completeness of the catalog were examined and discussed in detail using the most recent simulations of the eFEDS field. The X-ray luminosity function of the sample is found to agree well with the results obtained from other recent X-ray surveys, such as the XXL and the WARPS surveys. We find no significant evolution of the cluster XLF in the redshift range [0.01, 1.3].

Using the redshifts and spatial distribution of the clusters, we performed a search for superclusters in the eFEDS field and detected 19 supercluster candidates, most of which are located at redshifts between 0.1 and 0.5, including the one at $z\sim0.36$ that has been published in \citet{2021Ghirardini}. Another four high-$z$ supercluster candidates are detected at redshifts higher than 0.5, but need further spectroscopic confirmation as the errors on the photometric redshifts may bias the linking length calculations.

The eFEDS cluster and group catalog at the final eRASS equatorial depth provides a benchmark proof-of-concept for the eROSITA All-Sky Survey extended source detection and characterization. We confirm the excellent performance of eROSITA for cluster science and expect no significant deviations from our pre-launch expectations for the final all-sky survey.

\begin{acknowledgement}

This work is based on data from eROSITA, the soft X-ray instrument aboard SRG, a joint Russian-German science mission supported by the Russian Space Agency (Roskosmos), in the interests of the Russian Academy of Sciences represented by its Space Research Institute (IKI), and the Deutsches Zentrum f{\"{u}}r Luft- und Raumfahrt (DLR). The SRG spacecraft was built by Lavochkin Association (NPOL) and its subcontractors, and is operated by NPOL with support from the Max Planck Institute for Extraterrestrial Physics (MPE).
\\
The development and construction of the eROSITA X-ray instrument was led by MPE, with contributions from the Dr. Karl Remeis Observatory Bamberg \& ECAP (FAU Erlangen-Nuernberg), the University of Hamburg Observatory, the Leibniz Institute for Astrophysics Potsdam (AIP), and the Institute for Astronomy and Astrophysics of the University of T{\"{u}}bingen, with the support of DLR and the Max Planck Society. The Argelander Institute for Astronomy of the University of Bonn and the Ludwig Maximilians Universit{\"{a}}t Munich also participated in the science preparation for eROSITA.
\\
The eROSITA data shown here were processed using the \texttt{eSASS/NRTA} software system developed by the German eROSITA consortium.
\\
The Hyper Suprime-Cam (HSC) collaboration includes the astronomical communities of Japan and Taiwan, and Princeton University. The HSC instrumentation and software were developed by the National Astronomical Observatory of Japan (NAOJ), the Kavli Institute for the Physics and Mathematics of the Universe (Kavli IPMU), the University of Tokyo, the High Energy Accelerator Research Organization (KEK), the Academia Sinica Institute for Astronomy and Astrophysics in Taiwan (ASIAA), and Princeton University.  Funding was contributed by the FIRST program from the Japanese Cabinet Office, the Ministry of Education, Culture, Sports, Science and Technology (MEXT), the Japan Society for the Promotion of Science (JSPS), Japan Science and Technology Agency  (JST), the Toray Science  Foundation, NAOJ, Kavli IPMU, KEK, ASIAA, and Princeton University.
\\
This paper makes use of software developed for the Large Synoptic Survey Telescope. We thank the LSST Project for making their code available as free software at http://dm.lsst.org.
\\
This paper is based [in part] on data collected at the Subaru Telescope and retrieved from the HSC data archive system, which is operated by Subaru Telescope and Astronomy Data Center (ADC) at NAOJ. Data analysis was in part carried out with the cooperation of Center for Computational Astrophysics (CfCA), NAOJ.
\\
This work was supported in part by World Premier International Research Center Initiative (WPI Initiative), MEXT, Japan, and JSPS KAKENHI Grant Number JP19KK0076.
\\
This work was supported in part by CNES.
\\
This work was supported in part by the Fund for the Promotion of Joint International Research, JSPS KAKENHI Grant Number 16KK0101.
\\
The authors thank Dominique Eckert for useful discussions.
\\
DNH acknowledges support from the ERC through the grant ERC-Stg DRANOEL n. 714245.
\\
E.B. acknowledges financial support from the European Research Council (ERC) Consolidator Grant under the European Union’s Horizon 2020 research and innovation programme (grant agreement No 101002585).

\end{acknowledgement}

\bibliographystyle{aa}
\bibliography{efeds_ext_cat}

\appendix
%

\section{Properties of supercluster candidates}
\begin{figure*}
\begin{center}
\includegraphics[width=0.245\textwidth, trim=0 20 40 50, clip]{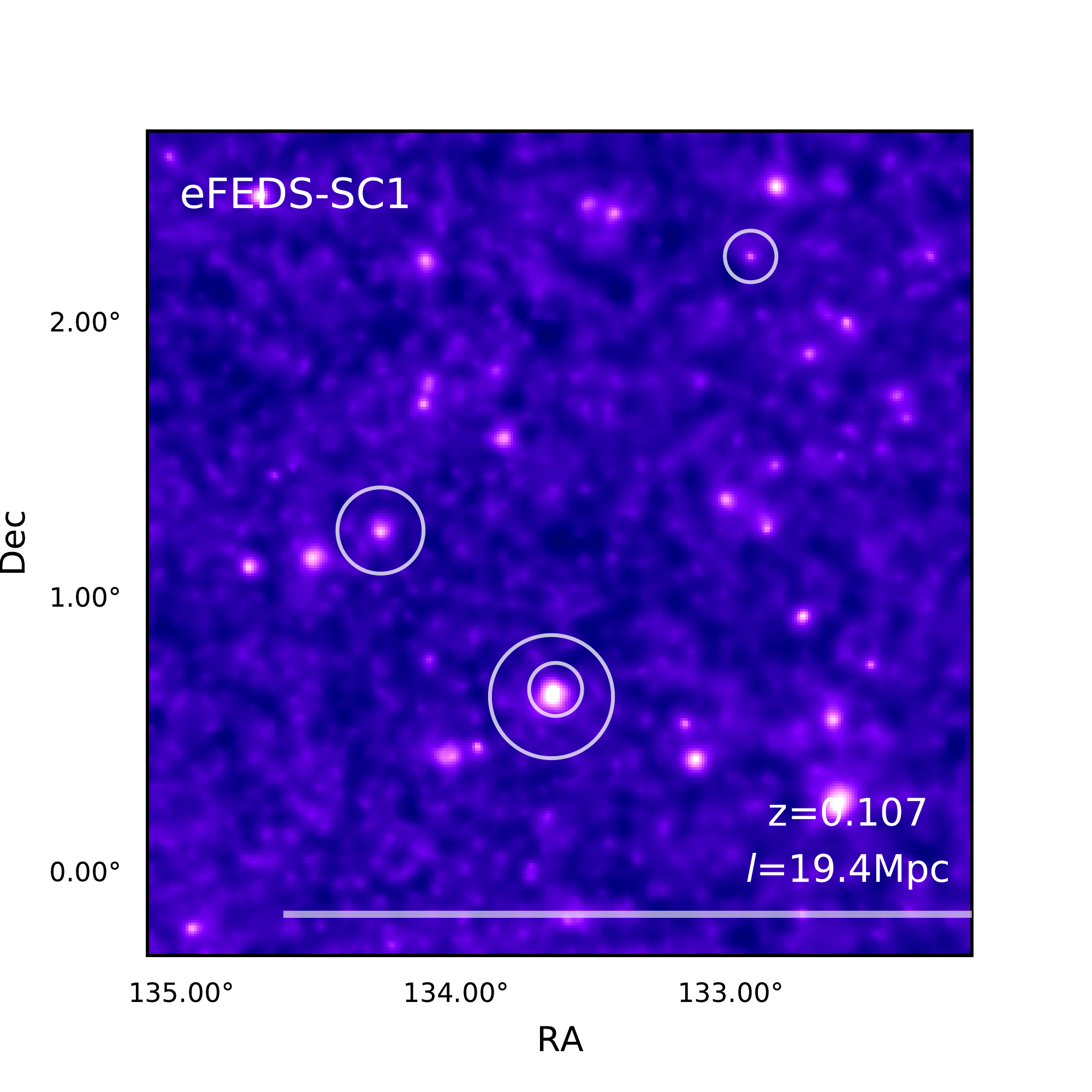}
\includegraphics[width=0.245\textwidth, trim=0 20 40 50, clip]{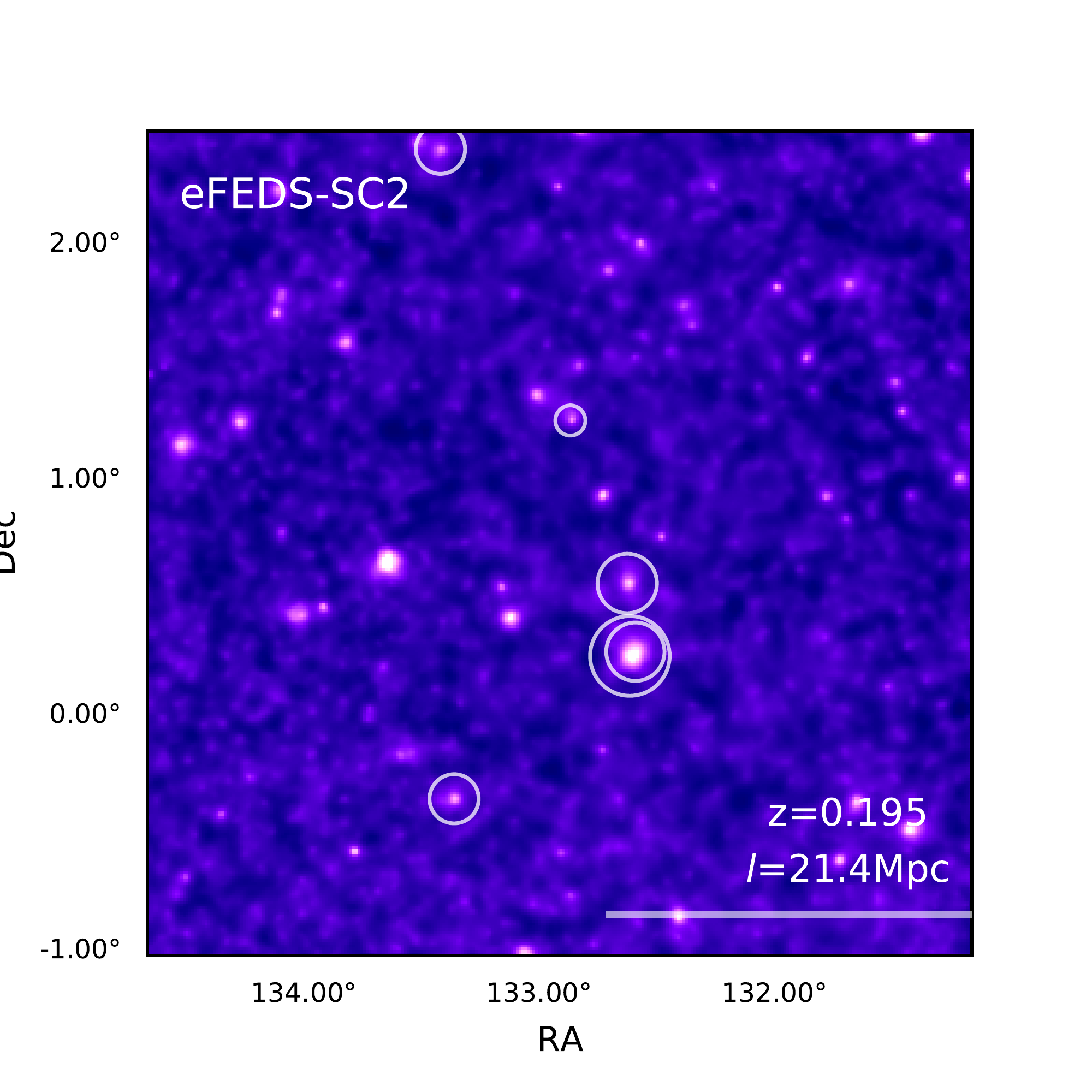}
\includegraphics[width=0.245\textwidth, trim=0 20 40 50, clip]{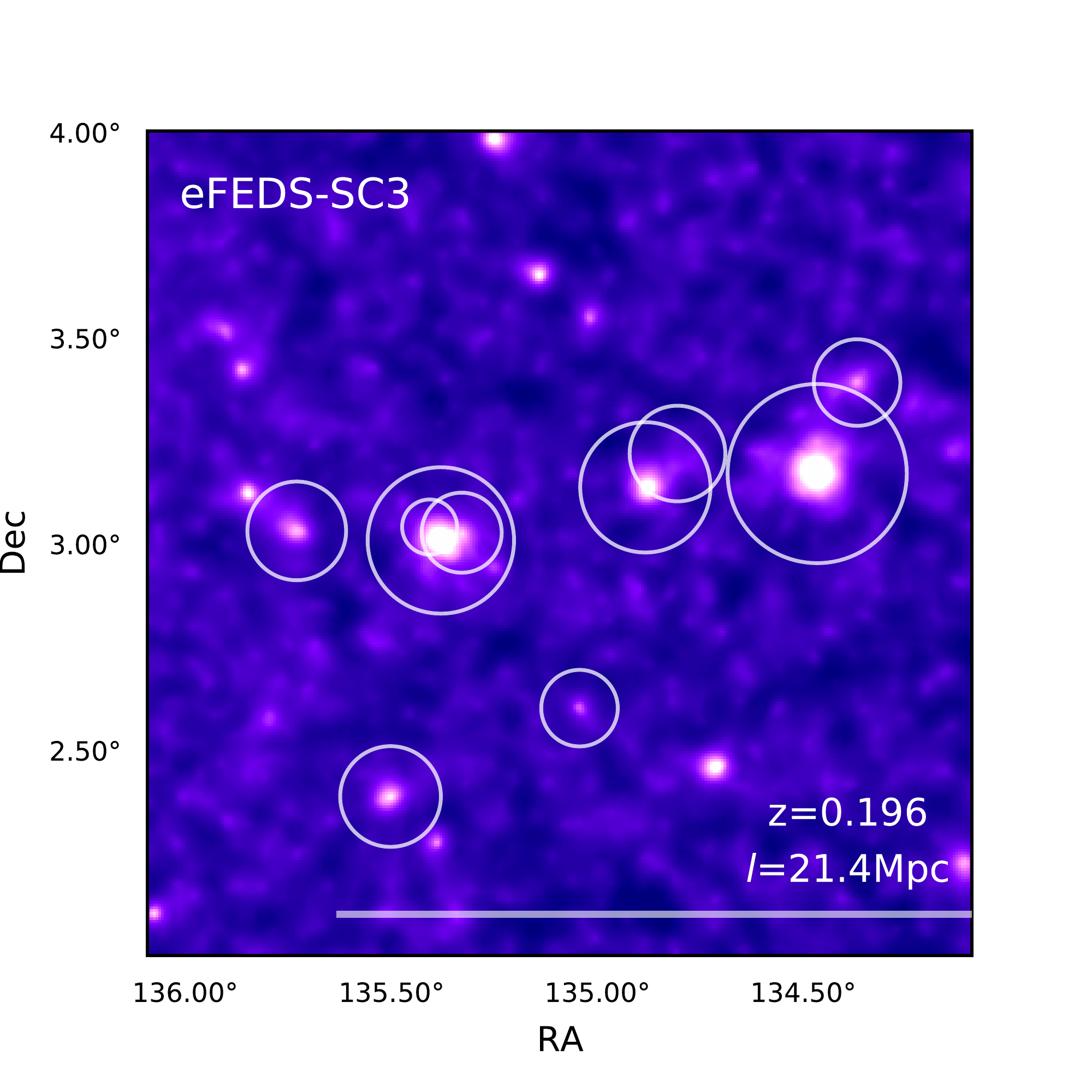}
\includegraphics[width=0.245\textwidth, trim=0 20 40 50, clip]{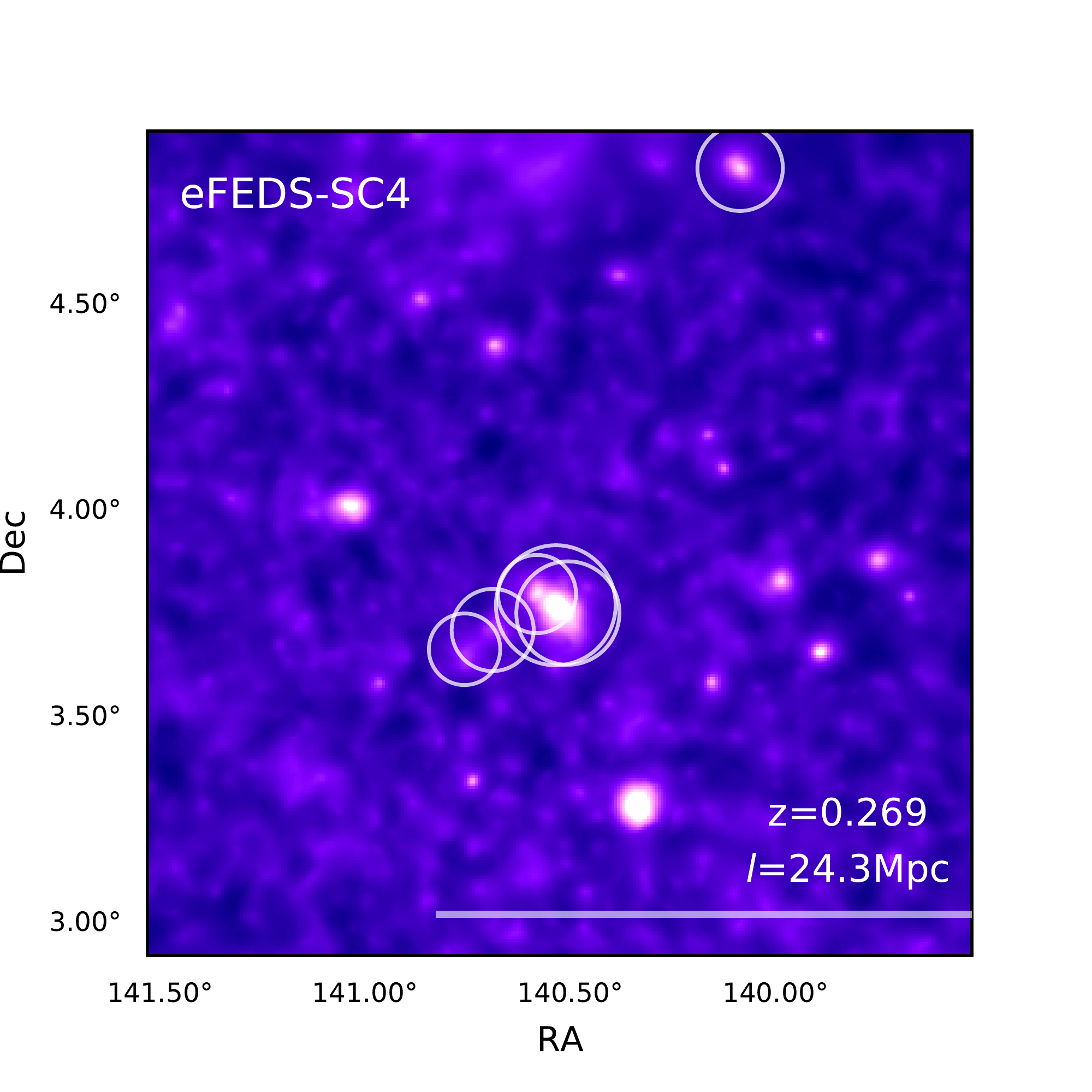}
\includegraphics[width=0.245\textwidth, trim=0 20 40 50, clip]{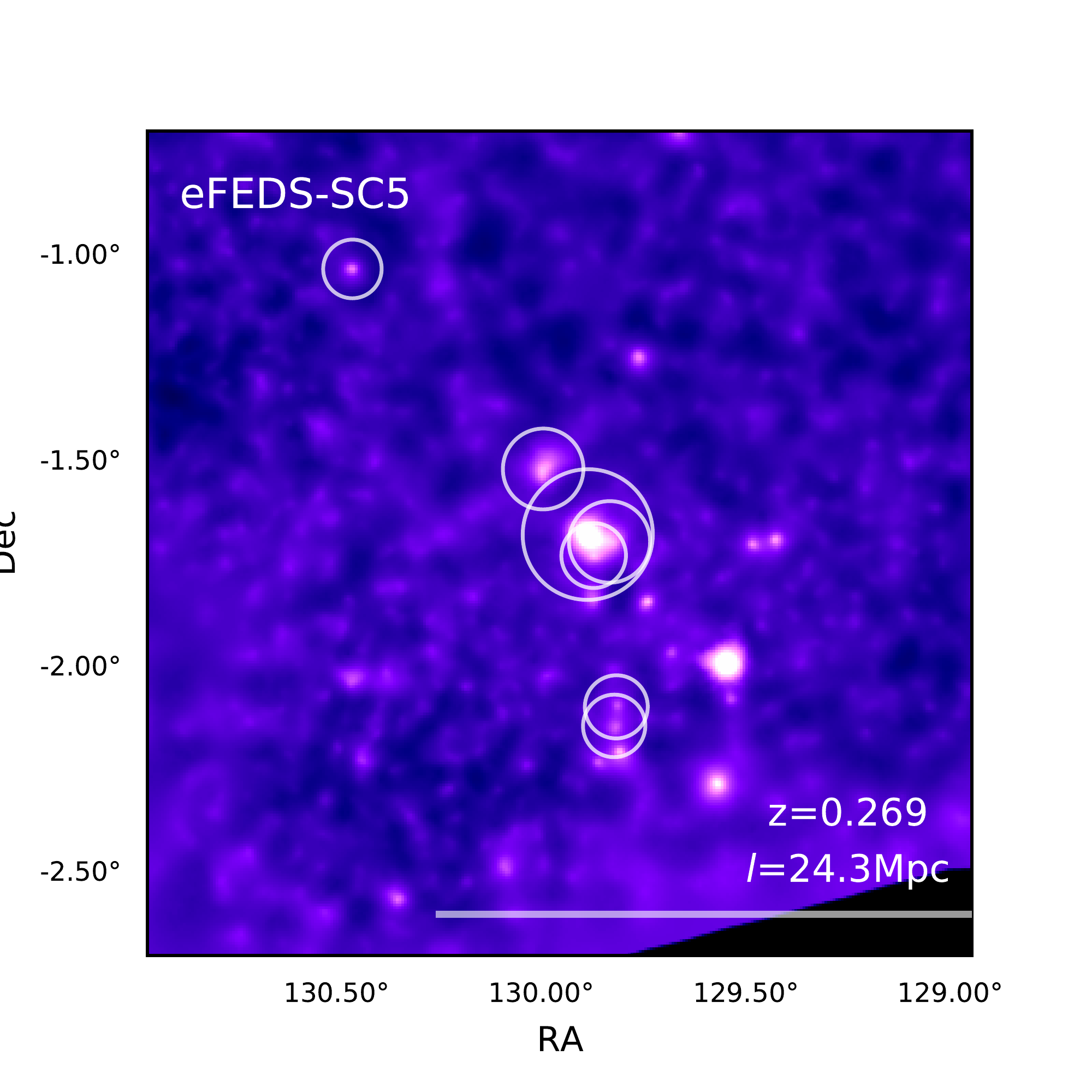}
\includegraphics[width=0.245\textwidth, trim=0 20 40 50, clip]{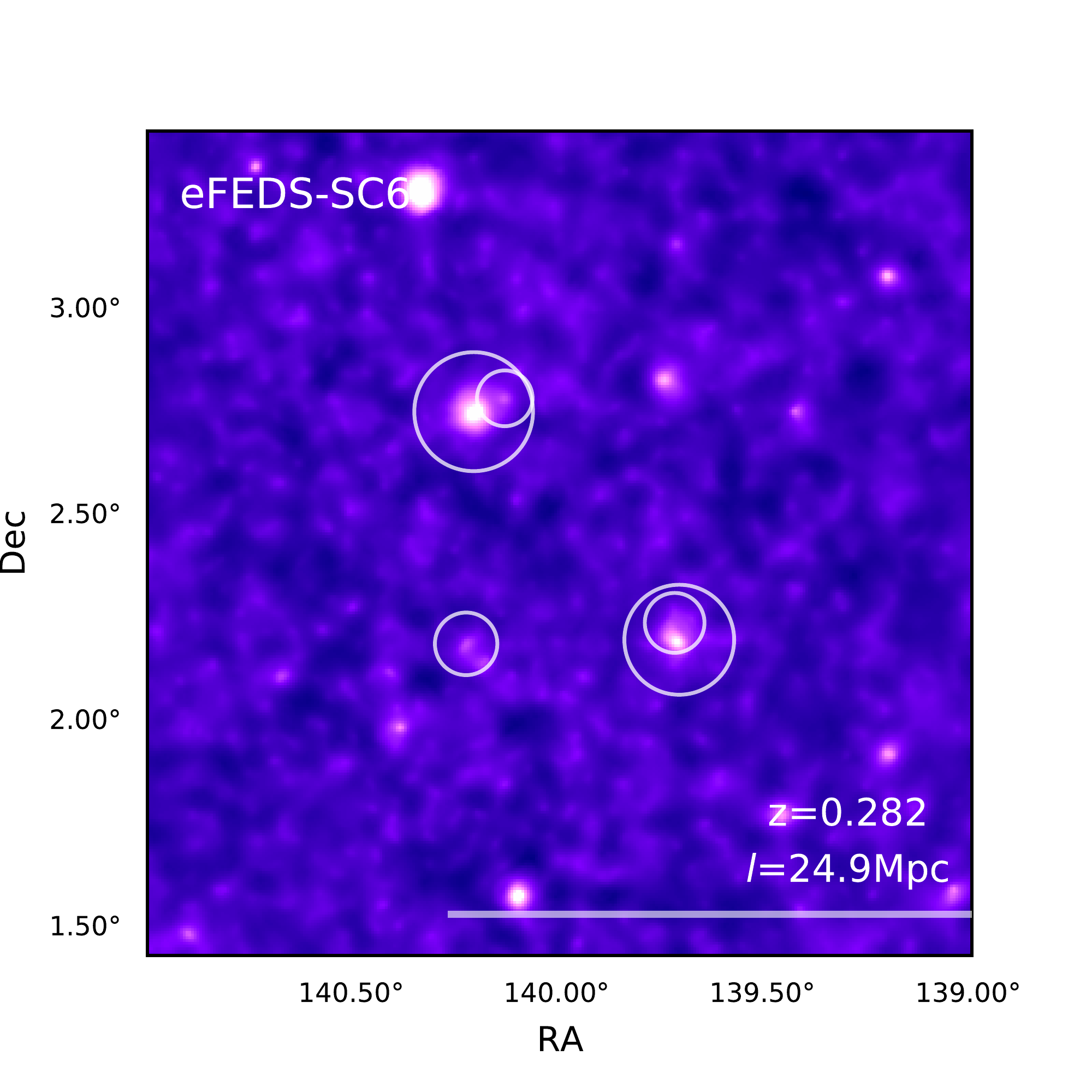}
\includegraphics[width=0.245\textwidth, trim=0 20 40 50, clip]{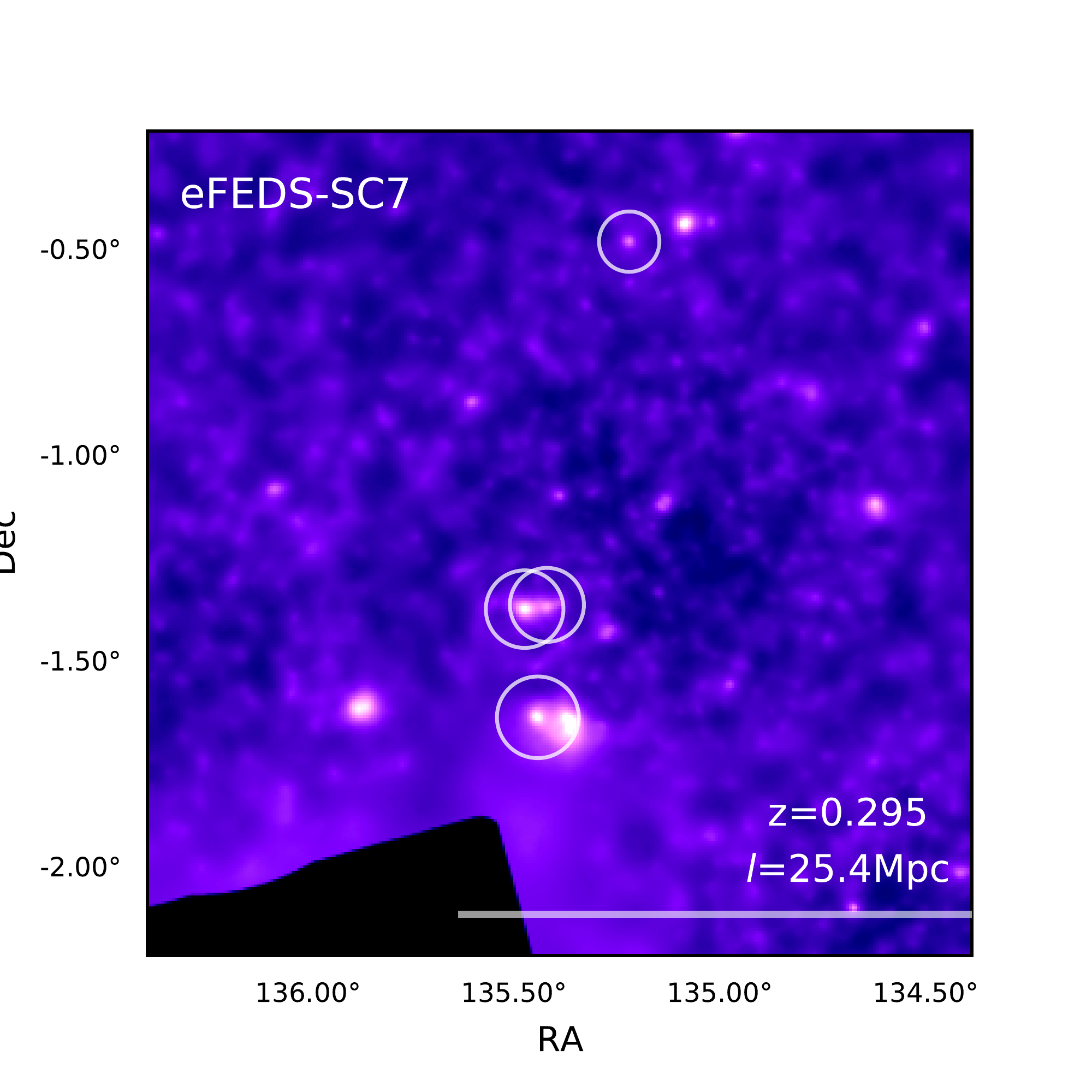}
\includegraphics[width=0.245\textwidth, trim=0 20 40 50, clip]{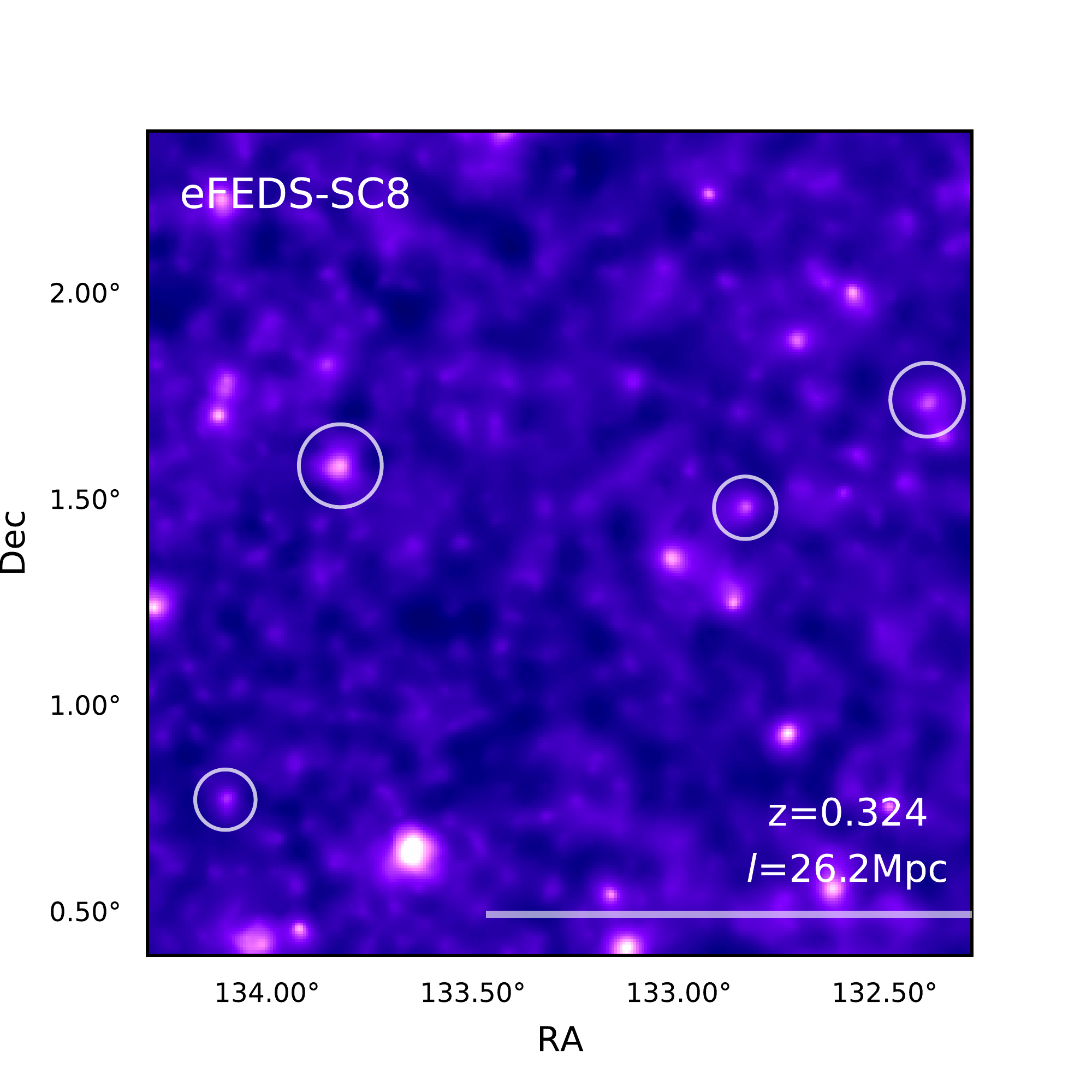}
\includegraphics[width=0.245\textwidth, trim=0 20 40 50, clip]{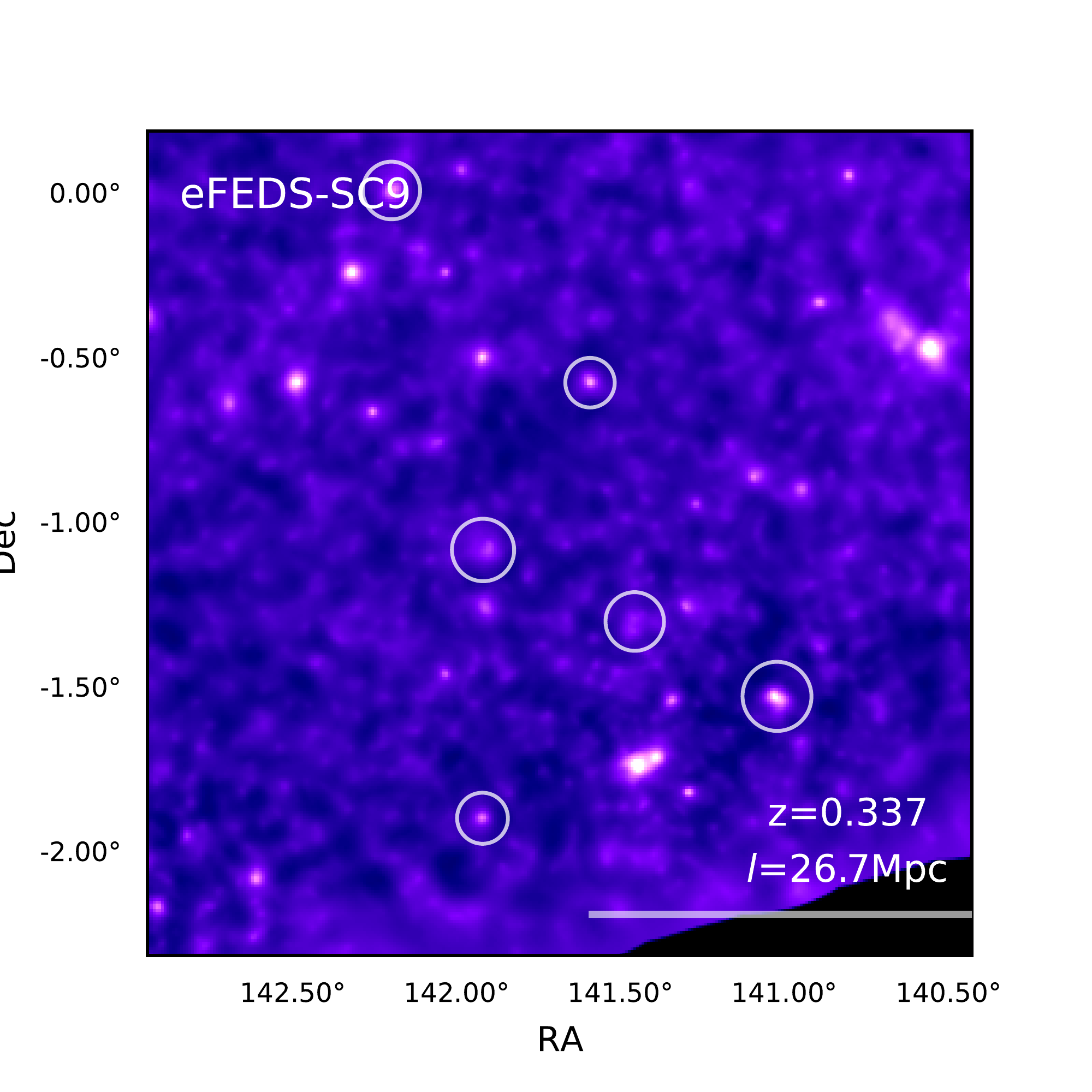}
\includegraphics[width=0.245\textwidth, trim=0 20 40 50, clip]{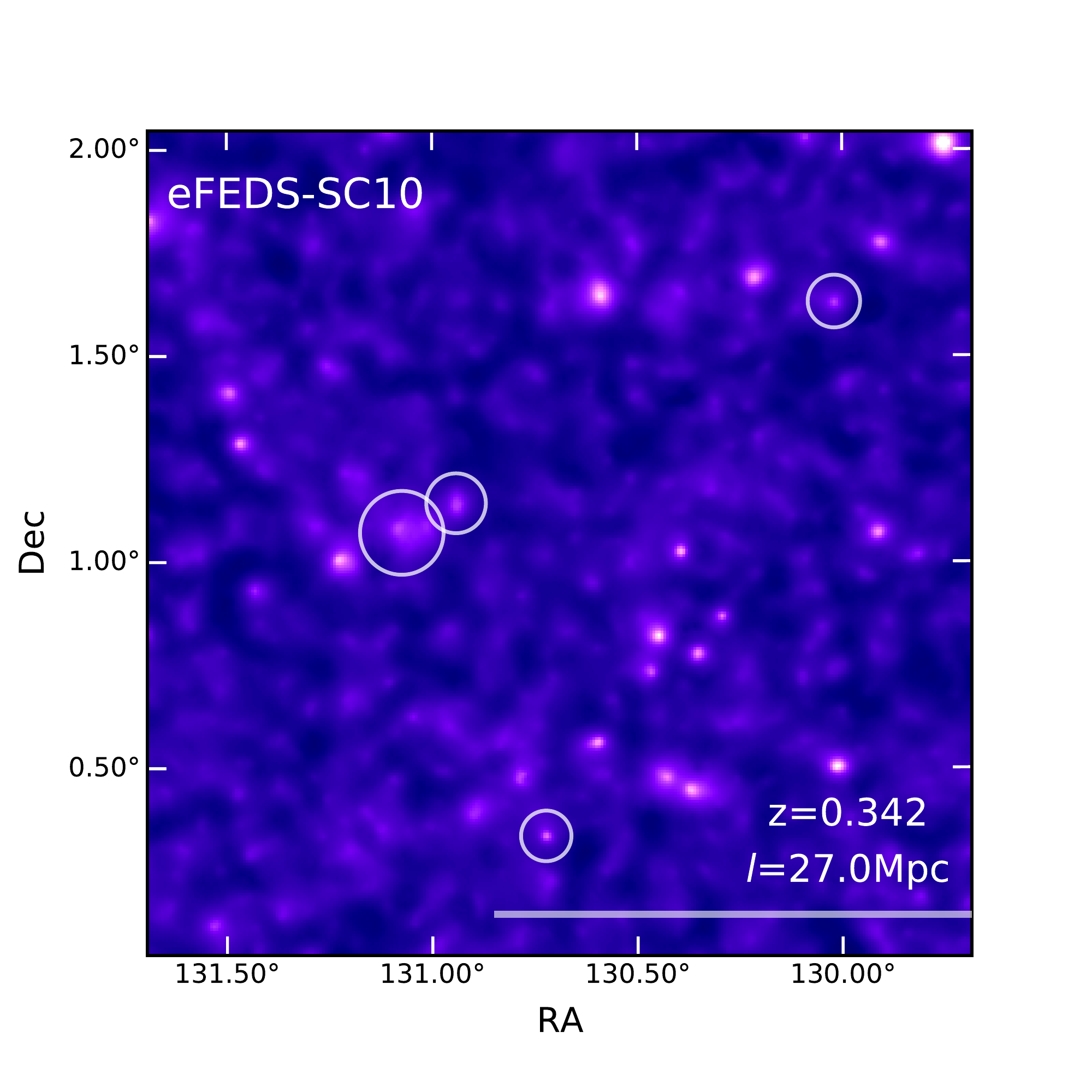}
\includegraphics[width=0.245\textwidth, trim=0 20 40 50, clip]{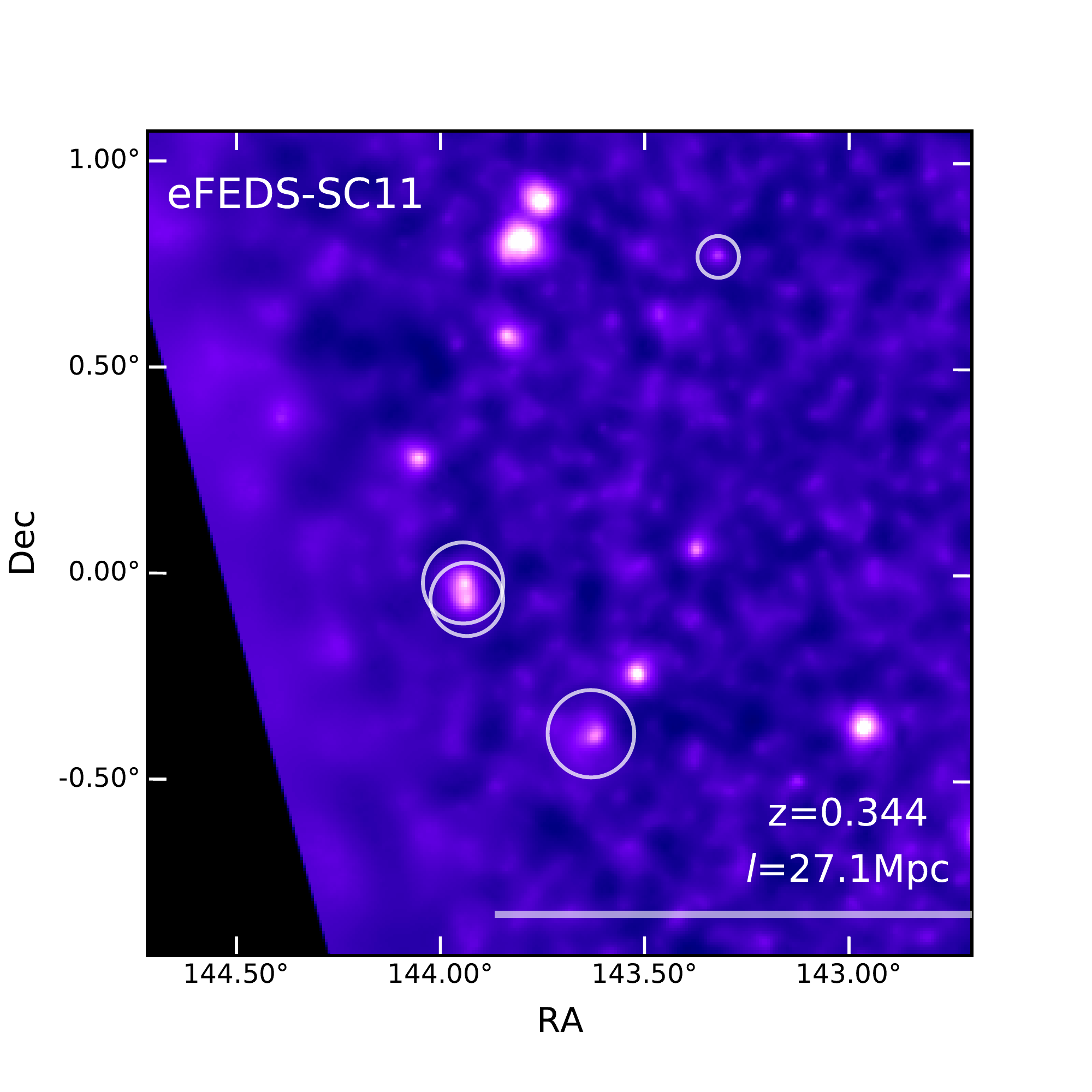}
\includegraphics[width=0.245\textwidth, trim=0 20 40 50, clip]{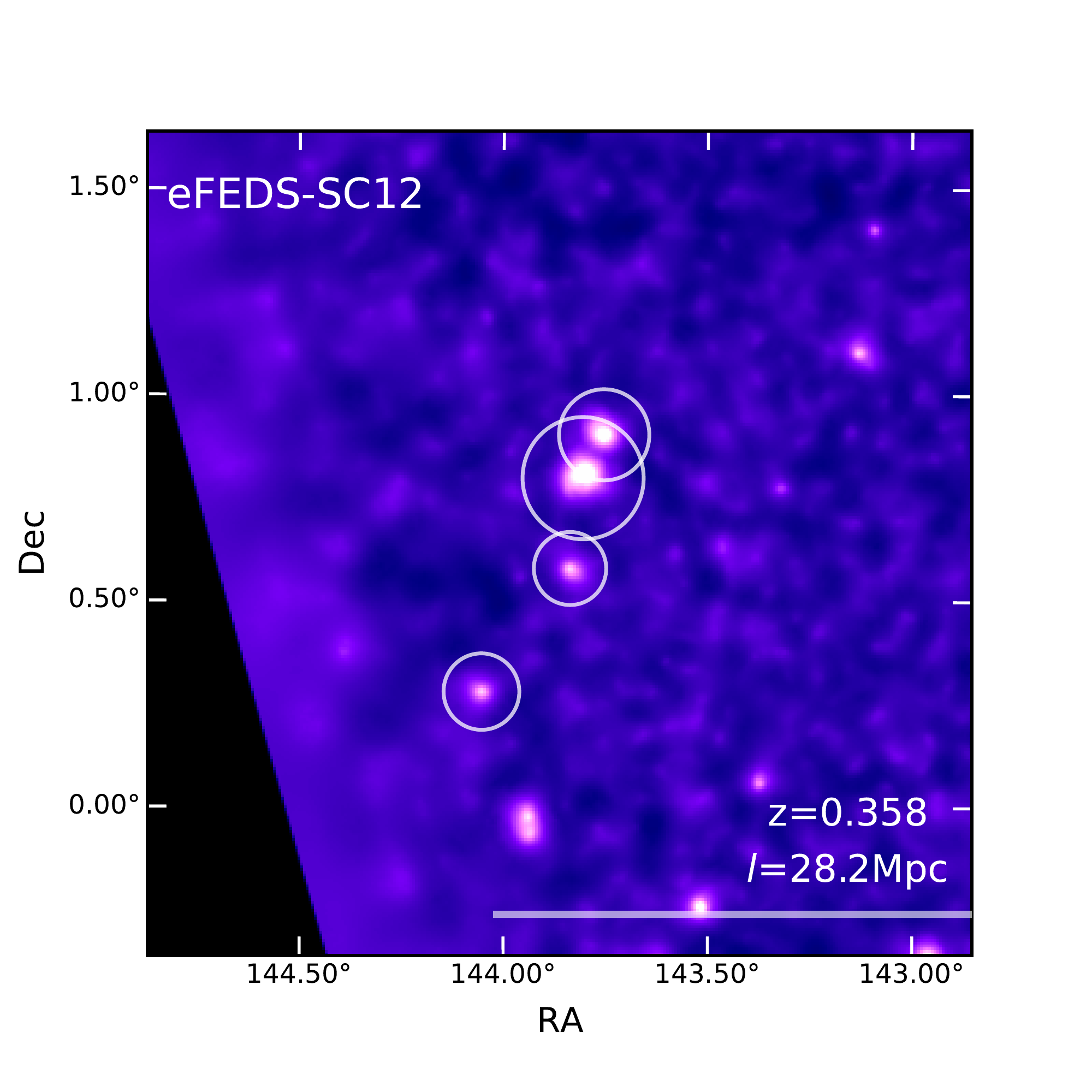}
\includegraphics[width=0.245\textwidth, trim=0 20 40 50, clip]{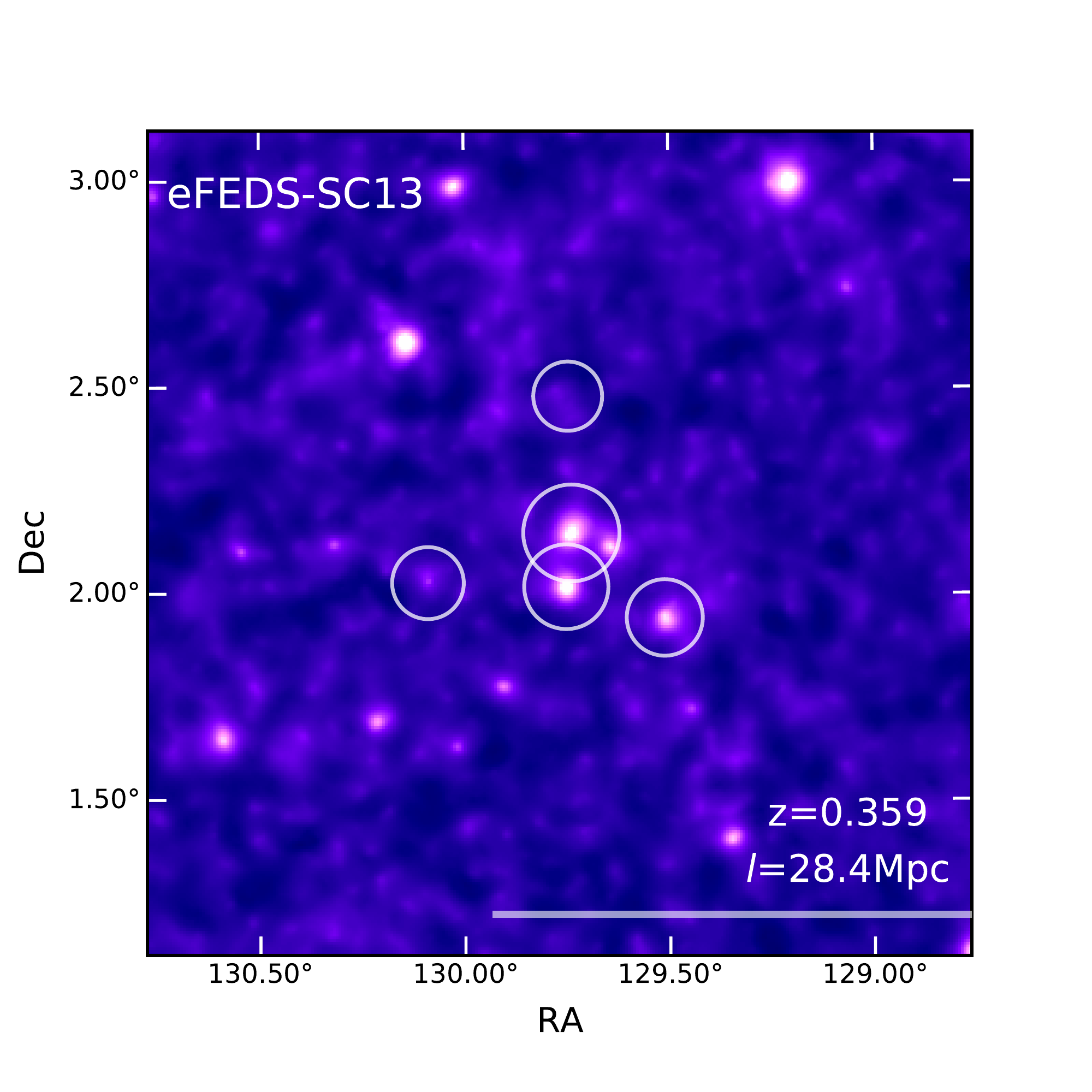}
\includegraphics[width=0.245\textwidth, trim=0 20 40 50, clip]{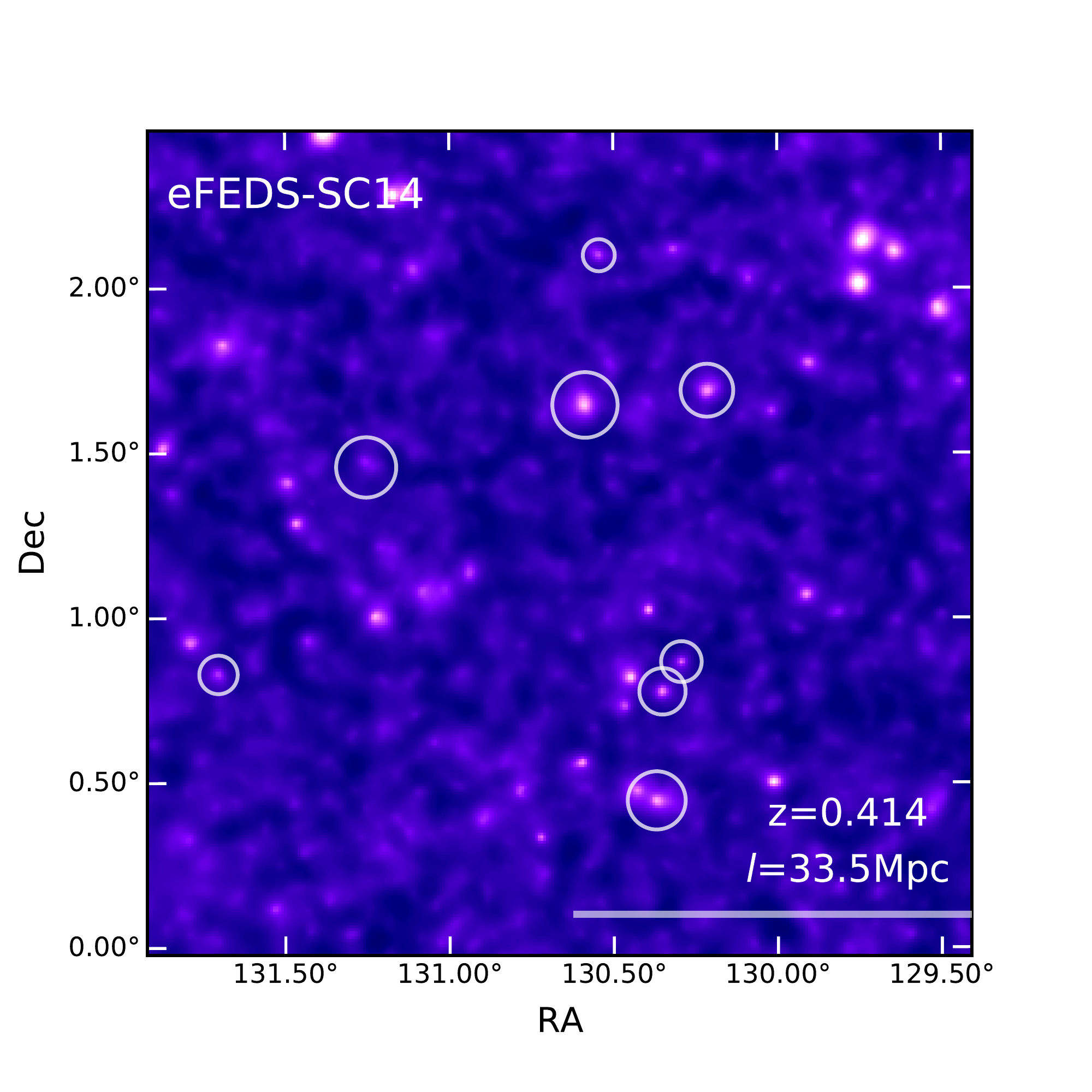}
\includegraphics[width=0.245\textwidth, trim=0 20 40 50, clip]{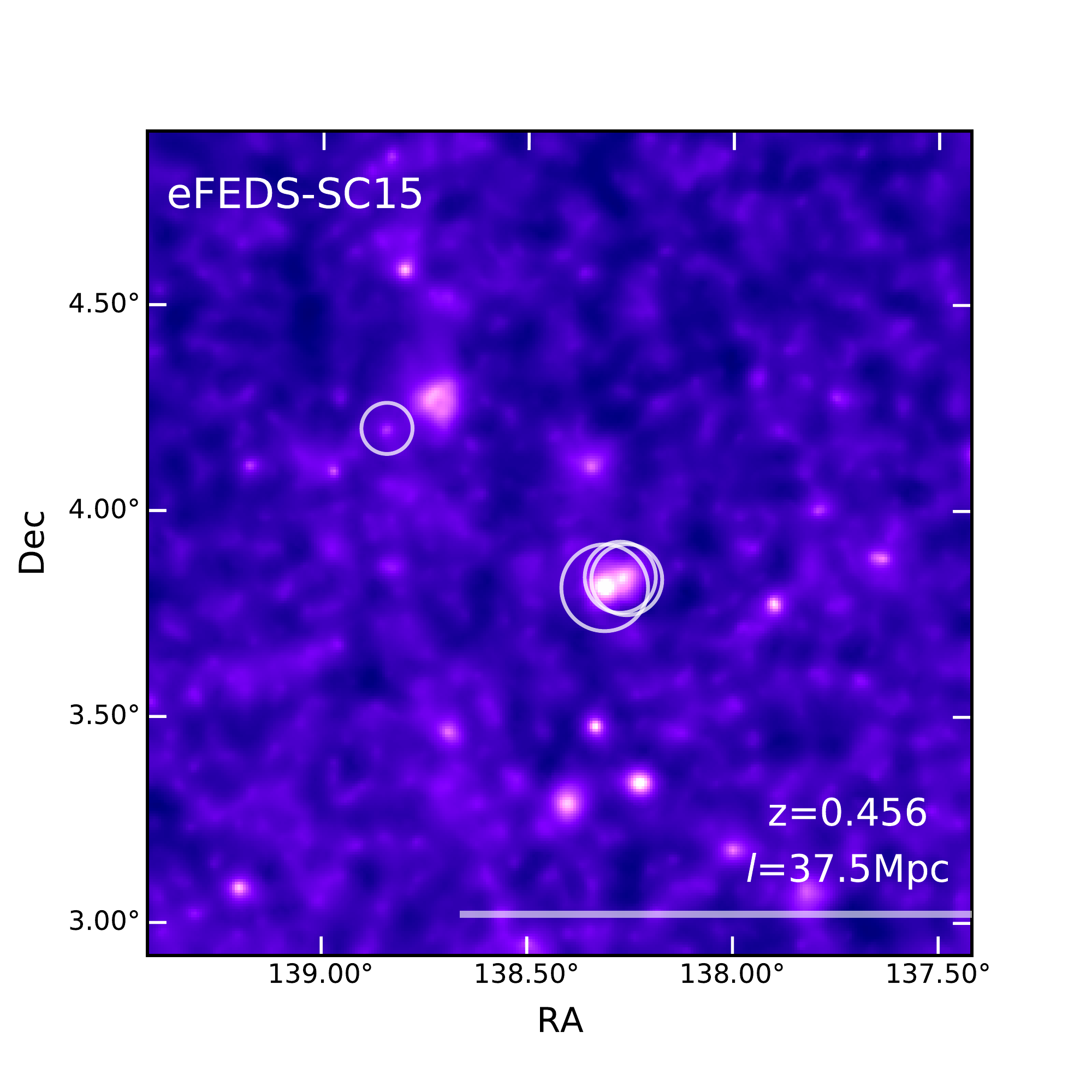}
\includegraphics[width=0.245\textwidth, trim=0 20 40 50, clip]{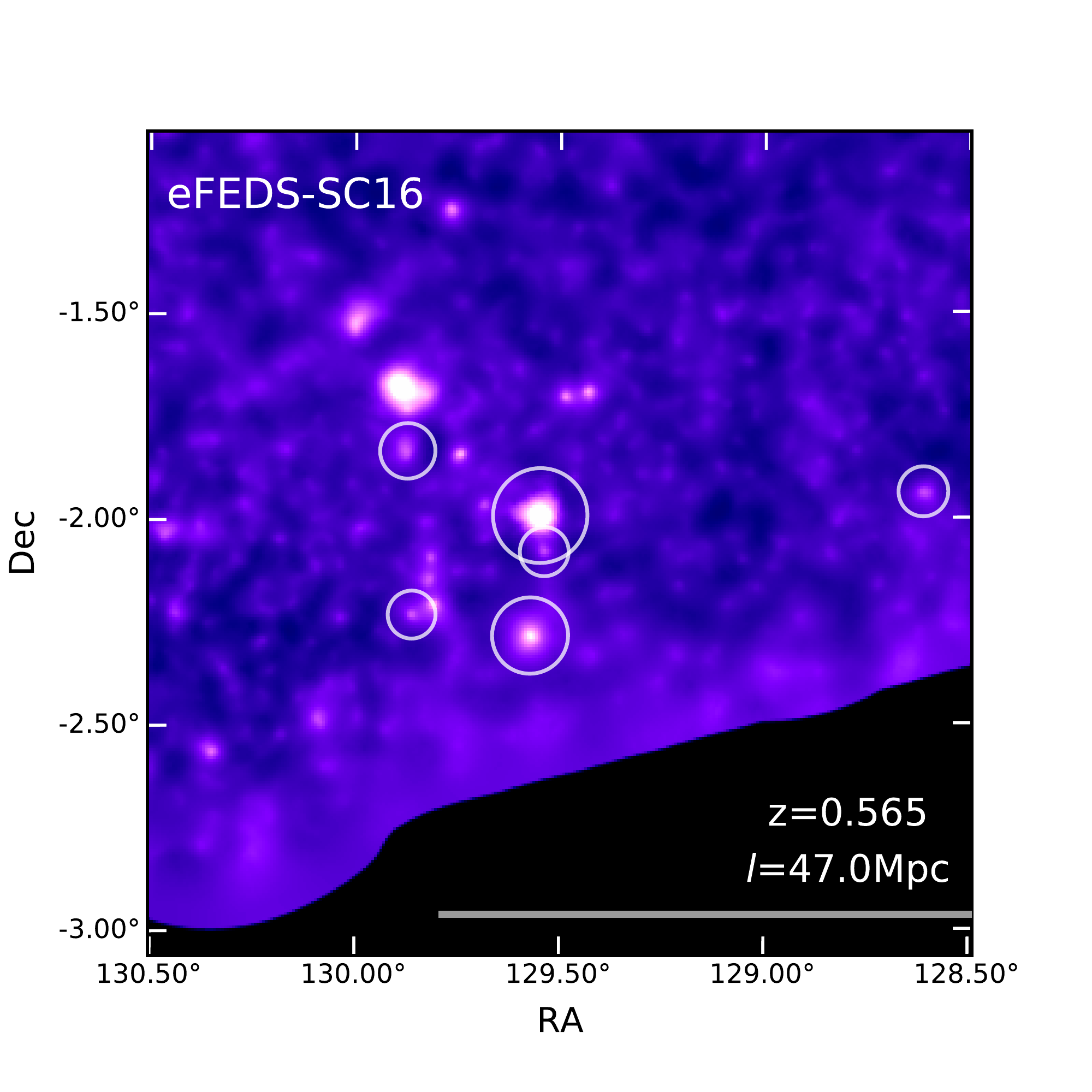}
\includegraphics[width=0.245\textwidth, trim=0 20 40 50, clip]{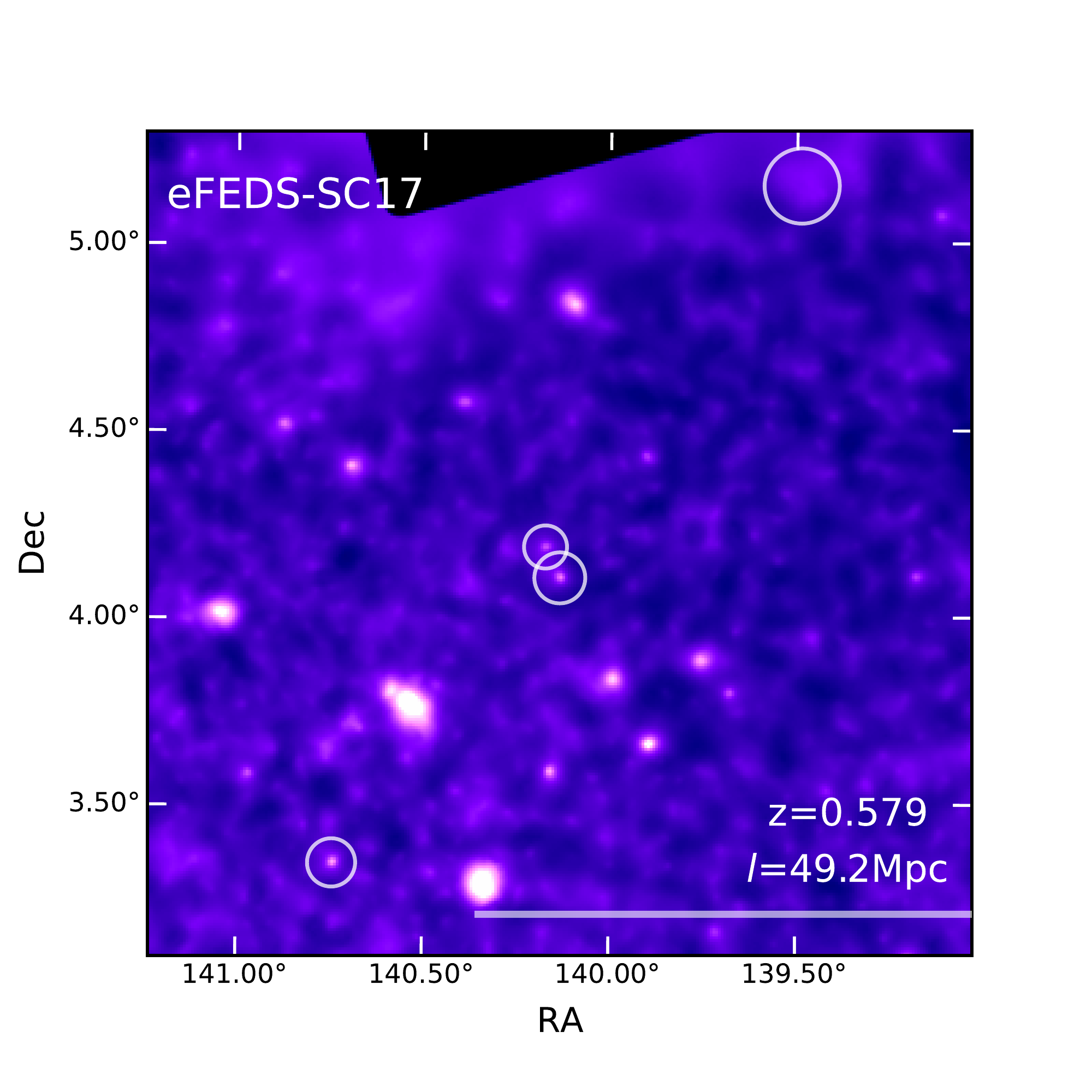}
\includegraphics[width=0.245\textwidth, trim=0 20 40 50, clip]{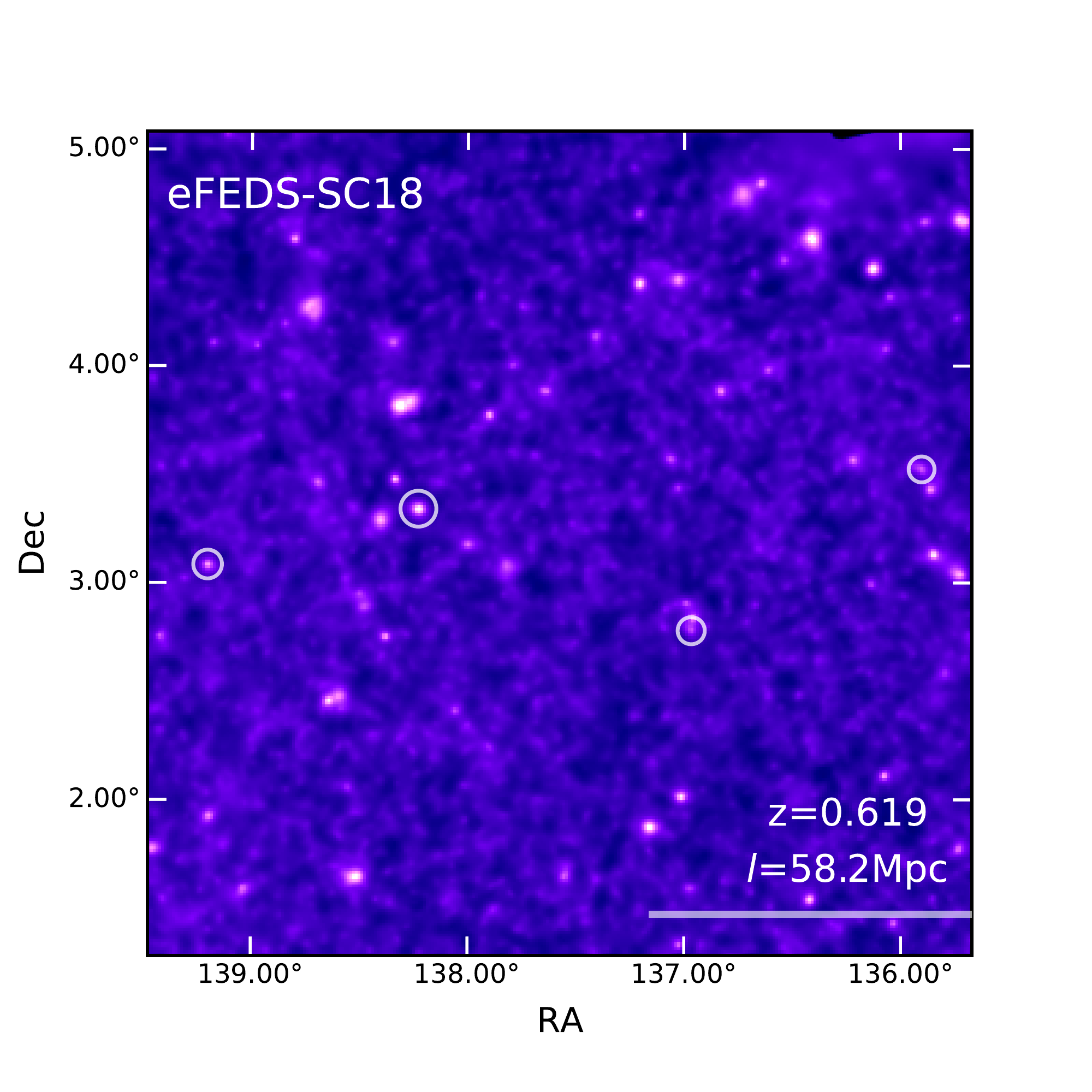}
\includegraphics[width=0.245\textwidth, trim=0 20 40 50, clip]{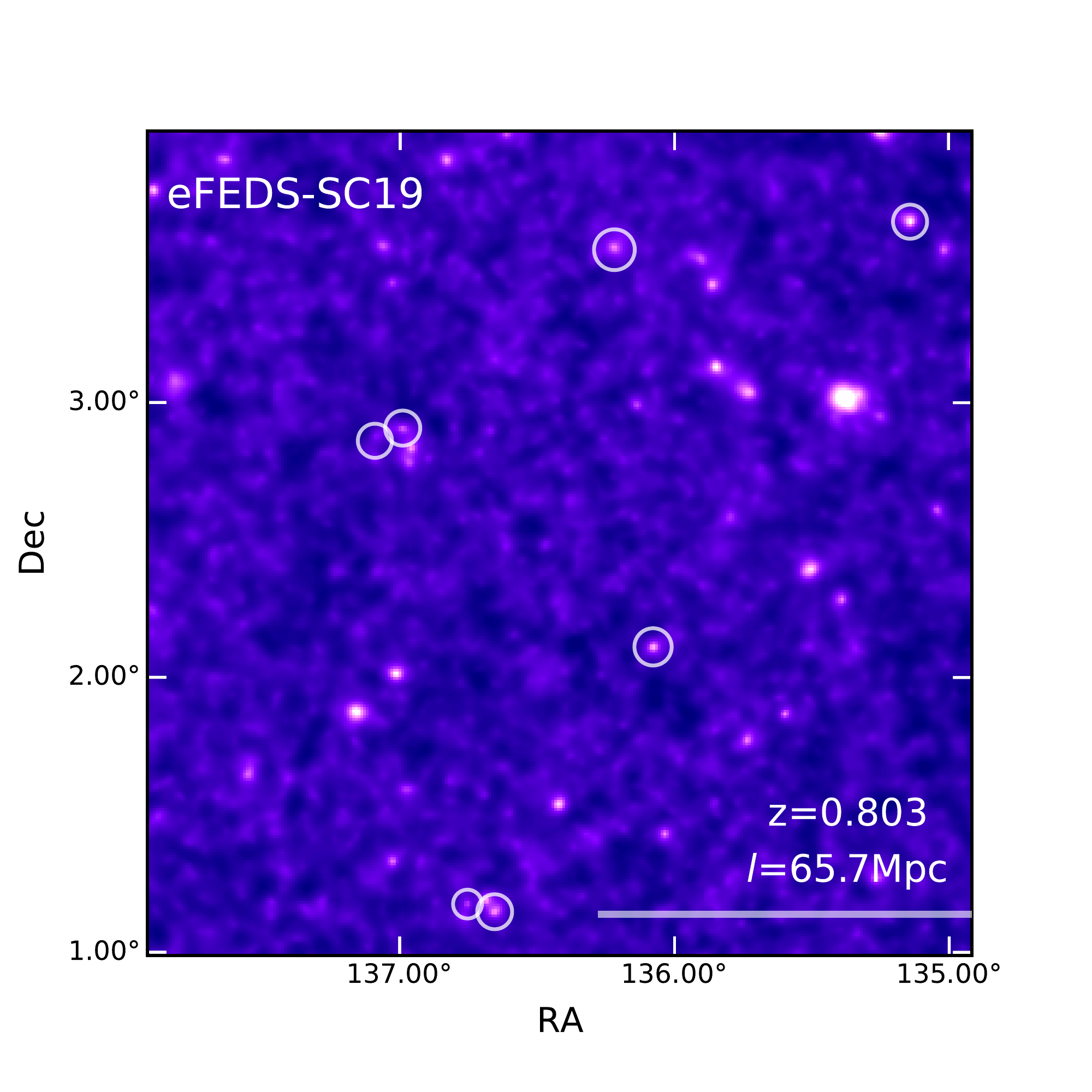}
\caption{X-ray images of the 19 superclusters we detected in the eFEDS field. The images are generated in the same way as for Fig.~\ref{fig:image_efeds_large}. The white circles mark the estimated virial radii of the member clusters of each supercluster. The thick bar in the lower right corner indicates the linking length at the supercluster redshift. }
\label{fig:sc_image}
\end{center}
\end{figure*}

\begin{figure*}
\begin{center}
\includegraphics[width=0.245\textwidth, trim=0 20 40 50, clip]{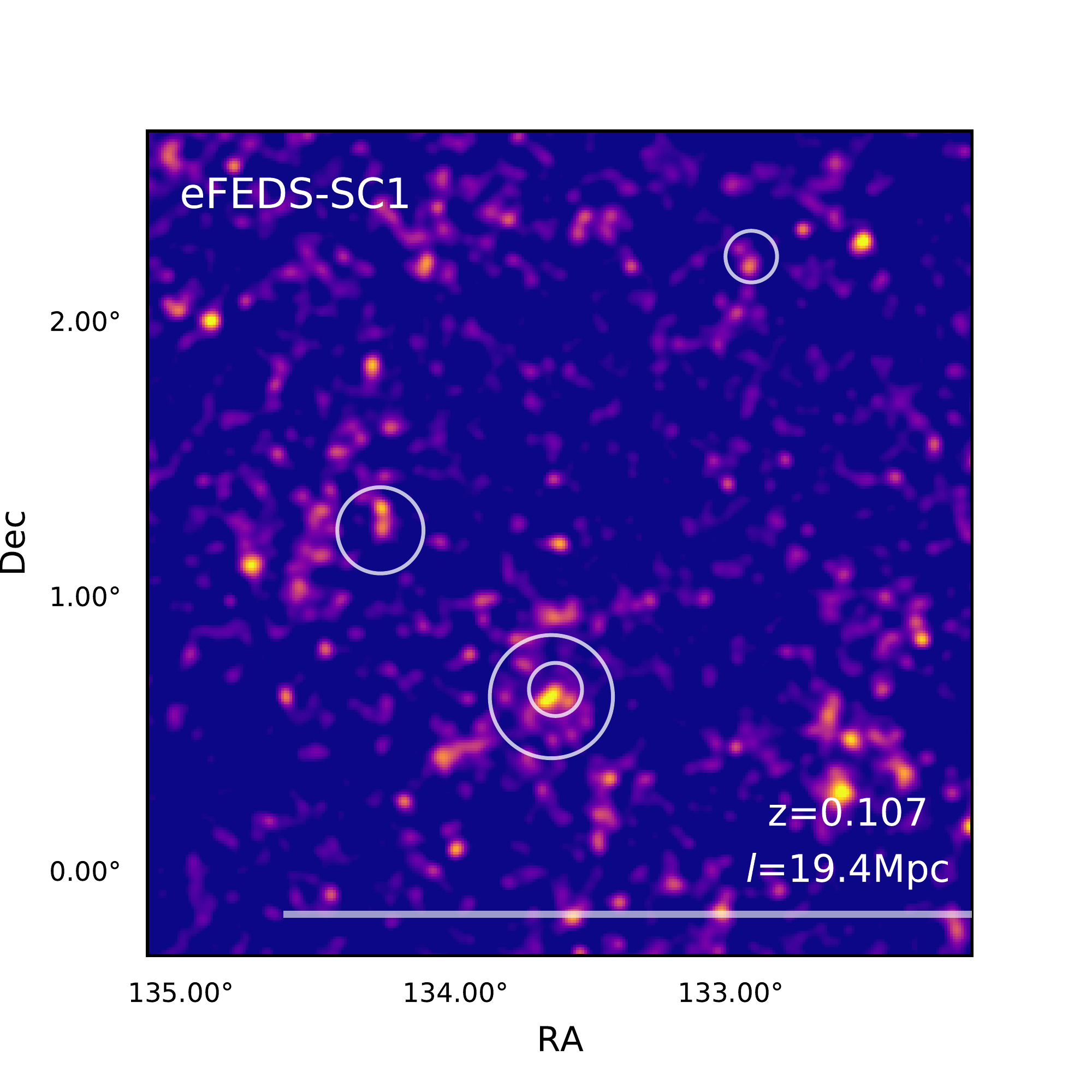}
\includegraphics[width=0.245\textwidth, trim=0 20 40 50, clip]{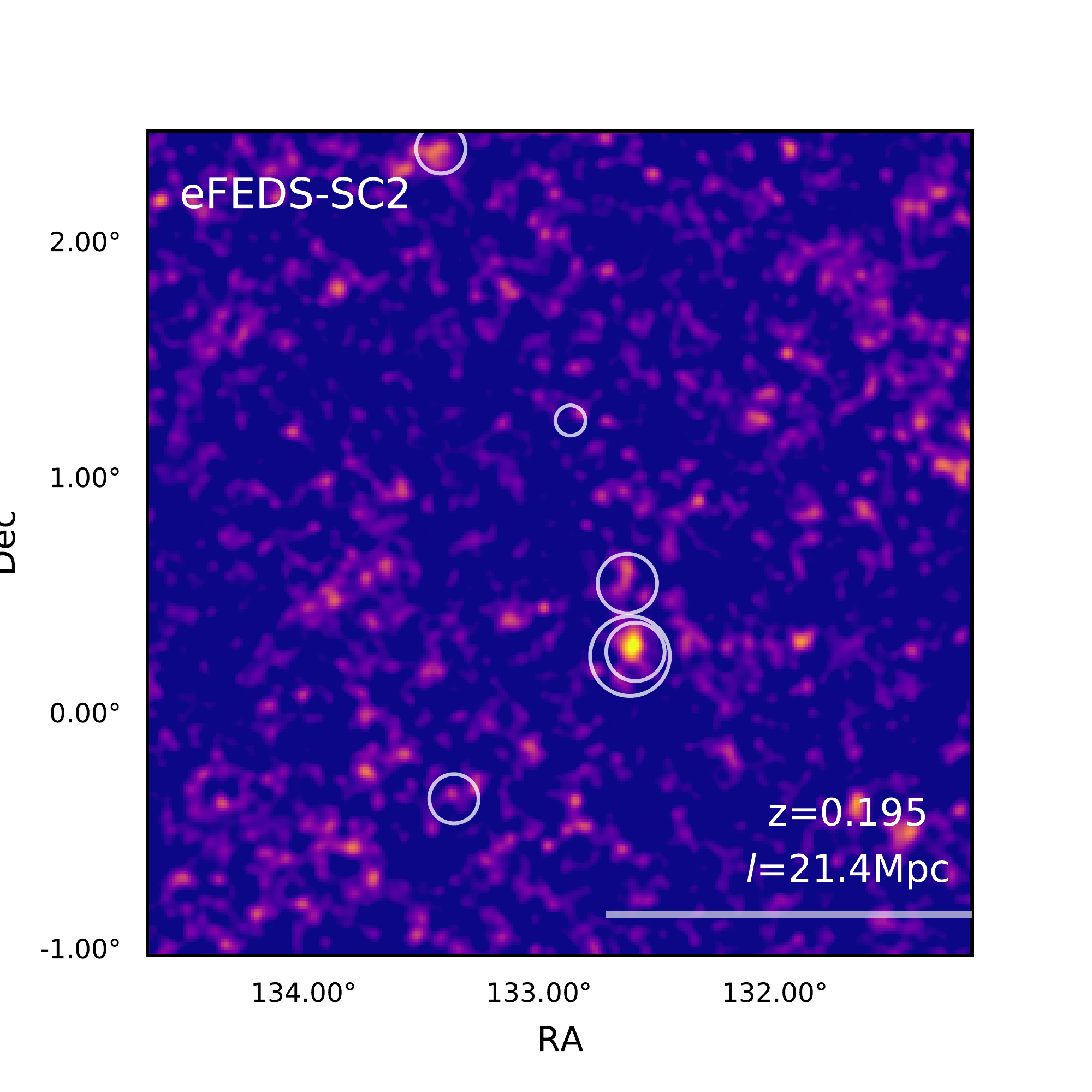}
\includegraphics[width=0.245\textwidth, trim=0 20 40 50, clip]{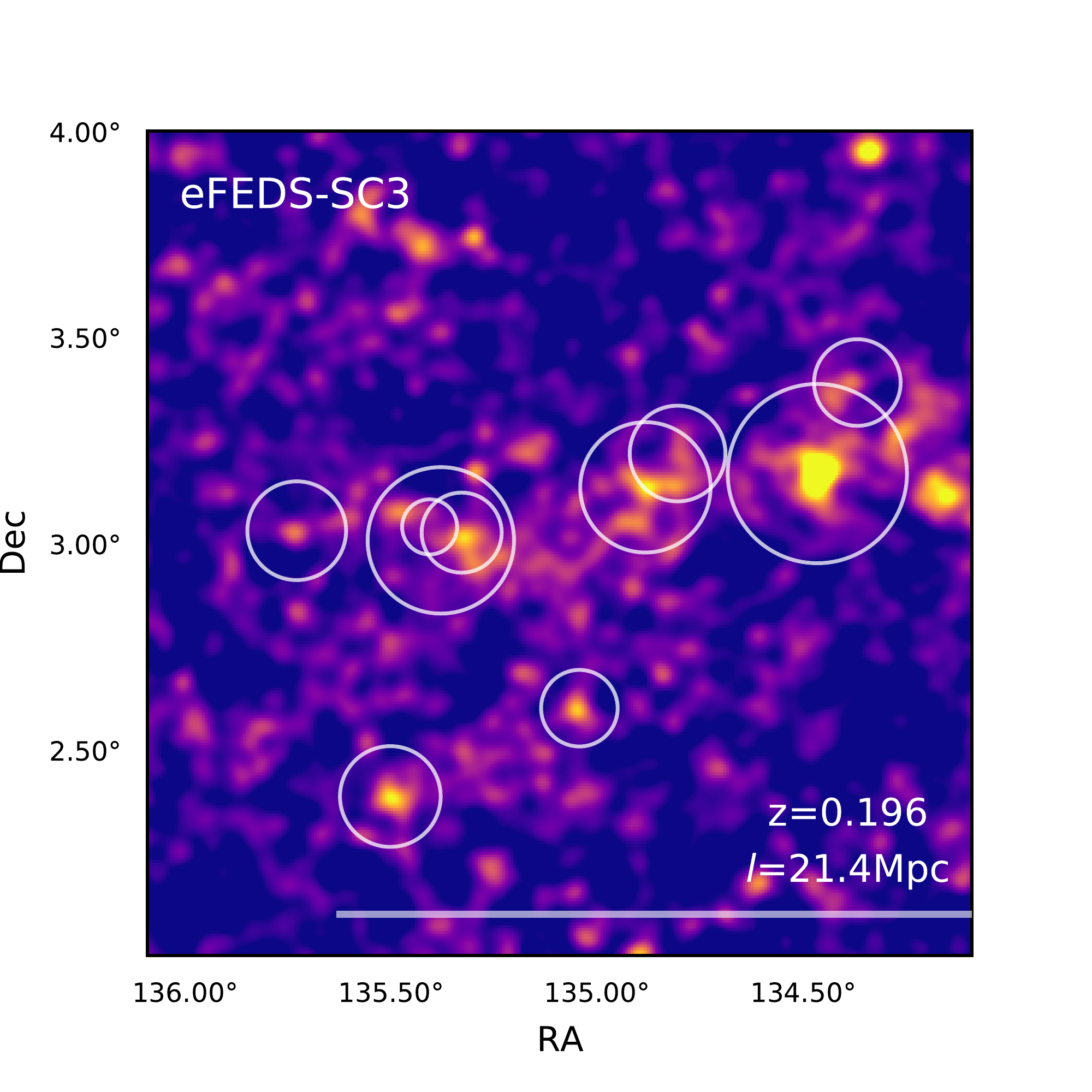}
\includegraphics[width=0.245\textwidth, trim=0 20 40 50, clip]{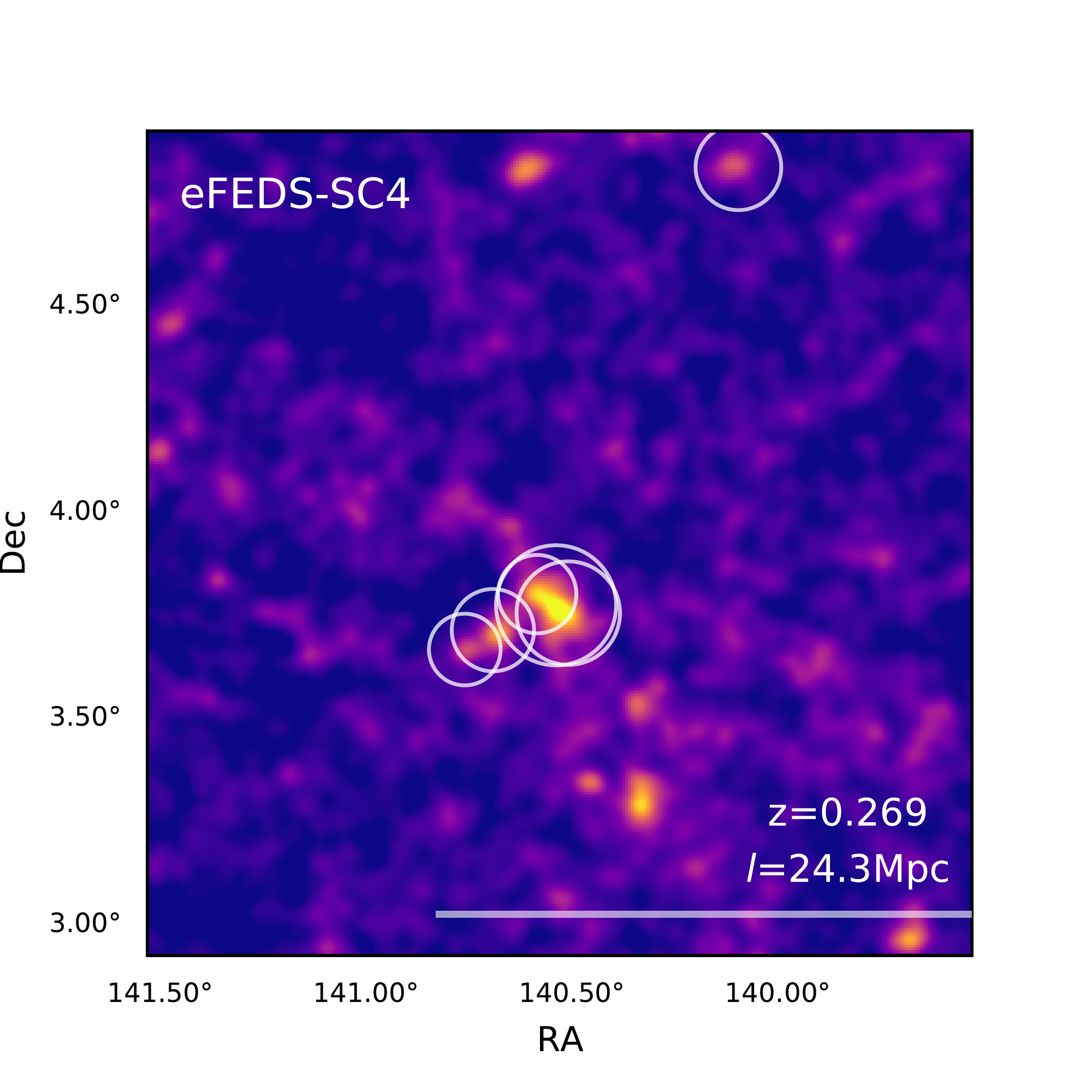}
\includegraphics[width=0.245\textwidth, trim=0 20 40 50, clip]{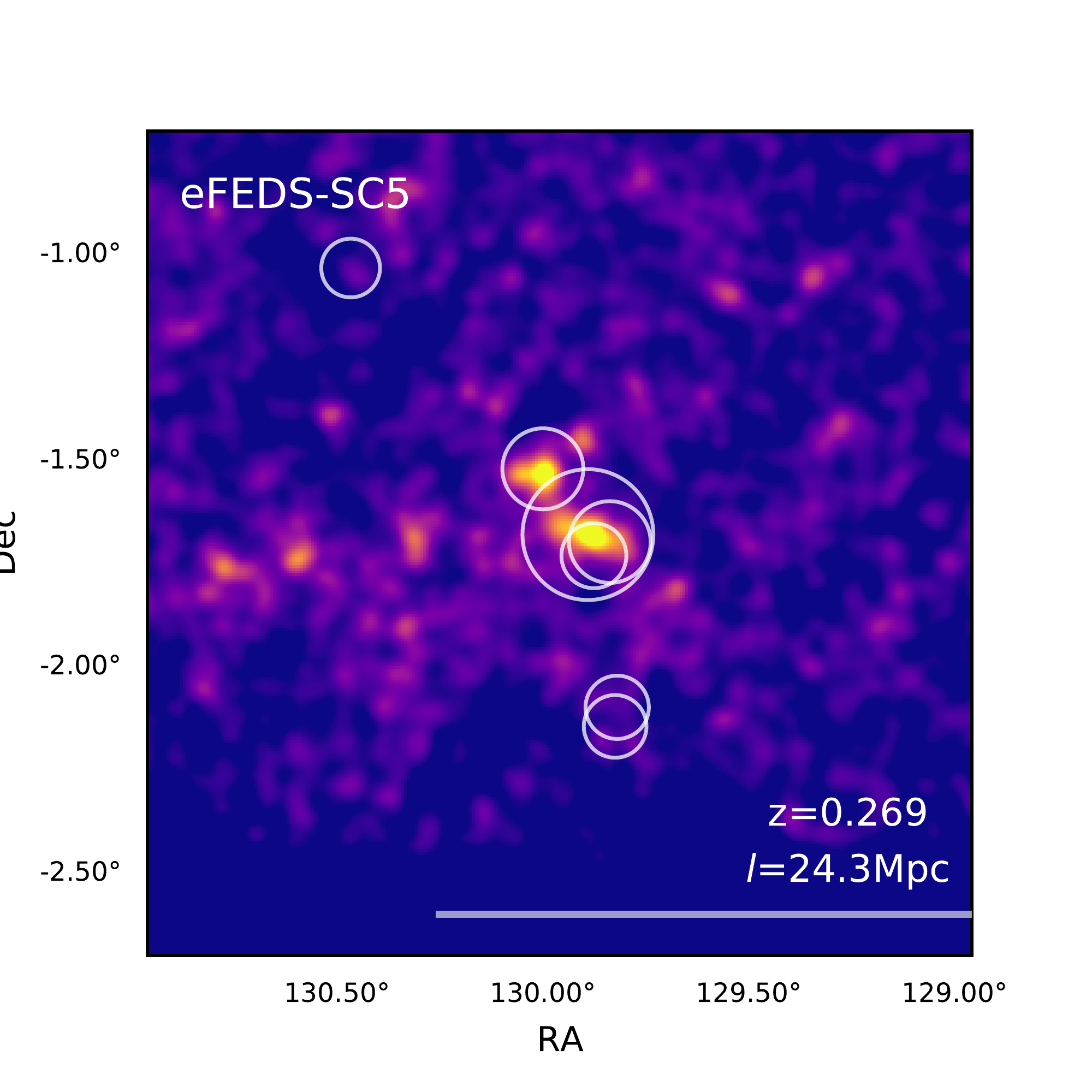}
\includegraphics[width=0.245\textwidth, trim=0 20 40 50, clip]{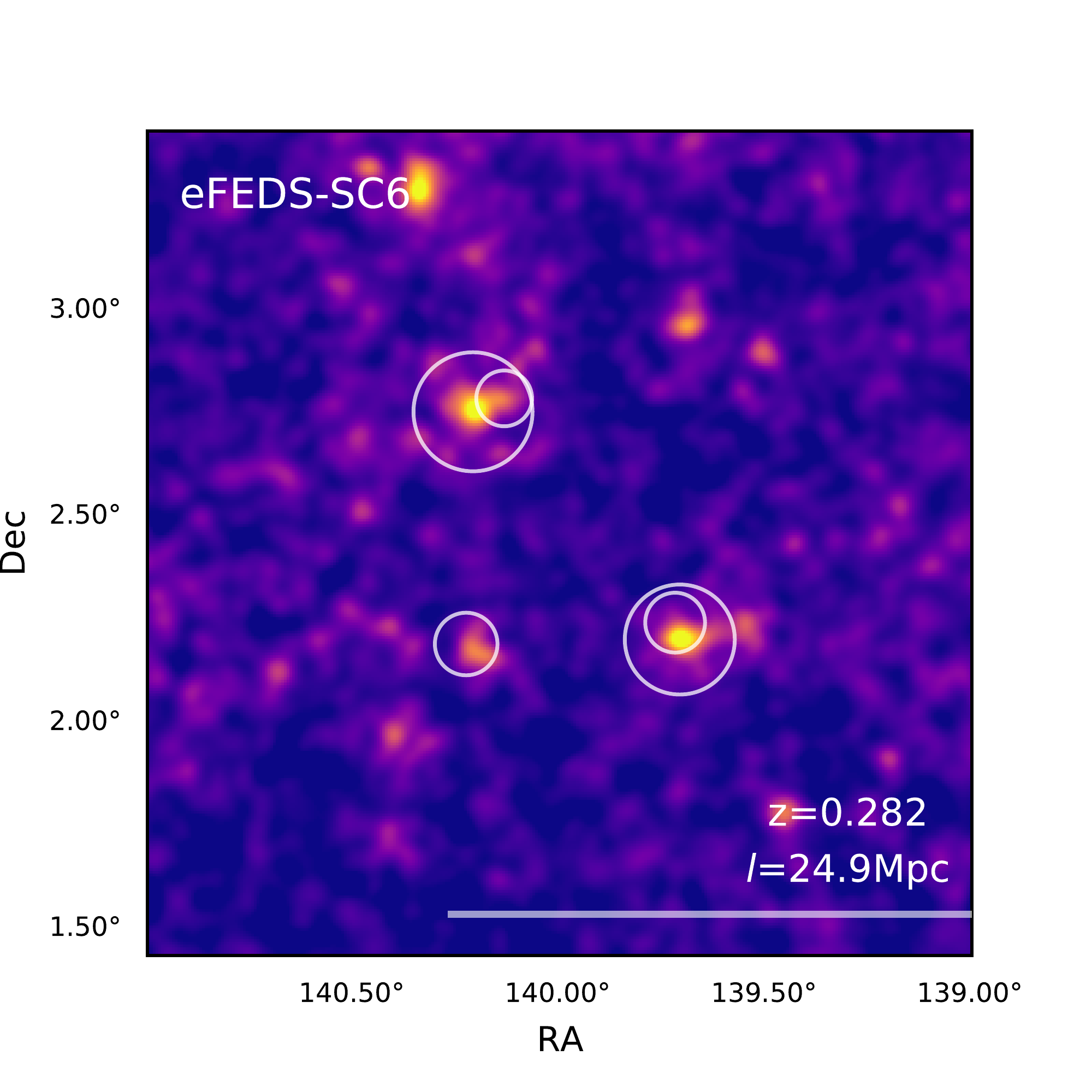}
\includegraphics[width=0.245\textwidth, trim=0 20 40 50, clip]{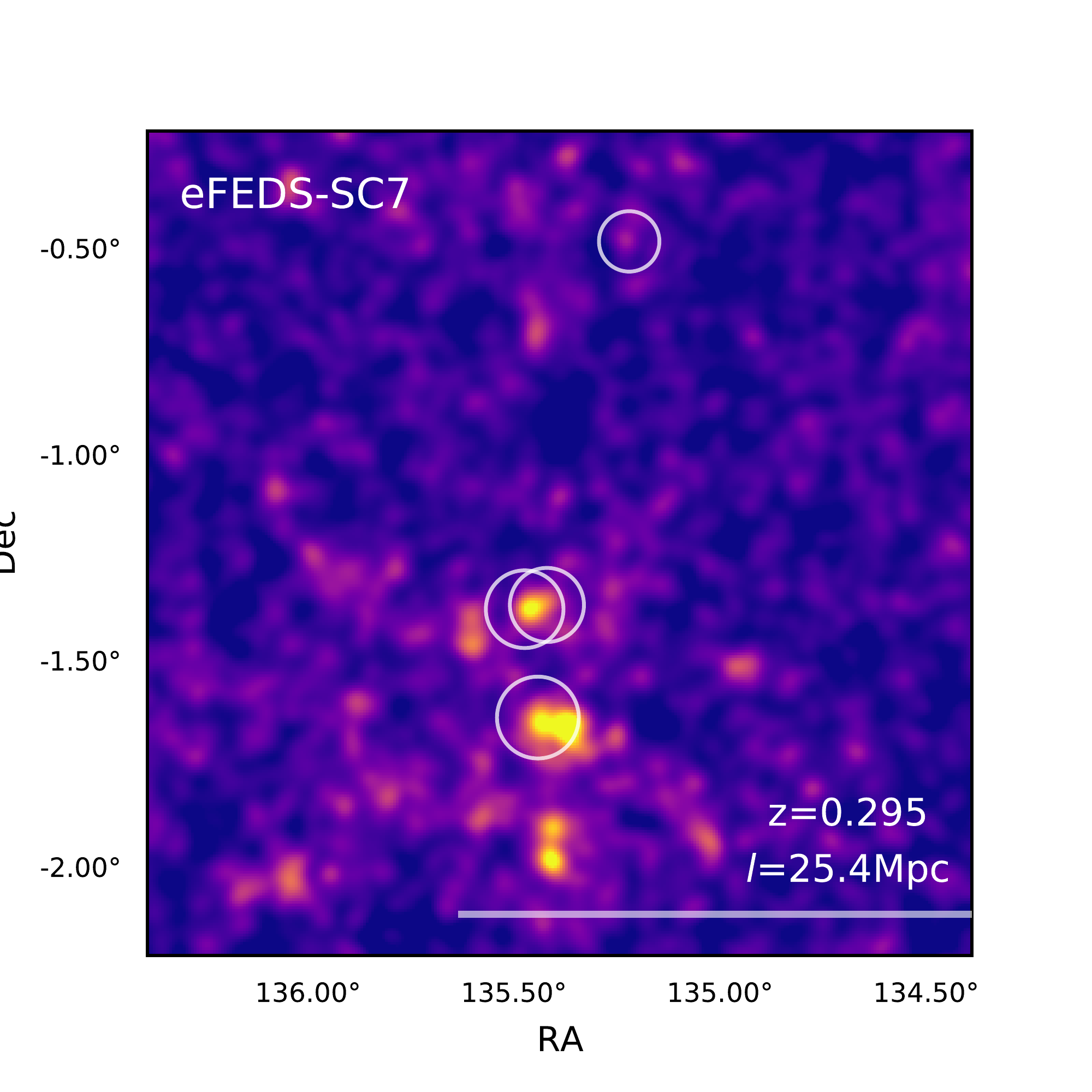}
\includegraphics[width=0.245\textwidth, trim=0 20 40 50, clip]{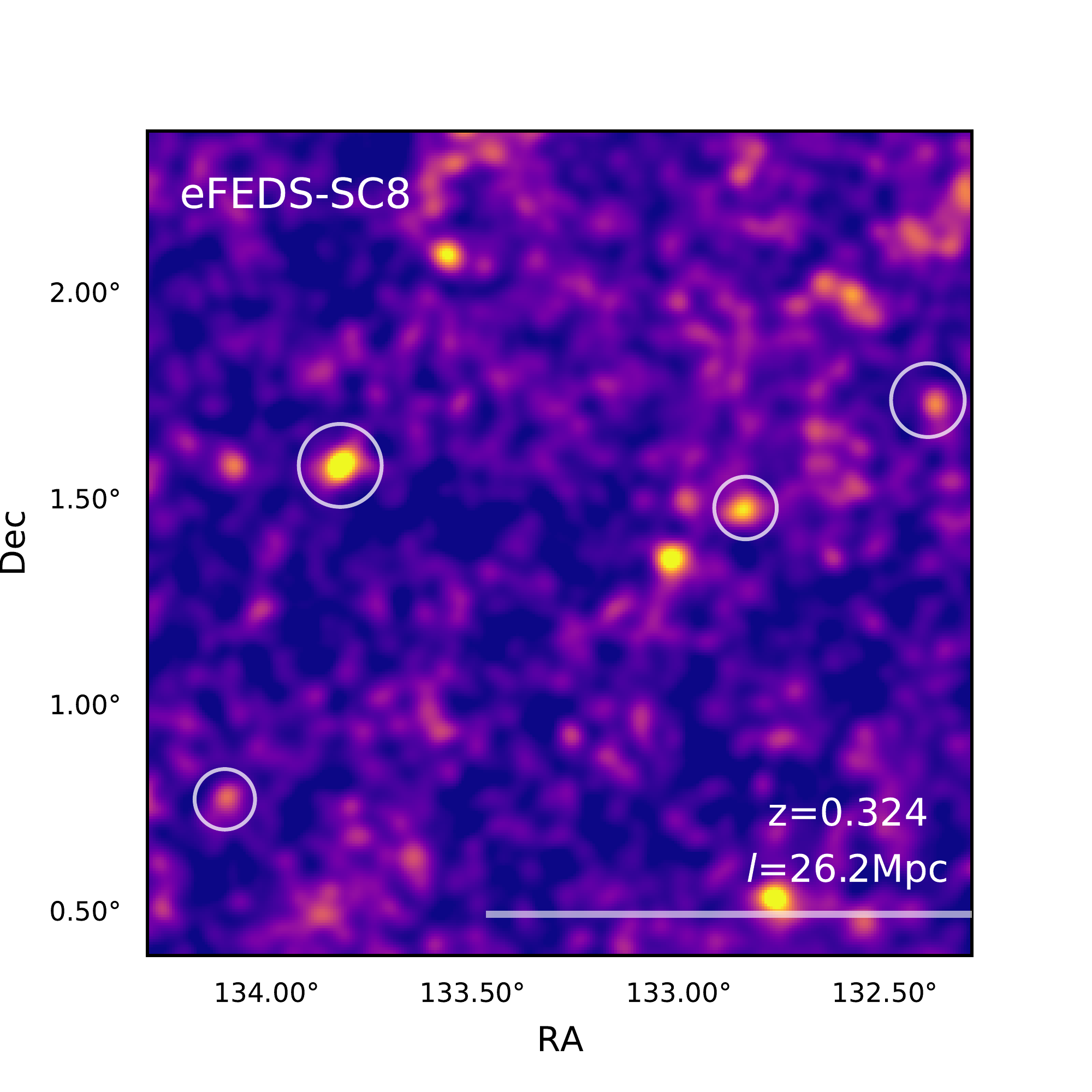}
\includegraphics[width=0.245\textwidth, trim=0 20 40 50, clip]{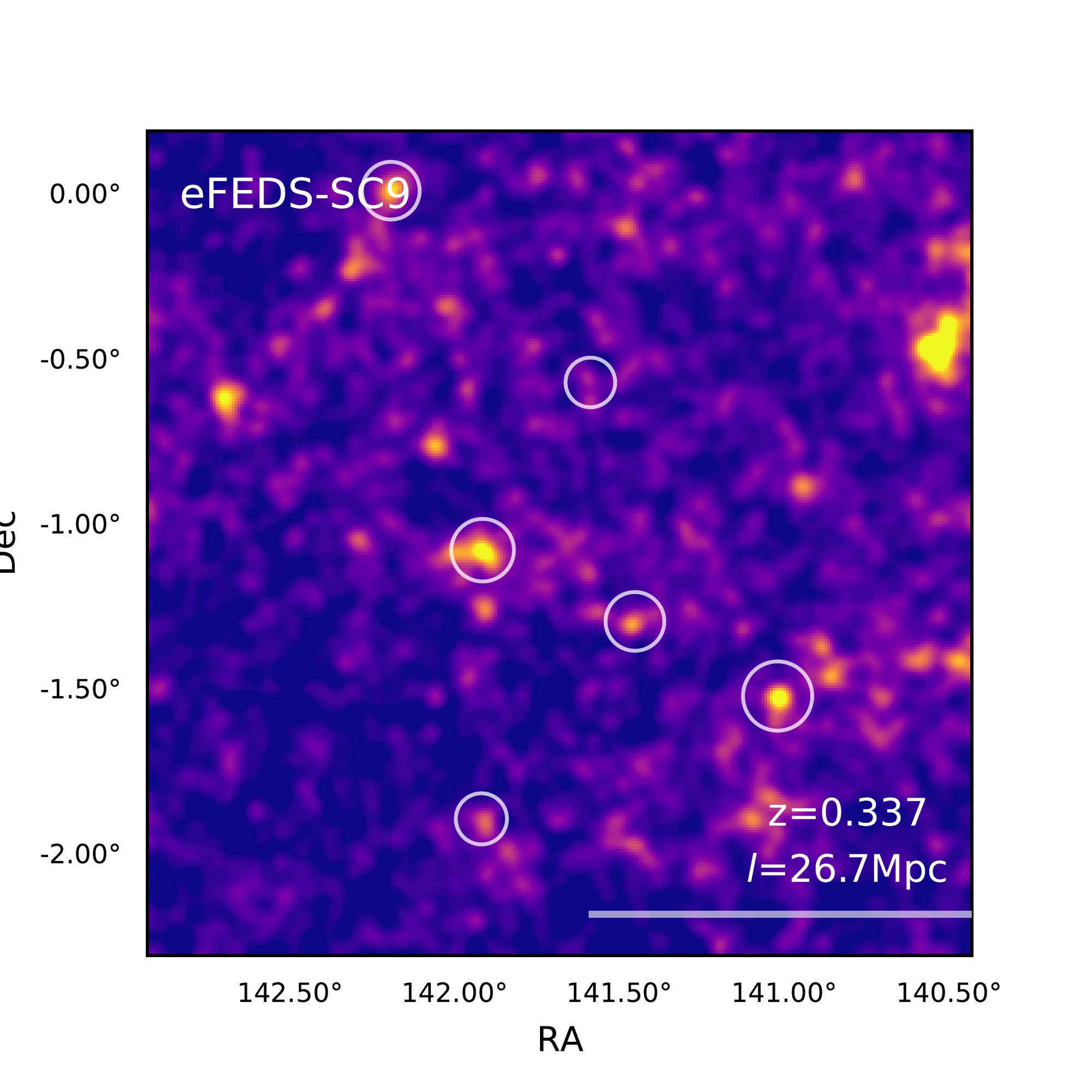}
\includegraphics[width=0.245\textwidth, trim=0 20 40 50, clip]{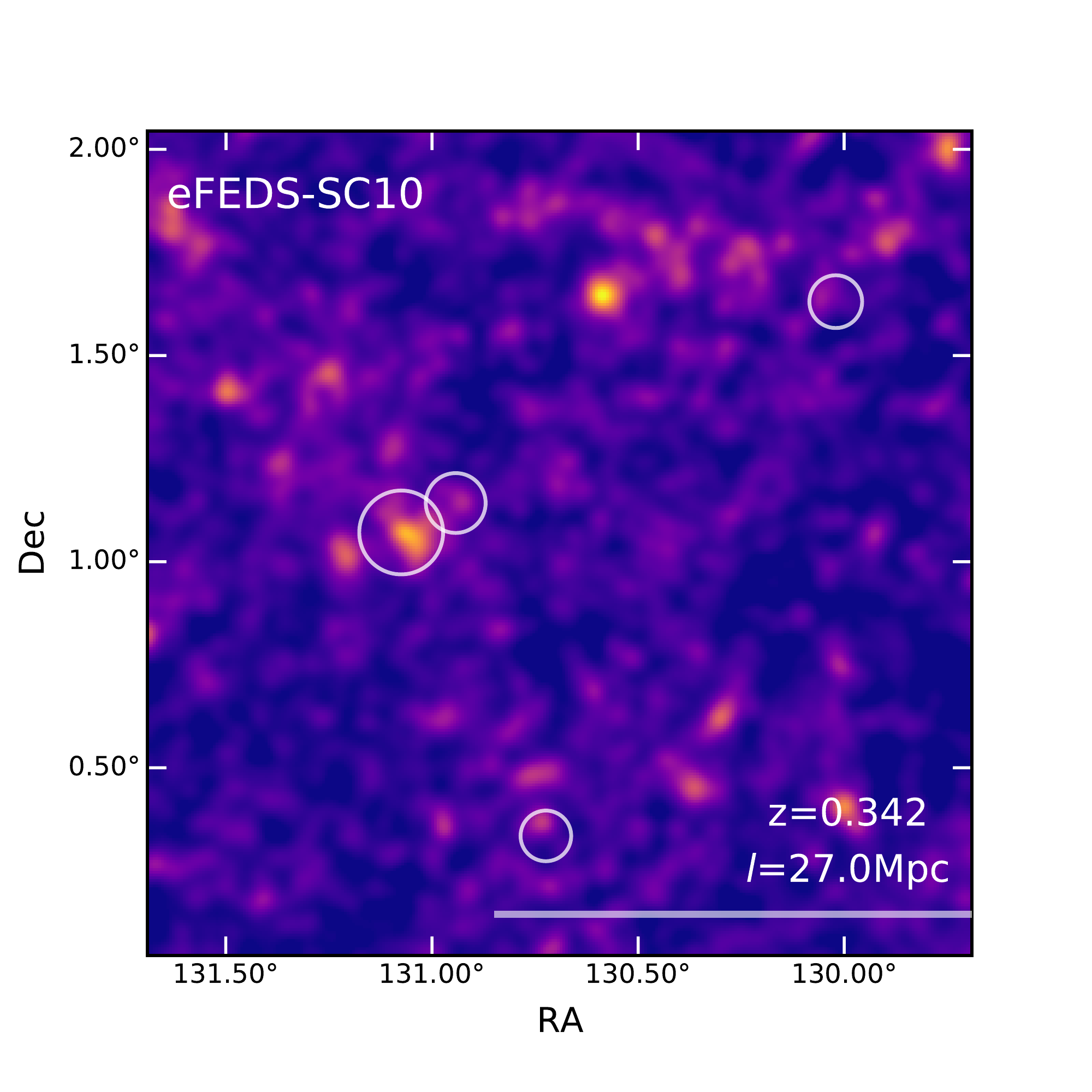}
\includegraphics[width=0.245\textwidth, trim=0 20 40 50, clip]{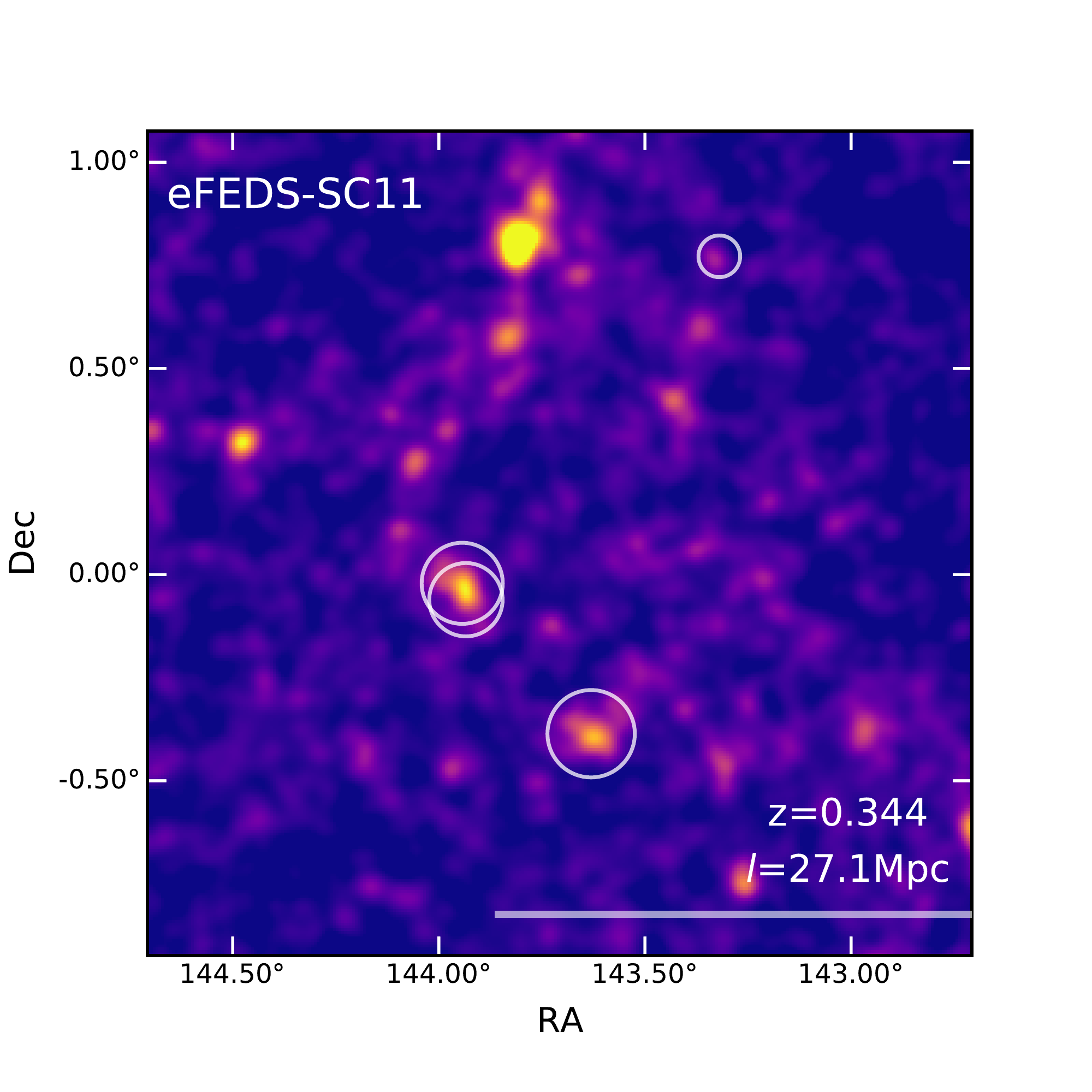}
\includegraphics[width=0.245\textwidth, trim=0 20 40 50, clip]{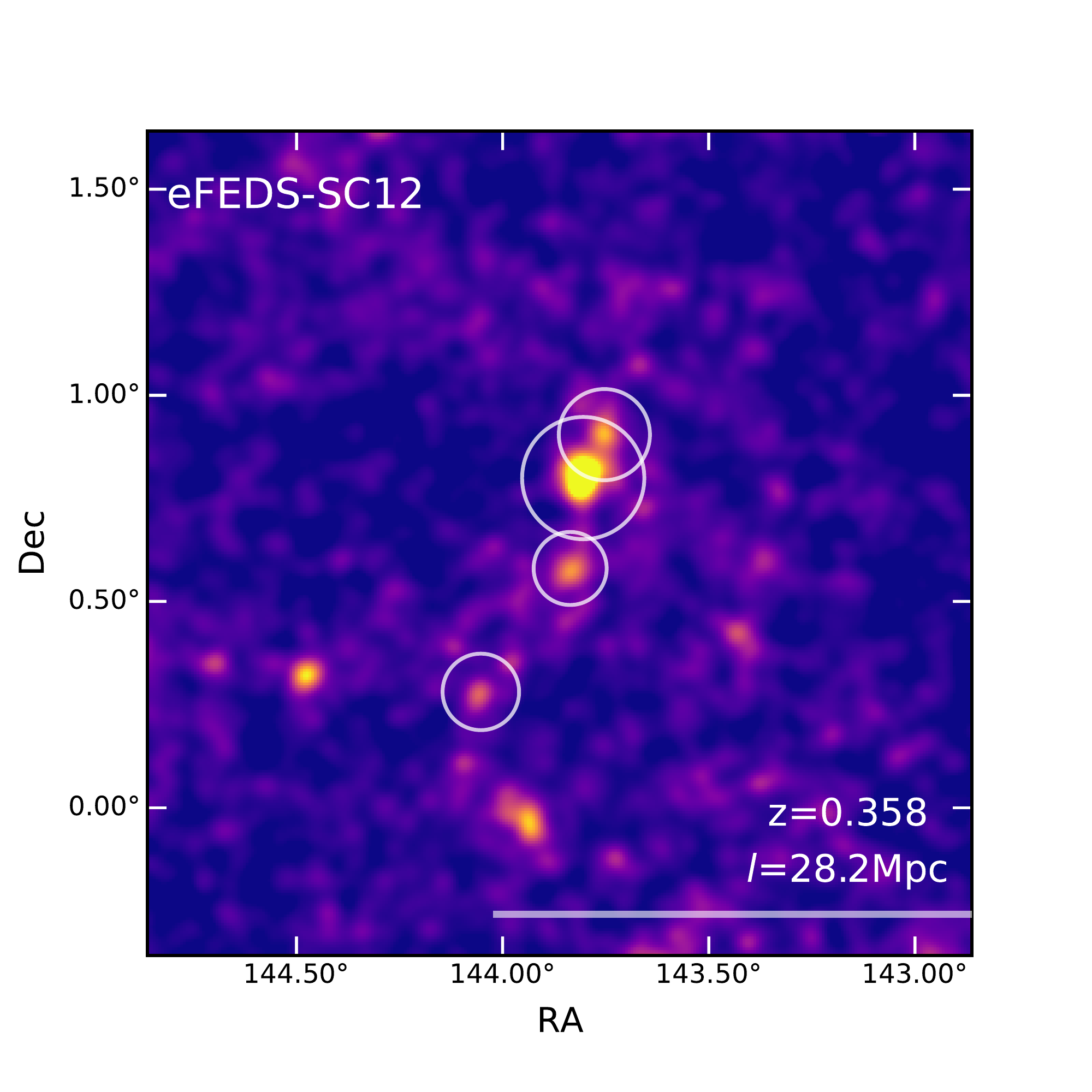}
\includegraphics[width=0.245\textwidth, trim=0 20 40 50, clip]{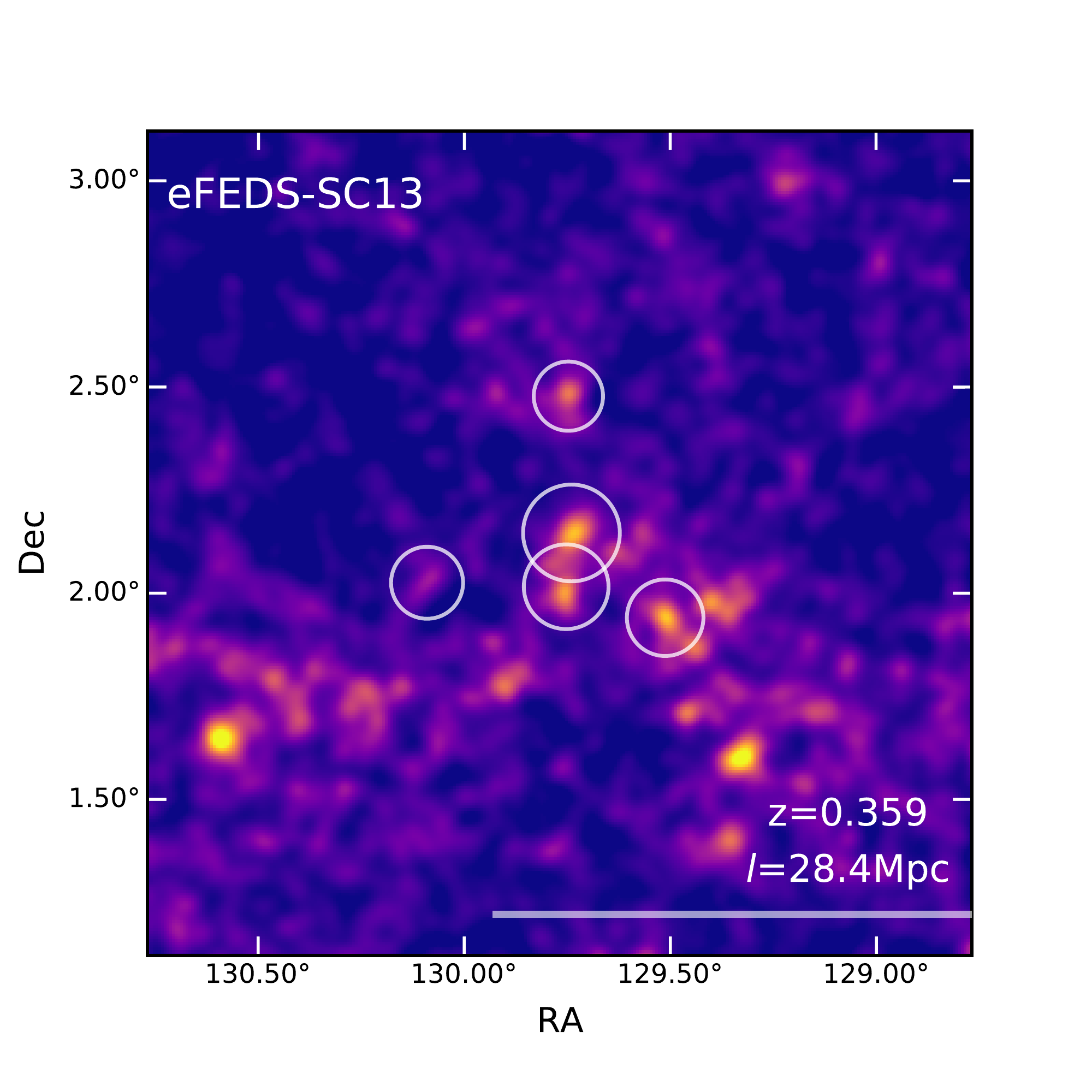}
\includegraphics[width=0.245\textwidth, trim=0 20 40 50, clip]{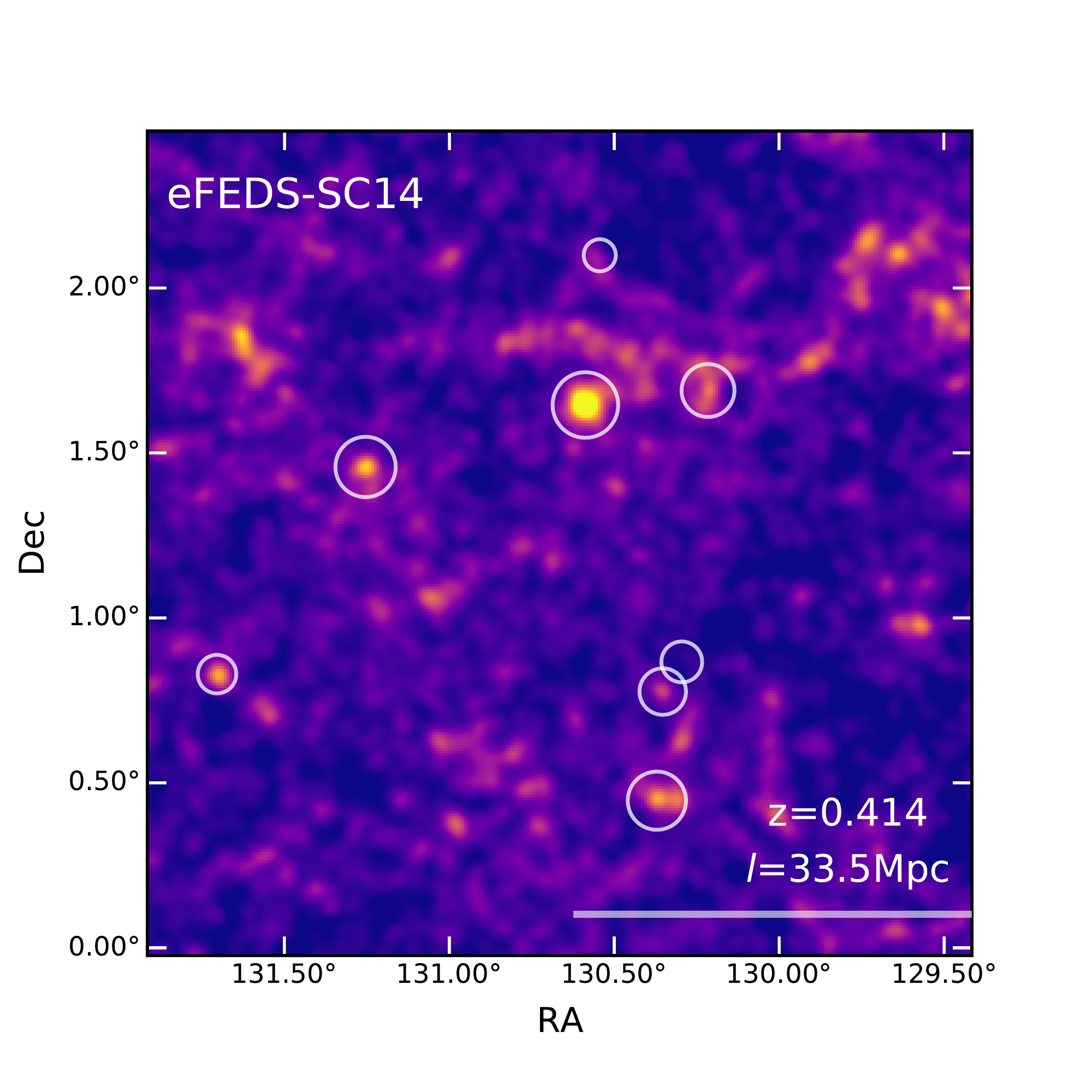}
\includegraphics[width=0.245\textwidth, trim=0 20 40 50, clip]{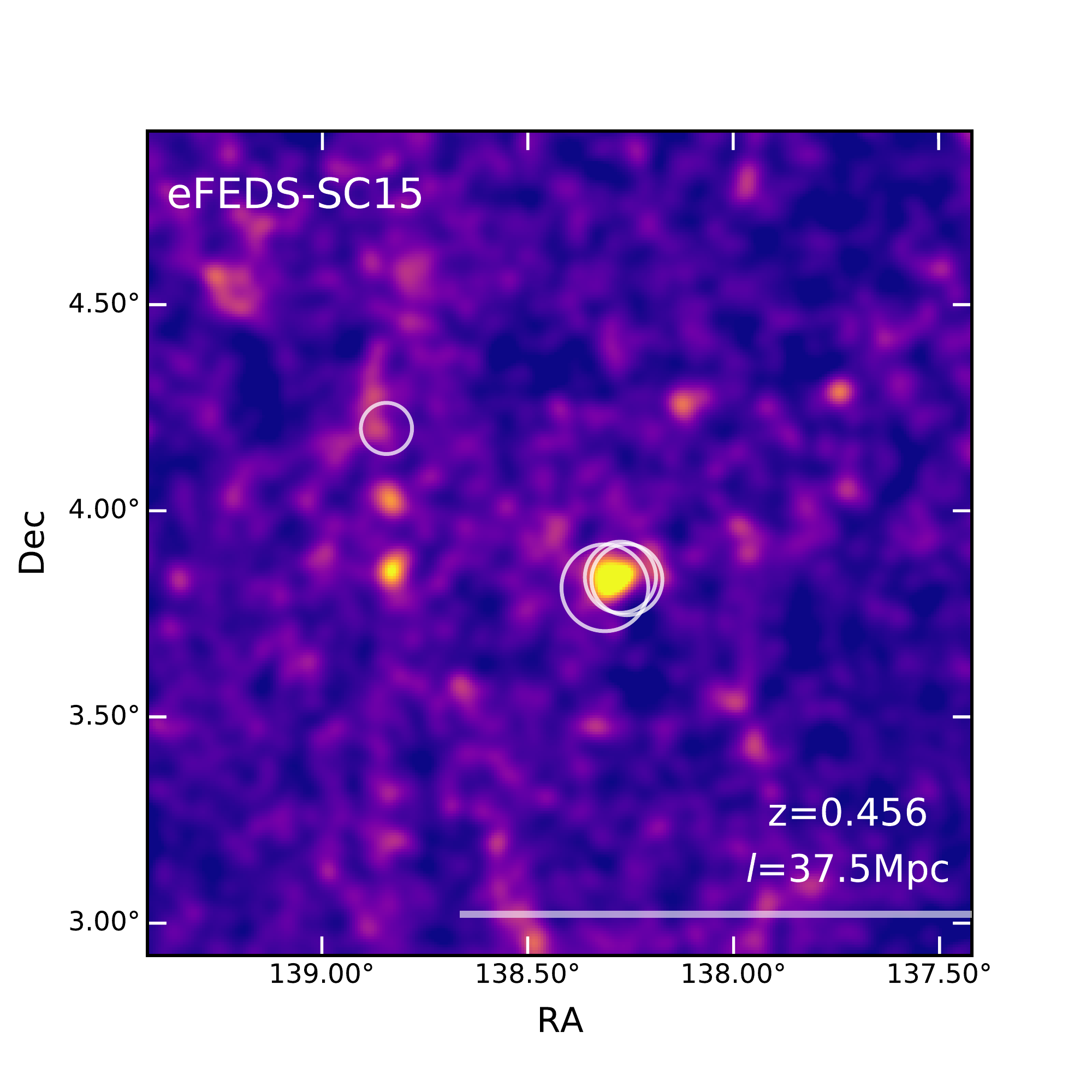}
\includegraphics[width=0.245\textwidth, trim=0 20 40 50, clip]{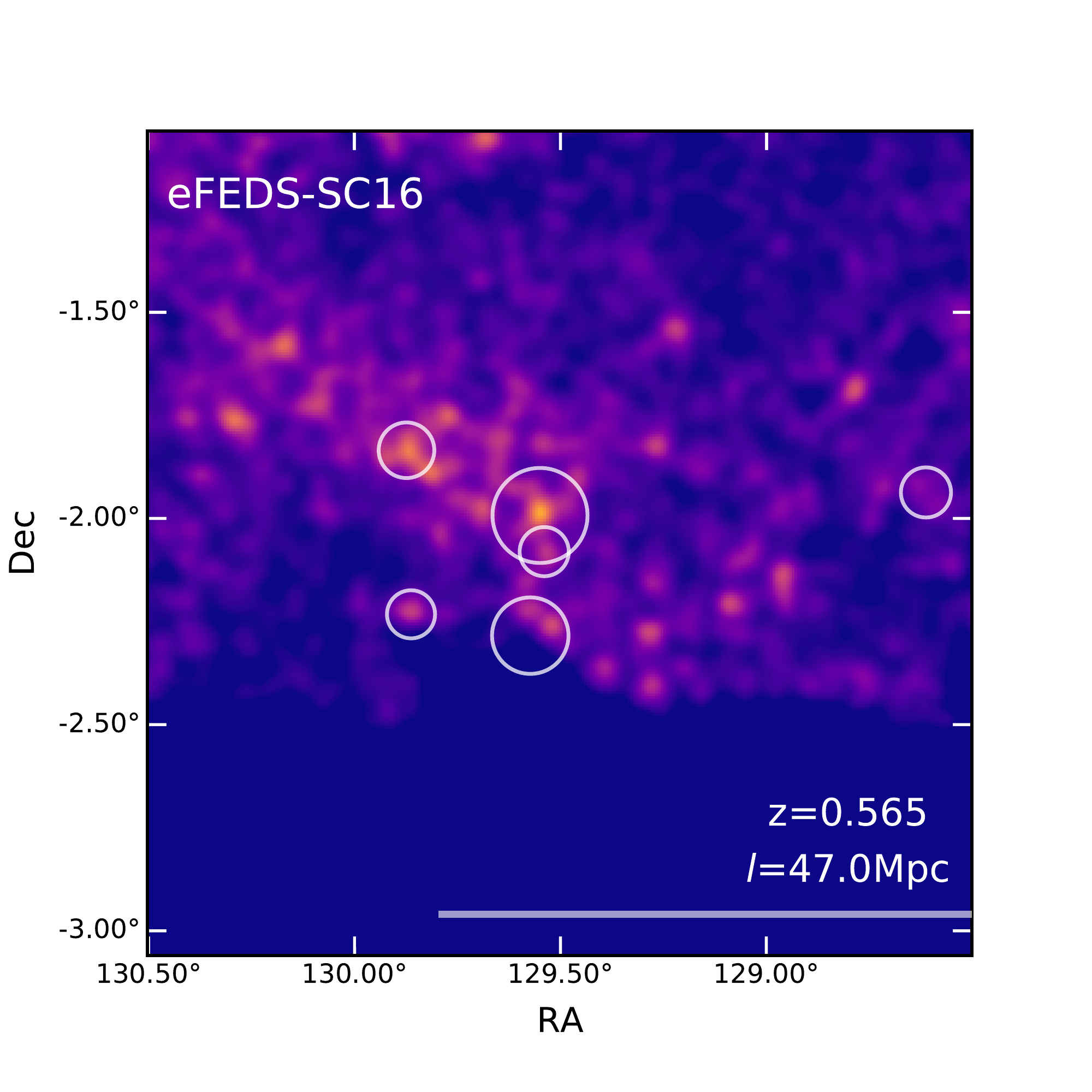}
\includegraphics[width=0.245\textwidth, trim=0 20 40 50, clip]{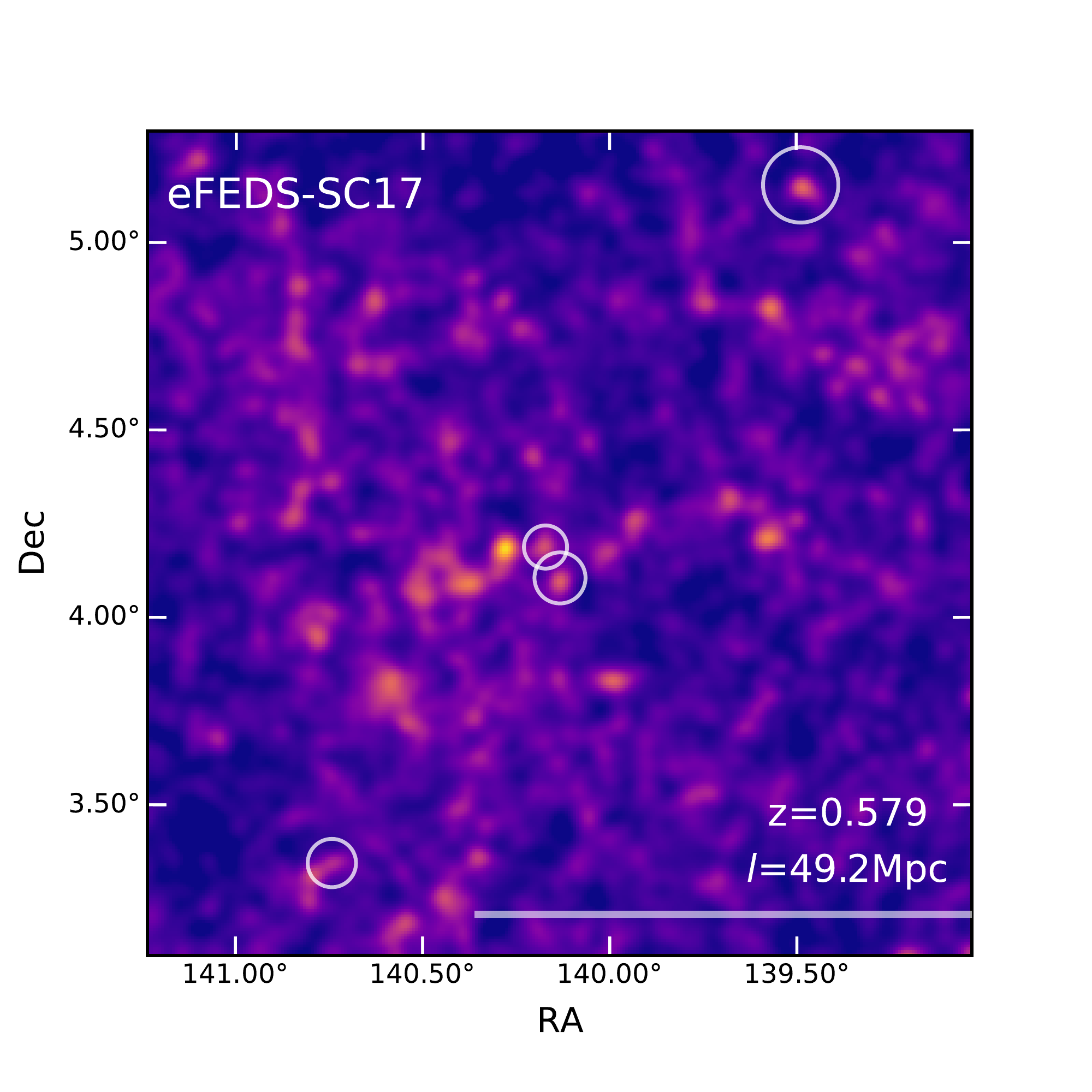}
\includegraphics[width=0.245\textwidth, trim=0 20 40 50, clip]{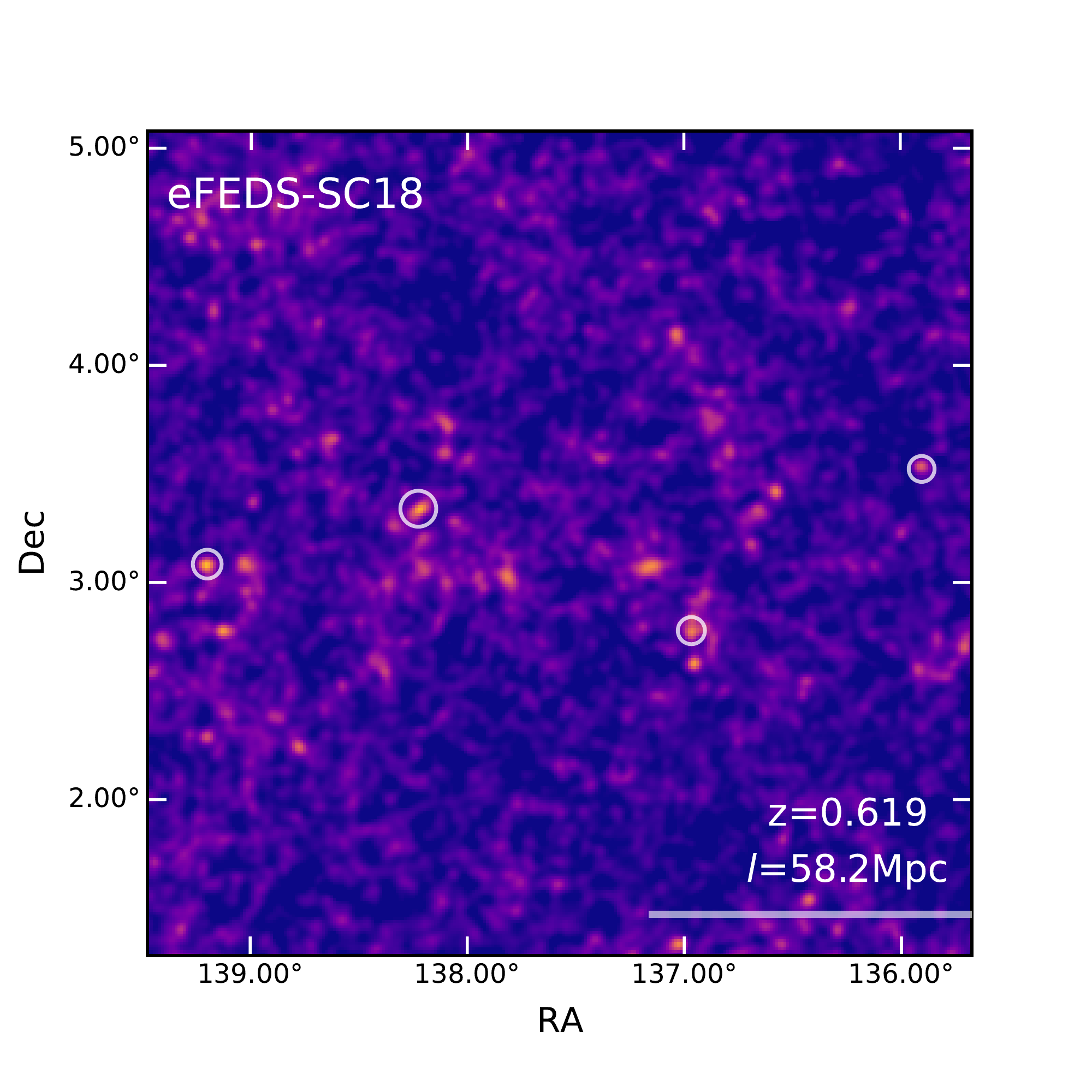}
\includegraphics[width=0.245\textwidth, trim=0 20 40 50, clip]{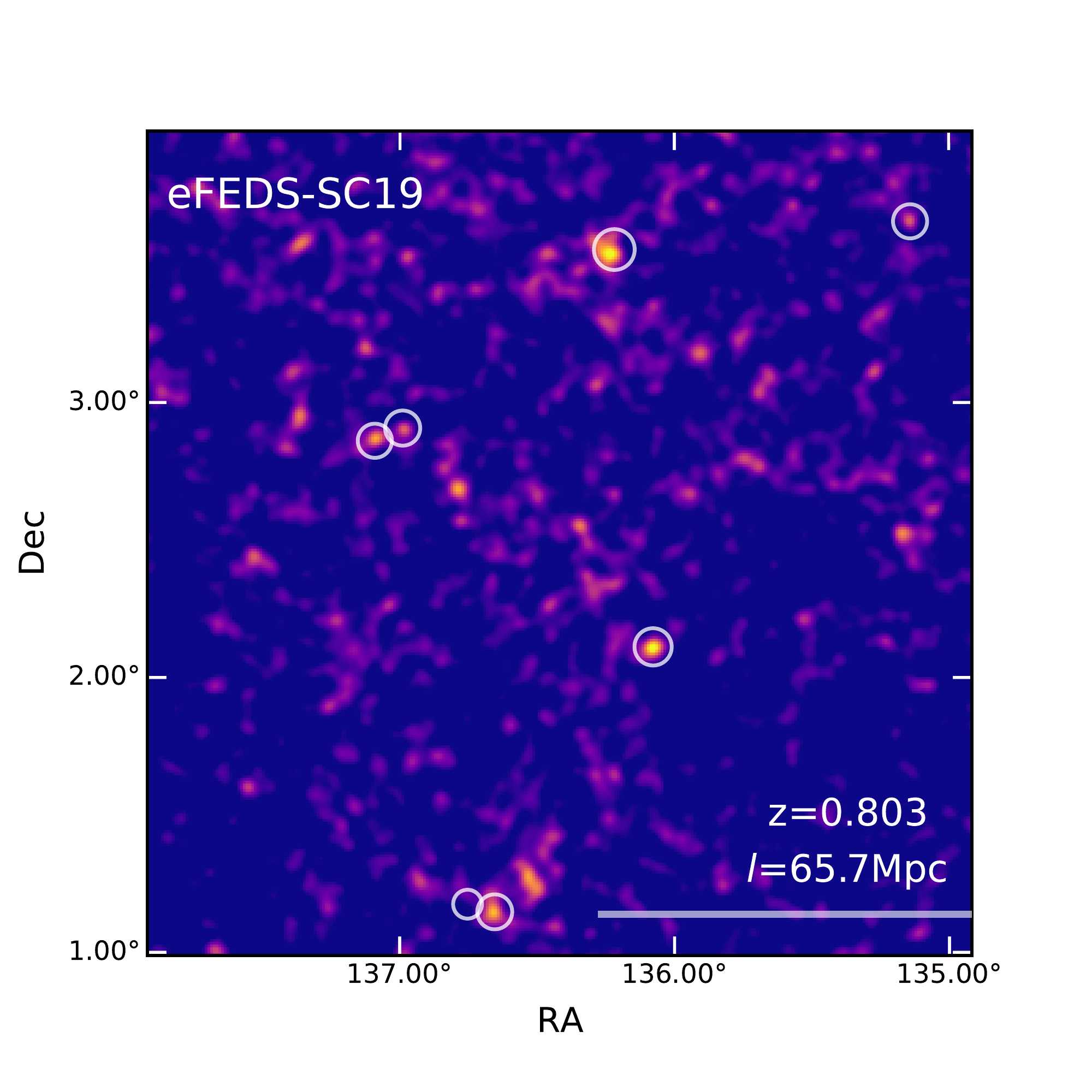}
\caption{Galaxy density map from HSC-SSP survey data at the locations and redshifts of the 19 supercluster candidates. They are obtained by integrating the full probability function of the {\sc MIZUKI} \citep{MIZUKI} photometric redshifts ($P(z)$) of each galaxy \citep[the S19A catalog;][]{2020arXiv200301511N} with a width of the redshift slice of $\Delta z=0.05$; $\int_{z-\Delta z}^{z+\Delta z} P(z)dz/ \int_0^\infty P(z)dz$, in a similar way as \citet{2021MNRAS.501.1701O}. The galaxies are brighter than 23 ABmag. The smoothing scale is FWHM$=3$ arcmin.}
\label{fig:sc_image_hsc}
\end{center}
\end{figure*}

We list in Table~\ref{tab:sc_mem} the member clusters of the supercluster candidates detected in the eFEDS field, as described in Sect.~\ref{sec:sc}. In Fig.~\ref{fig:sc_image} we show the X-ray images of these superclusters. The galaxy density maps from the HSC-SSP survey data at the corresponding redshifts for all the superclusters are shown in Fig.~\ref{fig:sc_image_hsc}.

\clearpage
\onecolumn

\begin{longtable}{llccccc}
\caption{\label{tab:sc_mem} The member clusters of the 19 superclusters in the eFEDS field.  } \\
\hline\hline
SC ID & Cluster ID & RA & Dec & $z$ & type & $M_{500}$ \\
  &  & [deg] & [deg] & & & [$10^{14}~M_{\odot}$] \\
\hline
\endfirsthead
\caption{continued.}\\
\hline\hline
SC ID & Cluster ID & RA & Dec & $z$ & type & $M_{500}$ \\
  &  & [deg] & [deg] & & & [$10^{14}~M_{\odot}$] \\
\hline
\endhead
\hline
\endfoot
eFEDS-SC1 & eFEDS~J085436.6+003835 & 133.6526 & +0.6431 & 0.106 & 1 & 1.2 \\
 & eFEDS~J085141.9+021438 & 132.9248 & +2.2440 & 0.107 & 1 & 0.1 \\
 & eFEDS~J085433.0+004009 & 133.6376 & +0.6693 & 0.109 & 1 & 0.1 \\
 & eFEDS~J085705.9+011453 & 134.2749 & +1.2482 & 0.106 & 1 & 0.4 \\
eFEDS-SC2 & eFEDS~J085030.5+003330 & 132.6272 & +0.5584 & 0.192 & 1 & 1.1 \\
 & eFEDS~J085327.2-002117 & 133.3634 & -0.3549 & 0.193 & 1 & 0.6 \\
 & eFEDS~J085022.2+001607 & 132.5927 & +0.2687 & 0.196 & 1 & 1.0 \\
 & eFEDS~J085340.5+022411 & 133.4191 & +2.4032 & 0.196 & 1 & 0.6 \\
 & eFEDS~J085027.8+001503 & 132.6160 & +0.2509 & 0.197 & 1 & 2.6 \\
 & eFEDS~J085128.4+011501 & 132.8685 & +1.2505 & 0.197 & 1 & 0.1 \\
eFEDS-SC3 & eFEDS~J090137.7+030253 & 135.4072 & +3.0483 & 0.188 & 0 & 0.2 \\
 & eFEDS~J085913.1+031334 & 134.8048 & +3.2263 & 0.189 & 1 & 0.9 \\
 & eFEDS~J090119.0+030204 & 135.3294 & +3.0345 & 0.193 & 1 & 0.5 \\
 & eFEDS~J090131.1+030056 & 135.3800 & +3.0157 & 0.193 & 1 & 3.1 \\
 & eFEDS~J085931.9+030839 & 134.8830 & +3.1443 & 0.196 & 1 & 2.2 \\
 & eFEDS~J085728.3+032354 & 134.3680 & +3.3984 & 0.200 & 1 & 0.6 \\
 & eFEDS~J090255.2+030220 & 135.7300 & +3.0389 & 0.200 & 1 & 1.0 \\
 & eFEDS~J090010.4+023631 & 135.0435 & +2.6086 & 0.200 & 1 & 0.4 \\
 & eFEDS~J085751.6+031039 & 134.4653 & +3.1775 & 0.201 & 1 & 5.7 \\
 & eFEDS~J090200.5+022339 & 135.5024 & +2.3943 & 0.202 & 1 & 1.0 \\
eFEDS-SC4 & eFEDS~J092209.3+034628 & 140.5391 & +3.7746 & 0.270 & 1 & 3.8 \\
 & eFEDS~J092202.2+034520 & 140.5095 & +3.7557 & 0.270 & 1 & 2.6 \\
 & eFEDS~J092022.8+045012 & 140.0954 & +4.8369 & 0.270 & 1 & 1.4 \\
 & eFEDS~J092220.4+034806 & 140.5852 & +3.8017 & 0.267 & 1 & 1.0 \\
 & eFEDS~J092246.2+034251 & 140.6928 & +3.7143 & 0.269 & 1 & 1.2 \\
 & eFEDS~J092302.6+034002 & 140.7611 & +3.6673 & 0.269 & 0 & 0.8 \\
eFEDS-SC5 & eFEDS~J084151.9-010156 & 130.4665 & -1.0323 & 0.270 & 1 & 0.4 \\
 & eFEDS~J083930.3-014348 & 129.8763 & -1.7302 & 0.271 & 0 & 0.6 \\
 & eFEDS~J083933.8-014044 & 129.8909 & -1.6790 & 0.272 & 1 & 4.9 \\
 & eFEDS~J084000.0-013109 & 130.0002 & -1.5194 & 0.266 & 1 & 1.2 \\
 & eFEDS~J083916.7-020552 & 129.8198 & -2.0979 & 0.269 & 0 & 0.6 \\
 & eFEDS~J083917.9-020839 & 129.8247 & -2.1442 & 0.269 & 0 & 0.5 \\
 & eFEDS~J083921.0-014149 & 129.8377 & -1.6970 & 0.269 & 1 & 1.2 \\
eFEDS-SC6 & eFEDS~J092031.3+024710 & 140.1306 & +2.7863 & 0.278 & 1 & 0.4 \\
 & eFEDS~J092053.4+021125 & 140.2228 & +2.1905 & 0.280 & 1 & 0.6 \\
 & eFEDS~J091851.7+021432 & 139.7155 & +2.2423 & 0.283 & 1 & 0.5 \\
 & eFEDS~J091849.0+021204 & 139.7042 & +2.2013 & 0.283 & 1 & 3.3 \\
 & eFEDS~J092049.5+024513 & 140.2063 & +2.7538 & 0.284 & 1 & 4.1 \\
eFEDS-SC7 & eFEDS~J090140.9-012132 & 135.4207 & -1.3591 & 0.295 & 1 & 1.1 \\
 & eFEDS~J090146.2-013756 & 135.4427 & -1.6322 & 0.295 & 1 & 1.5 \\
 & eFEDS~J090053.0-002837 & 135.2212 & -0.4772 & 0.295 & 1 & 0.6 \\
 & eFEDS~J090153.9-012209 & 135.4748 & -1.3694 & 0.295 & 1 & 1.3 \\
eFEDS-SC8 & eFEDS~J085517.2+013508 & 133.8219 & +1.5857 & 0.324 & 1 & 2.0 \\
 & eFEDS~J085624.3+004632 & 134.1016 & +0.7757 & 0.319 & 1 & 0.8 \\
 & eFEDS~J085121.2+012856 & 132.8384 & +1.4825 & 0.327 & 1 & 0.8 \\
 & eFEDS~J084934.9+014437 & 132.3958 & +1.7438 & 0.325 & 1 & 1.4 \\
eFEDS-SC9 & eFEDS~J092739.7-010427 & 141.9158 & -1.0743 & 0.329 & 1 & 1.8 \\
 & eFEDS~J092740.7-015320 & 141.9196 & -1.8889 & 0.332 & 1 & 1.0 \\
 & eFEDS~J092548.9-011725 & 141.4539 & -1.2903 & 0.337 & 0 & 1.5 \\
 & eFEDS~J092405.0-013059 & 141.0211 & -1.5165 & 0.337 & 1 & 2.4 \\
 & eFEDS~J092621.3-003356 & 141.5890 & -0.5657 & 0.340 & 0 & 0.9 \\
 & eFEDS~J092846.5+000056 & 142.1940 & +0.0157 & 0.344 & 1 & 1.4 \\
eFEDS-SC10 & eFEDS~J084253.7+002006 & 130.7238 & +0.3350 & 0.345 & 0 & 0.5 \\
 & eFEDS~J084417.9+010415 & 131.0748 & +1.0711 & 0.340 & 1 & 2.3 \\
 & eFEDS~J084346.2+010833 & 130.9425 & +1.1425 & 0.342 & 1 & 0.8 \\
 & eFEDS~J084004.8+013751 & 130.0203 & +1.6309 & 0.342 & 0 & 0.6 \\
eFEDS-SC11 & eFEDS~J093431.3-002309 & 143.6304 & -0.3860 & 0.342 & 1 & 2.6 \\
 & eFEDS~J093316.6+004619 & 143.3195 & +0.7721 & 0.347 & 0 & 0.3 \\
 & eFEDS~J093544.2-000339 & 143.9342 & -0.0609 & 0.347 & 1 & 1.6 \\
 & eFEDS~J093546.3-000115 & 143.9433 & -0.0211 & 0.339 & 1 & 2.1 \\
eFEDS-SC12 & eFEDS~J093500.7+005417 & 143.7532 & +0.9048 & 0.361 & 1 & 3.3 \\
 & eFEDS~J093612.7+001650 & 144.0529 & +0.2807 & 0.358 & 1 & 1.9 \\
 & eFEDS~J093520.7+003448 & 143.8363 & +0.5802 & 0.355 & 1 & 1.7 \\
 & eFEDS~J093513.0+004757 & 143.8046 & +0.7994 & 0.356 & 1 & 7.9 \\
eFEDS-SC13 & eFEDS~J084021.6+020132 & 130.0903 & +2.0256 & 0.357 & 1 & 1.3 \\
 & eFEDS~J083900.6+020057 & 129.7527 & +2.0159 & 0.359 & 1 & 2.6 \\
 & eFEDS~J083859.3+022841 & 129.7473 & +2.4782 & 0.359 & 1 & 1.4 \\
 & eFEDS~J083802.9+015626 & 129.5124 & +1.9407 & 0.360 & 1 & 2.0 \\
 & eFEDS~J083857.5+020846 & 129.7398 & +2.1464 & 0.360 & 1 & 4.0 \\
eFEDS-SC14 & eFEDS~J084124.7+004636 & 130.3530 & +0.7768 & 0.407 & 1 & 1.2 \\
 & eFEDS~J084051.7+014122 & 130.2156 & +1.6895 & 0.411 & 1 & 1.7 \\
 & eFEDS~J084110.8+005200 & 130.2953 & +0.8668 & 0.415 & 0 & 0.8 \\
 & eFEDS~J084649.0+004946 & 131.7045 & +0.8295 & 0.416 & 1 & 0.7 \\
 & eFEDS~J084501.0+012728 & 131.2542 & +1.4578 & 0.420 & 1 & 2.6 \\
 & eFEDS~J084220.9+013844 & 130.5875 & +1.6457 & 0.421 & 1 & 3.4 \\
 & eFEDS~J084210.5+020558 & 130.5439 & +2.0997 & 0.421 & 0 & 0.4 \\
 & eFEDS~J084129.0+002645 & 130.3708 & +0.4460 & 0.402 & 1 & 2.3 \\
eFEDS-SC15 & eFEDS~J091305.9+035021 & 138.2747 & +3.8394 & 0.456 & 1 & 2.7 \\
 & eFEDS~J091315.0+034850 & 138.3125 & +3.8139 & 0.453 & 1 & 4.9 \\
 & eFEDS~J091302.1+035000 & 138.2590 & +3.8336 & 0.455 & 0 & 2.7 \\
 & eFEDS~J091522.5+041201 & 138.8438 & +4.2003 & 0.460 & 1 & 1.0 \\
eFEDS-SC16 & eFEDS~J083809.4-020450 & 129.5394 & -2.0808 & 0.550 & 0 & 1.4 \\
 & eFEDS~J083811.8-015934 & 129.5496 & -1.9930 & 0.560 & 0 & 10.3 \\
 & eFEDS~J083817.5-021704 & 129.5732 & -2.2845 & 0.565 & 0 & 5.4 \\
 & eFEDS~J083927.0-021357 & 129.8627 & -2.2328 & 0.567 & 0 & 1.3 \\
 & eFEDS~J083427.0-015612 & 128.6129 & -1.9369 & 0.573 & 0 & 1.5 \\
 & eFEDS~J083929.6-015005 & 129.8736 & -1.8348 & 0.575 & 0 & 2.1 \\
eFEDS-SC17 & eFEDS~J092031.8+040621 & 140.1329 & +4.1059 & 0.575 & 1 & 2.2 \\
 & eFEDS~J092041.1+041117 & 140.1716 & +4.1883 & 0.580 & 0 & 1.4 \\
 & eFEDS~J091757.1+050915 & 139.4881 & +5.1542 & 0.586 & 0 & 7.3 \\
 & eFEDS~J092258.2+032041 & 140.7426 & +3.3449 & 0.575 & 0 & 1.9 \\
eFEDS-SC18 & eFEDS~J091254.4+032028 & 138.2270 & +3.3414 & 0.619 & 0 & 4.6 \\
 & eFEDS~J091648.1+030506 & 139.2007 & +3.0851 & 0.620 & 1 & 2.4 \\
 & eFEDS~J090336.7+033124 & 135.9033 & +3.5235 & 0.617 & 0 & 1.7 \\
 & eFEDS~J090751.9+024647 & 136.9666 & +2.7797 & 0.618 & 1 & 2.0 \\
eFEDS-SC19 & eFEDS~J090636.9+010852 & 136.6542 & +1.1479 & 0.786 & 0 & 3.5 \\
 & eFEDS~J090757.5+025427 & 136.9900 & +2.9077 & 0.798 & 1 & 3.6 \\
 & eFEDS~J090700.7+011032 & 136.7531 & +1.1757 & 0.799 & 0 & 2.0 \\
 & eFEDS~J090418.6+020642 & 136.0778 & +2.1117 & 0.808 & 0 & 4.3 \\
 & eFEDS~J090452.4+033326 & 136.2187 & +3.5574 & 0.808 & 0 & 5.7 \\
 & eFEDS~J090033.7+033932 & 135.1408 & +3.6591 & 0.809 & 0 & 3.3 \\
 & eFEDS~J090821.9+025141 & 137.0913 & +2.8615 & 0.812 & 0 & 3.4 \\
\hline
\end{longtable}

\tablefoot{Redshift type: 0 for photometric redshift and 1 for spectroscopic redshift. $M_{\rm 500}$ in the last column is estimated using the scaling relation in \citet{2015Lovisari}. }

\section{Main results of the X-ray analysis}
In this section we list the main results of the X-ray analysis. Instead of providing the results for all the 542 clusters, we specifically selected a subsample of 102 clusters with $>2\sigma$ temperature measurements within either 300~kpc or 500~kpc. The results of the clusters in the subsample are listed in Table~\ref{tab:main}.

\newpage
\onecolumn
\fontsize{7.0pt}{0cm}\selectfont
\begin{landscape}
\begin{longtable}{lccccccccccccccccccc}
\caption{\label{tab:main} Main results of the X-ray analysis for the 102 clusters in the subsample with $>2\sigma$ temperature measurements within either 300~kpc or 500~kpc. }\\
\hline\hline
ID (eFEDS~J+) & ID\_SRC & RA & Dec & $\mathcal{L}_{\rm ext}$ & $\mathcal{L}_{\rm det}$ & $z$ & $T_{\rm 300kpc}$ & $L_{\rm 300kpc}$ & $L_{\rm bol,300kpc}$ & $F_{\rm 300kpc}$ & $T_{\rm 500kpc}$ & $L_{\rm 500kpc}$ & $L_{\rm bol,500kpc}$ & $F_{\rm 500kpc}$ & SN$_{\rm max}$ & $R({\rm SN}_{\rm max})$  \\
\hline
\endfirsthead
\caption{continued.}\\
\hline\hline
ID (eFEDS~J+) & ID\_SRC & RA & Dec & $\mathcal{L}_{\rm ext}$ & $\mathcal{L}_{\rm det}$ & $z$ & $T_{\rm 300kpc}$ & $L_{\rm 300kpc}$ & $L_{\rm bol,300kpc}$ & $F_{\rm 300kpc}$ & $T_{\rm 500kpc}$ & $L_{\rm 500kpc}$ & $L_{\rm bol,500kpc}$ & $F_{\rm 500kpc}$ & SN$_{\rm max}$ & $R({\rm SN}_{\rm max})$  \\
\hline
\endhead
\hline
\endfoot
082808.7-001003 & 4800 & $ 127.0366 $ & $-0.1677 $ & 28.5 & 62.5 & 0.08 & $ 0.89_{-0.10}^{+0.11} $ & $ 0.26_{-0.04}^{+0.04} $ & $ 0.40_{-0.06}^{+0.06} $ & $ 18.02_{-2.59}^{+2.64} $ & $ 0.88_{-0.09}^{+0.09} $ & $ 0.28_{-0.05}^{+0.06} $ & $ 0.44_{-0.08}^{+0.09} $ & $ 19.76_{-3.52}^{+4.06} $ & 8.8 & 2.63 & \\
083204.4+041907 & 3593 & $ 128.0185 $ & $+4.3188 $ & 66.0 & 118.6 & 0.20 & $ 2.79_{-0.17}^{+0.34} $ & $ 2.08_{-0.20}^{+0.22} $ & $ 4.72_{-0.59}^{+0.70} $ & $ 19.25_{-1.86}^{+2.05} $ & $ 2.42_{-0.49}^{+1.07} $ & $ 3.33_{-0.28}^{+0.32} $ & $ 7.59_{-0.89}^{+1.01} $ & $ 31.00_{-2.70}^{+2.83} $ & 11.9 & 3.45 & \\
083330.4+050427 & 2528 & $ 128.3767 $ & $+5.0744 $ & 50.1 & 122.2 & 0.21 &  - & $ 1.12_{-0.14}^{+0.15} $ & $ 2.47_{-0.40}^{+0.56} $ & $ 8.69_{-1.07}^{+1.24} $ & $ 2.05_{-0.53}^{+1.34} $ & $ 1.54_{-0.21}^{+0.21} $ & $ 3.37_{-0.60}^{+0.74} $ & $ 11.89_{-1.59}^{+1.64} $ & 8.7 & 2.30 & \\
083651.3+030002 & 197 & $ 129.2138 $ & $+3.0006 $ & 153.5 & 747.9 & 0.192 & $ 2.79_{-0.35}^{+0.62} $ & $ 3.97_{-0.18}^{+0.20} $ & $ 9.19_{-0.63}^{+0.55} $ & $ 39.03_{-1.77}^{+1.99} $ & $ 2.57_{-0.37}^{+0.51} $ & $ 4.95_{-0.22}^{+0.21} $ & $ 11.43_{-0.75}^{+0.68} $ & $ 48.66_{-2.21}^{+2.10} $ & 23.1 & 2.89 & \\
083654.6+025954 & 16471 & $ 129.2277 $ & $+2.9984 $ & 10.9 & 14.0 & 0.191 & $ 2.55_{-0.44}^{+0.62} $ & $ 3.95_{-0.20}^{+0.20} $ & $ 9.28_{-0.77}^{+0.73} $ & $ 39.43_{-2.01}^{+1.92} $ & $ 2.65_{-0.45}^{+0.75} $ & $ 4.94_{-0.24}^{+0.20} $ & $ 11.59_{-0.91}^{+0.85} $ & $ 49.29_{-2.37}^{+2.04} $ & 23.1 & 2.87 & \\
083807.6+002501 & 19692 & $ 129.5321 $ & $+0.4171 $ & 6.4 & 15.1 & 0.080 &  - & $ 0.12_{-0.02}^{+0.02} $ & $ 0.19_{-0.03}^{+0.03} $ & $ 7.19_{-1.17}^{+1.21} $ & $ 0.48_{-0.03}^{+0.02} $ & $ 0.15_{-0.03}^{+0.03} $ & $ 0.25_{-0.05}^{+0.05} $ & $ 9.36_{-1.78}^{+1.95} $ & 6.0 & 3.38 & \\
083811.8-015934 & 53 & $ 129.5496 $ & $-1.9930 $ & 163.7 & 2105.1 & 0.56 & $ 4.27_{-0.74}^{+1.12} $ & $ 37.04_{-3.11}^{+3.37} $ & $ 112.88_{-11.43}^{+15.16} $ & $ 34.54_{-2.91}^{+2.94} $ & $ 4.72_{-0.94}^{+1.53} $ & $ 39.98_{-2.94}^{+3.45} $ & $ 122.00_{-11.01}^{+15.79} $ & $ 37.32_{-2.76}^{+2.98} $ & 23.4 & 1.25 & \\
083817.4+041821 & 6136 & $ 129.5727 $ & $+4.3060 $ & 21.1 & 56.6 & 0.211 &  - & $ 0.72_{-0.08}^{+0.10} $ & $ 1.32_{-0.19}^{+0.25} $ & $ 5.51_{-0.66}^{+0.73} $ & $ 1.41_{-0.33}^{+0.78} $ & $ 1.05_{-0.13}^{+0.14} $ & $ 1.90_{-0.26}^{+0.39} $ & $ 7.99_{-1.05}^{+1.12} $ & 7.8 & 2.33 & \\
083857.5+020846 & 1806 & $ 129.7398 $ & $+2.1464 $ & 92.8 & 192.1 & 0.360 & $ 2.81_{-0.60}^{+1.30} $ & $ 4.55_{-0.41}^{+0.42} $ & $ 11.30_{-1.15}^{+1.47} $ & $ 11.05_{-0.91}^{+1.04} $ & $ 3.01_{-0.66}^{+1.48} $ & $ 7.23_{-0.60}^{+0.72} $ & $ 18.03_{-1.79}^{+2.20} $ & $ 17.64_{-1.45}^{+1.71} $ & 14.6 & 2.74 & \\
083900.6+020057 & 736 & $ 129.7527 $ & $+2.0159 $ & 39.3 & 313.3 & 0.359 & $ 2.26_{-0.43}^{+0.67} $ & $ 5.24_{-0.52}^{+0.59} $ & $ 11.32_{-1.31}^{+1.49} $ & $ 12.31_{-1.20}^{+1.38} $ & $ 2.28_{-0.40}^{+0.82} $ & $ 5.84_{-0.57}^{+0.63} $ & $ 12.57_{-1.43}^{+1.62} $ & $ 13.75_{-1.30}^{+1.39} $ & 12.5 & 1.26 & \\
083933.8-014044 & 356 & $ 129.8909 $ & $-1.6790 $ & 245.8 & 663.7 & 0.272 & $ 4.14_{-0.82}^{+1.35} $ & $ 7.47_{-0.64}^{+0.80} $ & $ 20.31_{-2.06}^{+2.68} $ & $ 34.82_{-2.81}^{+3.96} $ & $ 3.74_{-0.67}^{+0.87} $ & $ 11.22_{-1.02}^{+1.18} $ & $ 30.50_{-3.13}^{+4.02} $ & $ 52.53_{-4.77}^{+5.58} $ & 24.5 & 2.98 & \\
083940.8+010416 & 2832 & $ 129.9201 $ & $+1.0714 $ & 27.3 & 84.0 & 0.157 & $ 1.44_{-0.27}^{+0.40} $ & $ 0.61_{-0.07}^{+0.08} $ & $ 1.03_{-0.12}^{+0.14} $ & $ 8.89_{-0.98}^{+1.17} $ & $ 1.22_{-0.11}^{+0.12} $ & $ 0.74_{-0.10}^{+0.11} $ & $ 1.26_{-0.18}^{+0.19} $ & $ 10.92_{-1.49}^{+1.63} $ & 9.1 & 2.07 & \\
084034.5+023638 & 250 & $ 130.1441 $ & $+2.6108 $ & 114.6 & 661.8 & 0.049 & $ 0.80_{-0.03}^{+0.02} $ & $ 0.11_{-0.01}^{+0.01} $ & $ 0.17_{-0.02}^{+0.02} $ & $ 18.86_{-2.02}^{+1.84} $ & $ 0.82_{-0.08}^{+0.06} $ & $ 0.11_{-0.01}^{+0.01} $ & $ 0.17_{-0.02}^{+0.02} $ & $ 18.96_{-2.10}^{+1.95} $ & 15.2 & 1.47 & \\
084105.5+031639 & 1906 & $ 130.2731 $ & $+3.2777 $ & 32.3 & 146.2 & 0.34 & $ 2.72_{-0.72}^{+1.65} $ & $ 2.78_{-0.32}^{+0.35} $ & $ 6.86_{-1.10}^{+1.80} $ & $ 7.85_{-0.82}^{+0.99} $ &  - & $ 3.56_{-0.36}^{+0.46} $ & $ 8.86_{-1.38}^{+2.43} $ & $ 10.14_{-1.00}^{+1.16} $ & 9.9 & 1.66 & \\
084220.9+013844 & 3295 & $ 130.5875 $ & $+1.6457 $ & 27.7 & 64.5 & 0.421 &  - & $ 3.37_{-0.42}^{+0.46} $ & $ 7.05_{-0.91}^{+1.03} $ & $ 5.42_{-0.67}^{+0.72} $ & $ 2.06_{-0.41}^{+0.97} $ & $ 5.65_{-0.66}^{+0.79} $ & $ 11.85_{-1.53}^{+1.70} $ & $ 9.09_{-1.08}^{+1.18} $ & 10.0 & 2.48 & \\
084417.9+010415 & 1812 & $ 131.0748 $ & $+1.0711 $ & 26.0 & 94.6 & 0.340 & $ 2.46_{-0.69}^{+1.24} $ & $ 3.31_{-0.38}^{+0.40} $ & $ 8.44_{-1.41}^{+2.75} $ & $ 9.26_{-1.02}^{+1.05} $ &  - & $ 4.40_{-0.47}^{+0.46} $ & $ 11.19_{-1.79}^{+3.80} $ & $ 12.32_{-1.25}^{+1.16} $ & 10.2 & 1.70 & \\
084528.6+032739 & 144 & $ 131.3695 $ & $+3.4609 $ & 316.7 & 1183.0 & 0.334 & $ 6.68_{-1.83}^{+3.74} $ & $ 12.99_{-0.61}^{+0.64} $ & $ 44.96_{-3.99}^{+3.45} $ & $ 39.72_{-1.83}^{+1.89} $ & $ 6.31_{-1.33}^{+1.93} $ & $ 18.49_{-0.78}^{+0.79} $ & $ 64.06_{-5.20}^{+4.57} $ & $ 56.55_{-2.26}^{+2.45} $ & 26.1 & 2.90 & \\
084531.6+022831 & 1039 & $ 131.3818 $ & $+2.4753 $ & 98.9 & 275.7 & 0.076 & $ 1.45_{-0.09}^{+0.12} $ & $ 0.41_{-0.03}^{+0.03} $ & $ 0.72_{-0.05}^{+0.06} $ & $ 28.55_{-2.04}^{+1.87} $ & $ 1.27_{-0.08}^{+0.11} $ & $ 0.43_{-0.03}^{+0.03} $ & $ 0.76_{-0.06}^{+0.07} $ & $ 30.09_{-2.30}^{+2.40} $ & 14.8 & 2.63 & \\
084544.3-002914 & 2214 & $ 131.4349 $ & $-0.4874 $ & 33.5 & 103.1 & 0.156 & $ 1.72_{-0.26}^{+0.57} $ & $ 1.18_{-0.11}^{+0.11} $ & $ 2.29_{-0.24}^{+0.25} $ & $ 17.92_{-1.75}^{+1.72} $ & $ 1.64_{-0.23}^{+0.37} $ & $ 1.39_{-0.13}^{+0.13} $ & $ 2.70_{-0.28}^{+0.28} $ & $ 21.17_{-2.09}^{+1.93} $ & 11.5 & 2.14 & \\
084645.6+014947 & 6215 & $ 131.6902 $ & $+1.8298 $ & 12.4 & 39.2 & 0.051 & $ 0.50_{-0.01}^{+0.01} $ & $ 0.09_{-0.01}^{+0.01} $ & $ 0.14_{-0.02}^{+0.02} $ & $ 14.12_{-2.18}^{+2.21} $ & $ 0.49_{-0.02}^{+0.03} $ & $ 0.11_{-0.02}^{+0.02} $ & $ 0.17_{-0.04}^{+0.04} $ & $ 17.08_{-3.82}^{+3.80} $ & 8.2 & 3.42 & \\
085027.8+001503 & 1023 & $ 132.6160 $ & $+0.2509 $ & 128.7 & 280.4 & 0.197 & $ 3.56_{-0.86}^{+1.65} $ & $ 2.88_{-0.17}^{+0.18} $ & $ 7.06_{-0.56}^{+0.69} $ & $ 27.13_{-1.63}^{+1.70} $ & $ 2.99_{-0.56}^{+0.93} $ & $ 4.36_{-0.24}^{+0.24} $ & $ 10.71_{-0.82}^{+0.91} $ & $ 41.04_{-2.18}^{+2.33} $ & 18.6 & 4.19 & \\
085119.9+022951 & 2101 & $ 132.8333 $ & $+2.4977 $ & 54.7 & 133.3 & 0.38 &  - & $ 5.90_{-0.74}^{+0.65} $ & $ 12.81_{-1.82}^{+1.75} $ & $ 12.09_{-1.50}^{+1.39} $ & $ 2.33_{-0.53}^{+1.09} $ & $ 8.43_{-0.94}^{+0.82} $ & $ 18.16_{-2.18}^{+2.57} $ & $ 17.28_{-1.87}^{+1.73} $ & 9.7 & 1.94 & \\
085419.5-000925 & 1294 & $ 133.5814 $ & $-0.1572 $ & 56.0 & 110.0 & 0.282 & $ 1.97_{-0.40}^{+0.78} $ & $ 2.64_{-0.20}^{+0.24} $ & $ 5.99_{-0.88}^{+1.36} $ & $ 11.08_{-0.93}^{+0.91} $ &  - & $ 3.75_{-0.28}^{+0.32} $ & $ 8.46_{-1.20}^{+1.89} $ & $ 15.63_{-1.13}^{+1.27} $ & 12.9 & 2.76 & \\
085436.6+003835 & 328 & $ 133.6526 $ & $+0.6431 $ & 242.8 & 688.6 & 0.106 & $ 3.39_{-0.67}^{+0.95} $ & $ 1.50_{-0.11}^{+0.11} $ & $ 3.55_{-0.32}^{+0.35} $ & $ 53.58_{-3.94}^{+3.87} $ & $ 2.42_{-0.33}^{+0.65} $ & $ 1.83_{-0.15}^{+0.14} $ & $ 4.35_{-0.41}^{+0.45} $ & $ 65.62_{-5.63}^{+4.77} $ & 24.1 & 3.87 & \\
085447.0-012132 & 2079 & $ 133.6962 $ & $-1.3591 $ & 81.5 & 153.9 & 0.353 & $ 1.79_{-0.27}^{+0.55} $ & $ 3.39_{-0.34}^{+0.32} $ & $ 6.64_{-0.67}^{+0.98} $ & $ 8.07_{-0.78}^{+0.81} $ & $ 1.71_{-0.26}^{+0.46} $ & $ 4.98_{-0.38}^{+0.46} $ & $ 9.81_{-0.88}^{+1.43} $ & $ 11.89_{-0.94}^{+1.13} $ & 11.4 & 2.02 & \\
085517.2+013508 & 6746 & $ 133.8219 $ & $+1.5857 $ & 36.4 & 55.0 & 0.324 &  - & $ 1.74_{-0.24}^{+0.24} $ & $ 4.07_{-0.66}^{+0.80} $ & $ 5.31_{-0.74}^{+0.76} $ & $ 2.67_{-0.72}^{+1.66} $ & $ 2.94_{-0.35}^{+0.37} $ & $ 6.94_{-1.05}^{+1.17} $ & $ 9.05_{-1.08}^{+1.07} $ & 9.3 & 2.54 & \\
085604.8+002520 & 8826 & $ 134.0202 $ & $+0.4223 $ & 25.8 & 43.6 & 0.168 & $ 1.87_{-0.37}^{+0.73} $ & $ 0.76_{-0.08}^{+0.09} $ & $ 1.54_{-0.18}^{+0.20} $ & $ 9.94_{-1.10}^{+1.14} $ & $ 1.75_{-0.29}^{+0.42} $ & $ 1.29_{-0.14}^{+0.12} $ & $ 2.57_{-0.29}^{+0.32} $ & $ 16.76_{-1.78}^{+1.54} $ & 11.7 & 3.63 & \\
085626.2+021348 & 5697 & $ 134.1094 $ & $+2.2302 $ & 21.5 & 51.9 & 0.125 & $ 1.03_{-0.36}^{+0.28} $ & $ 0.36_{-0.04}^{+0.04} $ & $ 0.58_{-0.06}^{+0.07} $ & $ 8.60_{-0.89}^{+0.92} $ & $ 1.07_{-0.27}^{+0.25} $ & $ 0.47_{-0.06}^{+0.06} $ & $ 0.76_{-0.09}^{+0.11} $ & $ 11.36_{-1.46}^{+1.34} $ & 8.4 & 2.69 & \\
085705.9+011453 & 3128 & $ 134.2749 $ & $+1.2482 $ & 53.7 & 119.3 & 0.106 & $ 1.10_{-0.10}^{+0.11} $ & $ 0.45_{-0.04}^{+0.04} $ & $ 0.70_{-0.06}^{+0.07} $ & $ 15.31_{-1.27}^{+1.44} $ & $ 0.87_{-0.19}^{+0.13} $ & $ 0.51_{-0.05}^{+0.05} $ & $ 0.79_{-0.08}^{+0.08} $ & $ 17.33_{-1.69}^{+1.77} $ & 11.7 & 2.51 & \\
085751.6+031039 & 108 & $ 134.4653 $ & $+3.1775 $ & 742.2 & 2053.9 & 0.201 & $ 5.79_{-0.85}^{+1.17} $ & $ 9.59_{-0.35}^{+0.44} $ & $ 31.45_{-1.75}^{+1.93} $ & $ 90.06_{-3.43}^{+4.13} $ & $ 5.65_{-0.85}^{+1.11} $ & $ 13.54_{-0.50}^{+0.56} $ & $ 44.44_{-2.54}^{+2.51} $ & $ 127.08_{-4.87}^{+5.18} $ & 40.3 & 4.37 & \\
085805.0+010906 & 4070 & $ 134.5209 $ & $+1.1517 $ & 39.7 & 90.8 & 0.071 & $ 1.29_{-0.12}^{+0.20} $ & $ 0.24_{-0.05}^{+0.04} $ & $ 0.40_{-0.08}^{+0.07} $ & $ 19.14_{-3.91}^{+3.32} $ & $ 1.12_{-0.11}^{+0.13} $ & $ 0.29_{-0.07}^{+0.06} $ & $ 0.48_{-0.12}^{+0.10} $ & $ 23.42_{-5.98}^{+4.85} $ & 11.5 & 2.93 & \\
085849.8+022800 & 715 & $ 134.7076 $ & $+2.4668 $ & 92.3 & 304.5 & 0.215 & $ 1.81_{-0.20}^{+0.38} $ & $ 2.14_{-0.32}^{+0.39} $ & $ 4.39_{-0.70}^{+0.84} $ & $ 15.97_{-2.35}^{+2.95} $ & $ 1.82_{-0.23}^{+0.44} $ & $ 2.71_{-0.35}^{+0.37} $ & $ 5.55_{-0.77}^{+0.87} $ & $ 20.23_{-2.47}^{+2.81} $ & 14.4 & 2.35 & \\
085901.5+010649 & 2757 & $ 134.7564 $ & $+1.1137 $ & 33.6 & 97.4 & 0.162 &  - & $ 0.56_{-0.08}^{+0.08} $ & $ 0.90_{-0.14}^{+0.13} $ & $ 7.58_{-1.02}^{+1.08} $ & $ 0.97_{-0.22}^{+0.40} $ & $ 0.61_{-0.09}^{+0.09} $ & $ 0.98_{-0.16}^{+0.14} $ & $ 8.22_{-1.22}^{+1.20} $ & 8.4 & 1.53 & \\
085913.1+031334 & 485 & $ 134.8048 $ & $+3.2263 $ & 15.9 & 16.8 & 0.189 & $ 1.27_{-0.19}^{+0.35} $ & $ 1.14_{-0.17}^{+0.15} $ & $ 1.84_{-0.29}^{+0.26} $ & $ 10.87_{-1.68}^{+1.43} $ & $ 1.11_{-0.10}^{+0.09} $ & $ 1.26_{-0.27}^{+0.22} $ & $ 2.04_{-0.43}^{+0.36} $ & $ 12.07_{-2.57}^{+2.09} $ & 12.3 & 1.34 & \\
085931.9+030839 & 360 & $ 134.8830 $ & $+3.1443 $ & 175.1 & 617.4 & 0.196 & $ 2.62_{-0.36}^{+0.56} $ & $ 3.27_{-0.17}^{+0.16} $ & $ 7.27_{-0.49}^{+0.52} $ & $ 30.69_{-1.69}^{+1.50} $ & $ 2.37_{-0.31}^{+0.43} $ & $ 4.12_{-0.21}^{+0.20} $ & $ 9.16_{-0.62}^{+0.64} $ & $ 38.61_{-2.01}^{+1.85} $ & 20.2 & 2.65 & \\
085948.9+041120 & 1350 & $ 134.9539 $ & $+4.1889 $ & 22.6 & 43.0 & 0.208 & $ 1.63_{-0.28}^{+0.49} $ & $ 1.25_{-0.12}^{+0.13} $ & $ 2.25_{-0.23}^{+0.25} $ & $ 9.93_{-0.99}^{+0.97} $ & $ 1.37_{-0.18}^{+0.26} $ & $ 1.47_{-0.15}^{+0.15} $ & $ 2.63_{-0.29}^{+0.28} $ & $ 11.60_{-1.24}^{+1.17} $ & 10.7 & 1.66 & \\
090104.4+011643 & 3171 & $ 135.2686 $ & $+1.2788 $ & 38.0 & 62.8 & 0.251 &  - & $ 1.65_{-0.18}^{+0.20} $ & $ 3.21_{-0.46}^{+0.68} $ & $ 8.71_{-0.99}^{+1.04} $ & $ 1.61_{-0.36}^{+0.90} $ & $ 2.03_{-0.20}^{+0.22} $ & $ 3.94_{-0.53}^{+0.84} $ & $ 10.76_{-1.09}^{+1.12} $ & 9.7 & 1.68 & \\
090131.1+030056 & 152 & $ 135.3800 $ & $+3.0157 $ & 373.9 & 1266.7 & 0.193 & $ 3.49_{-0.60}^{+0.98} $ & $ 4.91_{-0.24}^{+0.34} $ & $ 12.52_{-0.80}^{+1.11} $ & $ 48.53_{-2.33}^{+3.36} $ & $ 3.18_{-0.43}^{+0.77} $ & $ 6.33_{-0.39}^{+0.54} $ & $ 16.18_{-1.21}^{+1.70} $ & $ 62.60_{-3.88}^{+5.46} $ & 26.1 & 3.15 & \\
090140.9-012132 & 2412 & $ 135.4207 $ & $-1.3591 $ & 20.4 & 43.8 & 0.295 &  - & $ 0.92_{-0.14}^{+0.15} $ & $ 1.56_{-0.23}^{+0.31} $ & $ 3.24_{-0.48}^{+0.51} $ & $ 1.21_{-0.35}^{+0.55} $ & $ 1.57_{-0.23}^{+0.25} $ & $ 2.66_{-0.41}^{+0.45} $ & $ 5.54_{-0.79}^{+0.79} $ & 9.0 & 2.26 & \\
090153.9-012209 & 2048 & $ 135.4748 $ & $-1.3694 $ & 87.9 & 163.9 & 0.295 & $ 2.38_{-0.56}^{+1.10} $ & $ 1.58_{-0.19}^{+0.21} $ & $ 3.46_{-0.48}^{+0.57} $ & $ 5.87_{-0.69}^{+0.77} $ & $ 2.06_{-0.43}^{+0.96} $ & $ 2.01_{-0.27}^{+0.24} $ & $ 4.41_{-0.64}^{+0.73} $ & $ 7.50_{-0.98}^{+0.90} $ & 10.5 & 1.54 & \\
090255.2+030220 & 5489 & $ 135.7300 $ & $+3.0389 $ & 48.9 & 66.8 & 0.200 &  - & $ 0.90_{-0.11}^{+0.11} $ & $ 1.74_{-0.22}^{+0.25} $ & $ 7.87_{-0.94}^{+0.98} $ & $ 1.60_{-0.20}^{+0.37} $ & $ 1.29_{-0.14}^{+0.13} $ & $ 2.47_{-0.28}^{+0.35} $ & $ 11.21_{-1.23}^{+1.22} $ & 9.1 & 2.41 & \\
090430.7+042648 & 1075 & $ 136.1282 $ & $+4.4468 $ & 37.7 & 209.8 & 0.457 & $ 2.79_{-0.58}^{+1.45} $ & $ 5.89_{-0.82}^{+0.83} $ & $ 12.56_{-1.72}^{+2.01} $ & $ 7.86_{-1.08}^{+1.07} $ & $ 2.23_{-0.40}^{+0.62} $ & $ 6.57_{-0.87}^{+0.92} $ & $ 14.20_{-2.07}^{+2.12} $ & $ 8.77_{-1.18}^{+1.25} $ & 10.3 & 1.04 & \\
090540.0+043440 & 354 & $ 136.4171 $ & $+4.5780 $ & 231.4 & 678.2 & 0.24 & $ 3.87_{-0.89}^{+1.62} $ & $ 6.76_{-0.44}^{+0.45} $ & $ 18.20_{-1.91}^{+2.94} $ & $ 43.49_{-2.89}^{+2.81} $ & $ 3.87_{-0.92}^{+1.86} $ & $ 8.85_{-0.55}^{+0.51} $ & $ 23.74_{-2.31}^{+4.06} $ & $ 56.87_{-3.57}^{+3.11} $ & 23.8 & 2.65 & \\
090601.0+000055 & 3259 & $ 136.5043 $ & $+0.0154 $ & 18.6 & 92.2 & 0.200 &  - & $ 0.85_{-0.11}^{+0.10} $ & $ 1.92_{-0.29}^{+0.42} $ & $ 7.68_{-0.98}^{+0.90} $ & $ 2.15_{-0.48}^{+1.22} $ & $ 1.20_{-0.18}^{+0.15} $ & $ 2.70_{-0.47}^{+0.63} $ & $ 10.81_{-1.65}^{+1.37} $ & 9.1 & 2.55 & \\
090628.9-012938 & 2424 & $ 136.6207 $ & $-1.4941 $ & 46.1 & 119.0 & 0.431 & $ 2.46_{-0.54}^{+1.49} $ & $ 2.80_{-0.33}^{+0.37} $ & $ 6.94_{-1.02}^{+2.00} $ & $ 4.51_{-0.50}^{+0.56} $ &  - & $ 3.90_{-0.40}^{+0.47} $ & $ 9.71_{-1.38}^{+2.70} $ & $ 6.28_{-0.61}^{+0.75} $ & 10.6 & 1.56 & \\
090656.3+044717 & 12153 & $ 136.7347 $ & $+4.7882 $ & 29.0 & 32.2 & 0.12 & $ 1.58_{-0.21}^{+0.41} $ & $ 0.52_{-0.05}^{+0.06} $ & $ 0.95_{-0.10}^{+0.10} $ & $ 13.76_{-1.41}^{+1.51} $ & $ 1.40_{-0.16}^{+0.23} $ & $ 0.86_{-0.08}^{+0.08} $ & $ 1.57_{-0.15}^{+0.15} $ & $ 22.74_{-2.07}^{+2.04} $ & 10.6 & 4.28 & \\
090723.8-011210 & 2680 & $ 136.8494 $ & $-1.2029 $ & 32.5 & 107.9 & 0.251 & $ 2.15_{-0.47}^{+1.25} $ & $ 1.14_{-0.12}^{+0.14} $ & $ 2.57_{-0.32}^{+0.42} $ & $ 6.19_{-0.63}^{+0.74} $ & $ 2.31_{-0.56}^{+1.01} $ & $ 1.55_{-0.16}^{+0.17} $ & $ 3.50_{-0.44}^{+0.54} $ & $ 8.43_{-0.85}^{+0.93} $ & 9.6 & 1.96 & \\
090838.0+015226 & 1479 & $ 137.1585 $ & $+1.8741 $ & 42.9 & 183.6 & 0.264 &  - & $ 2.06_{-0.19}^{+0.17} $ & $ 4.64_{-0.54}^{+0.55} $ & $ 9.97_{-0.97}^{+0.82} $ & $ 2.49_{-0.58}^{+1.06} $ & $ 2.79_{-0.25}^{+0.22} $ & $ 6.30_{-0.73}^{+0.73} $ & $ 13.46_{-1.17}^{+1.09} $ & 11.6 & 2.24 & \\
090913.8-001214 & 885 & $ 137.3079 $ & $-0.2040 $ & 67.6 & 243.9 & 0.310 & $ 3.12_{-0.78}^{+1.37} $ & $ 3.98_{-0.32}^{+0.33} $ & $ 9.59_{-1.09}^{+1.47} $ & $ 13.63_{-1.14}^{+1.05} $ & $ 2.84_{-0.71}^{+1.44} $ & $ 5.50_{-0.43}^{+0.44} $ & $ 13.25_{-1.46}^{+1.95} $ & $ 18.78_{-1.43}^{+1.46} $ & 14.5 & 2.24 & \\
091033.8+005100 & 510 & $ 137.6409 $ & $+0.8501 $ & 45.6 & 416.8 & 0.366 & $ 2.31_{-0.38}^{+0.64} $ & $ 3.74_{-0.50}^{+0.54} $ & $ 8.55_{-1.22}^{+1.26} $ & $ 8.53_{-1.12}^{+1.24} $ & $ 2.59_{-0.46}^{+0.52} $ & $ 4.36_{-0.63}^{+0.66} $ & $ 9.95_{-1.55}^{+1.48} $ & $ 9.95_{-1.44}^{+1.51} $ & 11.4 & 1.32 & \\
091117.1+030441 & 15328 & $ 137.8216 $ & $+3.0782 $ & 15.2 & 19.2 & 0.213 &  - & $ 0.57_{-0.08}^{+0.10} $ & $ 1.34_{-0.21}^{+0.29} $ & $ 4.48_{-0.63}^{+0.79} $ & $ 2.53_{-0.60}^{+1.01} $ & $ 1.06_{-0.13}^{+0.14} $ & $ 2.47_{-0.34}^{+0.45} $ & $ 8.30_{-1.00}^{+1.08} $ & 8.1 & 3.55 & \\
091215.3-021743 & 626 & $ 138.0641 $ & $-2.2956 $ & 120.3 & 435.9 & 0.16 & $ 3.11_{-0.76}^{+1.28} $ & $ 4.70_{-0.39}^{+0.38} $ & $ 11.15_{-1.21}^{+1.32} $ & $ 69.85_{-6.03}^{+5.46} $ & $ 2.70_{-0.56}^{+0.76} $ & $ 6.29_{-0.51}^{+0.41} $ & $ 14.84_{-1.48}^{+1.64} $ & $ 93.23_{-7.45}^{+6.10} $ & 20.5 & 3.51 & \\
091315.0+034850 & 593 & $ 138.3125 $ & $+3.8139 $ & 88.3 & 393.5 & 0.453 &  - & $ 8.26_{-0.78}^{+0.71} $ & $ 24.23_{-3.00}^{+6.03} $ & $ 12.40_{-1.10}^{+1.01} $ & $ 4.39_{-1.26}^{+2.72} $ & $ 11.23_{-1.05}^{+0.95} $ & $ 33.02_{-4.14}^{+8.14} $ & $ 16.85_{-1.44}^{+1.43} $ & 14.6 & 2.12 & \\
091322.9+040617 & 9801 & $ 138.3454 $ & $+4.1050 $ & 9.6 & 26.9 & 0.088 & $ 0.83_{-0.14}^{+0.15} $ & $ 0.17_{-0.03}^{+0.03} $ & $ 0.29_{-0.07}^{+0.07} $ & $ 8.42_{-1.74}^{+1.72} $ &  - & $ 0.24_{-0.05}^{+0.06} $ & $ 0.41_{-0.11}^{+0.12} $ & $ 12.16_{-2.74}^{+2.73} $ & 7.3 & 3.88 & \\
091336.6+031723 & 5541 & $ 138.4026 $ & $+3.2899 $ & 54.9 & 82.7 & 0.142 & $ 1.84_{-0.40}^{+0.76} $ & $ 0.50_{-0.07}^{+0.07} $ & $ 0.83_{-0.12}^{+0.13} $ & $ 9.12_{-1.24}^{+1.40} $ & $ 1.13_{-0.17}^{+0.30} $ & $ 0.68_{-0.09}^{+0.09} $ & $ 1.12_{-0.15}^{+0.18} $ & $ 12.46_{-1.67}^{+1.75} $ & 9.4 & 2.71 & \\
091351.1-004507 & 1319 & $ 138.4632 $ & $-0.7520 $ & 64.6 & 113.1 & 0.294 & $ 1.58_{-0.19}^{+0.35} $ & $ 3.01_{-0.26}^{+0.28} $ & $ 6.47_{-0.72}^{+1.03} $ & $ 11.24_{-0.92}^{+1.01} $ & $ 2.02_{-0.36}^{+0.73} $ & $ 4.05_{-0.33}^{+0.31} $ & $ 8.71_{-0.90}^{+1.35} $ & $ 15.17_{-1.18}^{+1.15} $ & 12.4 & 1.96 & \\
091412.6+001856 & 1844 & $ 138.5528 $ & $+0.3158 $ & 55.0 & 87.3 & 0.165 & $ 1.81_{-0.31}^{+0.54} $ & $ 0.93_{-0.10}^{+0.10} $ & $ 2.04_{-0.26}^{+0.36} $ & $ 12.78_{-1.38}^{+1.45} $ & $ 2.01_{-0.35}^{+0.67} $ & $ 1.26_{-0.14}^{+0.14} $ & $ 2.75_{-0.36}^{+0.47} $ & $ 17.30_{-1.96}^{+1.92} $ & 11.8 & 2.61 & \\
091417.7+031159 & 396 & $ 138.5740 $ & $+3.1998 $ & 16.7 & 26.9 & 0.231 &  - & $ 1.12_{-0.25}^{+0.28} $ & $ 1.78_{-0.43}^{+0.45} $ & $ 6.78_{-1.53}^{+1.77} $ & $ 0.96_{-0.34}^{+0.34} $ & $ 1.15_{-0.27}^{+0.33} $ & $ 1.81_{-0.47}^{+0.52} $ & $ 6.96_{-1.69}^{+1.99} $ & 13.3 & 0.70 & \\
091453.6+041613 & 372 & $ 138.7235 $ & $+4.2704 $ & 123.9 & 398.6 & 0.143 & $ 2.81_{-0.35}^{+0.37} $ & $ 2.23_{-0.27}^{+0.35} $ & $ 4.91_{-0.63}^{+0.81} $ & $ 41.80_{-5.07}^{+6.65} $ & $ 2.15_{-0.24}^{+0.32} $ & $ 2.66_{-0.27}^{+0.32} $ & $ 5.83_{-0.60}^{+0.77} $ & $ 49.81_{-5.06}^{+6.02} $ & 22.7 & 2.91 & \\
091509.5+051521 & 6244 & $ 138.7897 $ & $+5.2560 $ & 48.0 & 70.0 & 0.25 &  - & $ 2.37_{-0.32}^{+0.31} $ & $ 4.90_{-0.73}^{+0.91} $ & $ 12.89_{-1.68}^{+1.70} $ & $ 1.83_{-0.32}^{+0.53} $ & $ 3.96_{-0.41}^{+0.44} $ & $ 8.18_{-1.07}^{+1.29} $ & $ 21.41_{-2.15}^{+2.47} $ & 10.7 & 3.19 & \\
091610.1-002348 & 534 & $ 139.0423 $ & $-0.3969 $ & 282.2 & 623.3 & 0.322 & $ 3.83_{-0.93}^{+1.77} $ & $ 8.66_{-0.44}^{+0.45} $ & $ 25.51_{-2.42}^{+3.43} $ & $ 28.22_{-1.42}^{+1.41} $ & $ 4.45_{-1.02}^{+1.58} $ & $ 15.51_{-0.62}^{+0.64} $ & $ 45.50_{-4.08}^{+6.41} $ & $ 50.53_{-1.97}^{+2.01} $ & 27.1 & 3.04 & \\
091722.4+010118 & 1526 & $ 139.3435 $ & $+1.0218 $ & 56.4 & 187.8 & 0.359 &  - & $ 3.68_{-0.41}^{+0.43} $ & $ 8.44_{-1.12}^{+1.38} $ & $ 8.85_{-0.95}^{+1.04} $ & $ 2.61_{-0.63}^{+1.38} $ & $ 4.67_{-0.46}^{+0.49} $ & $ 10.67_{-1.24}^{+1.59} $ & $ 11.21_{-1.05}^{+1.08} $ & 10.9 & 1.60 & \\
091849.0+021204 & 994 & $ 139.7042 $ & $+2.2013 $ & 94.8 & 245.3 & 0.283 & $ 3.64_{-1.02}^{+1.69} $ & $ 3.68_{-0.31}^{+0.33} $ & $ 9.78_{-1.05}^{+1.90} $ & $ 15.74_{-1.33}^{+1.29} $ & $ 3.63_{-0.90}^{+1.60} $ & $ 5.89_{-0.50}^{+0.45} $ & $ 15.56_{-1.66}^{+3.07} $ & $ 25.17_{-2.11}^{+1.84} $ & 17.8 & 3.06 & \\
091858.0+024946 & 7194 & $ 139.7418 $ & $+2.8295 $ & 26.3 & 46.3 & 0.199 &  - & $ 0.76_{-0.10}^{+0.09} $ & $ 1.31_{-0.18}^{+0.18} $ & $ 6.58_{-0.87}^{+0.74} $ & $ 1.28_{-0.21}^{+0.30} $ & $ 1.03_{-0.14}^{+0.12} $ & $ 1.79_{-0.25}^{+0.24} $ & $ 8.97_{-1.20}^{+1.08} $ & 8.1 & 2.08 & \\
092023.2+013444 & 222 & $ 140.0967 $ & $+1.5789 $ & 88.4 & 719.6 & 0.71 & $ 2.26_{-0.36}^{+0.58} $ & $ 25.58_{-2.45}^{+2.70} $ & $ 55.97_{-5.56}^{+6.12} $ & $ 11.97_{-1.16}^{+1.21} $ & $ 2.44_{-0.35}^{+0.50} $ & $ 30.67_{-2.98}^{+2.82} $ & $ 66.88_{-6.51}^{+6.85} $ & $ 14.30_{-1.33}^{+1.38} $ & 16.5 & 1.39 & \\
092037.0-011506 & 5739 & $ 140.1545 $ & $-1.2519 $ & 34.3 & 63.0 & 0.154 & $ 1.01_{-0.13}^{+0.15} $ & $ 0.48_{-0.06}^{+0.06} $ & $ 0.81_{-0.10}^{+0.11} $ & $ 7.40_{-0.86}^{+0.89} $ & $ 1.16_{-0.20}^{+0.30} $ & $ 0.61_{-0.08}^{+0.08} $ & $ 1.03_{-0.14}^{+0.15} $ & $ 9.39_{-1.18}^{+1.26} $ & 8.6 & 2.15 & \\
092046.2+002849 & 1915 & $ 140.1926 $ & $+0.4803 $ & 40.0 & 69.6 & 0.413 &  - & $ 3.71_{-0.36}^{+0.38} $ & $ 7.70_{-0.87}^{+0.93} $ & $ 6.21_{-0.57}^{+0.65} $ & $ 1.97_{-0.36}^{+0.72} $ & $ 5.56_{-0.50}^{+0.49} $ & $ 11.54_{-1.18}^{+1.32} $ & $ 9.31_{-0.81}^{+0.86} $ & 11.4 & 2.64 & \\
092049.5+024513 & 489 & $ 140.2063 $ & $+2.7538 $ & 142.1 & 514.2 & 0.284 & $ 2.88_{-0.50}^{+0.72} $ & $ 5.63_{-0.30}^{+0.34} $ & $ 14.65_{-0.99}^{+1.27} $ & $ 23.87_{-1.27}^{+1.38} $ & $ 3.39_{-0.64}^{+0.87} $ & $ 8.39_{-0.41}^{+0.40} $ & $ 21.77_{-1.31}^{+1.96} $ & $ 35.62_{-1.80}^{+1.61} $ & 23.0 & 3.08 & \\
092121.2+031726 & 100 & $ 140.3385 $ & $+3.2906 $ & 478.6 & 1729.7 & 0.333 & $ 4.81_{-0.83}^{+1.21} $ & $ 15.28_{-0.63}^{+0.78} $ & $ 47.50_{-3.72}^{+3.96} $ & $ 46.65_{-1.95}^{+2.14} $ & $ 5.25_{-0.87}^{+1.31} $ & $ 21.43_{-0.77}^{+0.76} $ & $ 66.28_{-4.78}^{+5.35} $ & $ 65.20_{-2.15}^{+2.28} $ & 31.6 & 2.80 & \\
092202.2+034520 & 5347 & $ 140.5095 $ & $+3.7557 $ & 43.4 & 66.6 & 0.270 &  - & $ 2.48_{-0.44}^{+0.41} $ & $ 7.72_{-1.52}^{+1.76} $ & $ 8.31_{-1.47}^{+1.32} $ & $ 5.42_{-1.49}^{+3.46} $ & $ 4.22_{-0.84}^{+0.98} $ & $ 13.28_{-2.93}^{+3.63} $ & $ 14.09_{-2.77}^{+3.36} $ & 13.2 & 2.60 & \\
092209.3+034628 & 367 & $ 140.5391 $ & $+3.7746 $ & 369.0 & 783.1 & 0.270 & $ 4.25_{-0.83}^{+1.25} $ & $ 5.76_{-0.52}^{+0.53} $ & $ 18.11_{-2.44}^{+2.28} $ & $ 28.01_{-2.54}^{+2.45} $ & $ 5.39_{-1.30}^{+1.80} $ & $ 7.98_{-1.01}^{+1.02} $ & $ 24.93_{-3.68}^{+3.98} $ & $ 38.73_{-4.82}^{+5.01} $ & 23.1 & 2.80 & \\
092220.4+034806 & 4014 & $ 140.5852 $ & $+3.8017 $ & 24.1 & 48.9 & 0.267 &  - & $ 1.24_{-0.19}^{+0.19} $ & $ 2.87_{-0.51}^{+0.62} $ & $ 5.90_{-0.86}^{+0.91} $ & $ 2.31_{-0.55}^{+1.25} $ & $ 1.53_{-0.26}^{+0.33} $ & $ 3.52_{-0.67}^{+0.98} $ & $ 7.27_{-1.27}^{+1.51} $ & 8.5 & 1.48 & \\
092235.8-002443 & 3133 & $ 140.6494 $ & $-0.4120 $ & 30.2 & 49.4 & 0.055 & $ 1.90_{-0.32}^{+0.59} $ & $ 0.15_{-0.03}^{+0.04} $ & $ 0.33_{-0.06}^{+0.08} $ & $ 20.84_{-3.68}^{+5.01} $ & $ 2.23_{-0.39}^{+0.56} $ & $ 0.16_{-0.03}^{+0.05} $ & $ 0.34_{-0.07}^{+0.11} $ & $ 21.67_{-4.31}^{+6.25} $ & 11.9 & 2.16 & \\
092241.9+020719 & 1535 & $ 140.6749 $ & $+2.1222 $ & 32.5 & 52.2 & 0.198 &  - & $ 0.79_{-0.13}^{+0.14} $ & $ 1.37_{-0.22}^{+0.25} $ & $ 6.91_{-1.10}^{+1.19} $ & $ 1.29_{-0.20}^{+0.37} $ & $ 0.93_{-0.15}^{+0.13} $ & $ 1.61_{-0.25}^{+0.27} $ & $ 8.15_{-1.29}^{+1.19} $ & 8.6 & 1.60 & \\
092339.0+052654 & 1894 & $ 140.9129 $ & $+5.4486 $ & 30.3 & 158.3 & 0.37 & $ 2.52_{-0.38}^{+0.52} $ & $ 35.48_{-10.16}^{+13.00} $ & $ 98.68_{-28.12}^{+35.92} $ & $ 81.89_{-23.26}^{+29.79} $ & $ 3.93_{-0.28}^{+0.46} $ & $ 36.83_{-9.88}^{+12.25} $ & $ 101.61_{-27.28}^{+34.04} $ & $ 84.78_{-22.60}^{+28.09} $ & 16.3 & 1.02 & \\
092405.0-013059 & 298 & $ 141.0211 $ & $-1.5165 $ & 196.6 & 576.5 & 0.337 &  - & $ 3.71_{-0.41}^{+0.49} $ & $ 10.24_{-1.39}^{+2.25} $ & $ 10.80_{-1.20}^{+1.36} $ & $ 3.97_{-1.01}^{+2.54} $ & $ 4.82_{-0.43}^{+0.53} $ & $ 13.25_{-1.56}^{+2.90} $ & $ 14.06_{-1.24}^{+1.47} $ & 16.7 & 1.84 & \\
092409.4+040057 & 1437 & $ 141.0393 $ & $+4.0160 $ & 122.1 & 238.5 & 0.084 &  - & $ 0.36_{-0.03}^{+0.03} $ & $ 0.56_{-0.04}^{+0.04} $ & $ 20.24_{-1.43}^{+1.55} $ & $ 0.92_{-0.07}^{+0.08} $ & $ 0.40_{-0.03}^{+0.03} $ & $ 0.63_{-0.05}^{+0.05} $ & $ 22.72_{-1.92}^{+1.73} $ & 14.8 & 2.81 & \\
092533.6-014205 & 1483 & $ 141.3903 $ & $-1.7016 $ & 51.0 & 150.6 & 0.23 &  - & $ 1.03_{-0.10}^{+0.11} $ & $ 2.29_{-0.26}^{+0.31} $ & $ 6.82_{-0.63}^{+0.73} $ & $ 2.30_{-0.44}^{+0.78} $ & $ 1.22_{-0.13}^{+0.14} $ & $ 2.71_{-0.36}^{+0.36} $ & $ 8.01_{-0.85}^{+0.95} $ & 11.2 & 1.54 & \\
092546.3-014347 & 566 & $ 141.4432 $ & $-1.7298 $ & 197.1 & 480.2 & 0.23 &  - & $ 2.23_{-0.20}^{+0.24} $ & $ 5.21_{-0.56}^{+0.61} $ & $ 15.01_{-1.31}^{+1.55} $ & $ 2.64_{-0.45}^{+0.84} $ & $ 2.98_{-0.24}^{+0.32} $ & $ 6.96_{-0.69}^{+0.82} $ & $ 20.04_{-1.55}^{+2.19} $ & 18.3 & 2.24 & \\
092629.3+032614 & 2050 & $ 141.6225 $ & $+3.4374 $ & 19.7 & 117.5 & 0.088 & $ 0.90_{-0.07}^{+0.07} $ & $ 0.24_{-0.03}^{+0.03} $ & $ 0.38_{-0.06}^{+0.05} $ & $ 12.24_{-1.79}^{+1.56} $ & $ 0.94_{-0.07}^{+0.07} $ & $ 0.30_{-0.05}^{+0.05} $ & $ 0.47_{-0.07}^{+0.08} $ & $ 15.24_{-2.38}^{+2.64} $ & 9.3 & 2.49 & \\
092647.5+050032 & 137 & $ 141.6980 $ & $+5.0091 $ & 505.4 & 1395.7 & 0.45 &  - & $ 18.64_{-1.76}^{+1.82} $ & $ 65.64_{-8.77}^{+8.92} $ & $ 28.64_{-2.66}^{+2.82} $ & $ 6.72_{-1.68}^{+2.91} $ & $ 25.68_{-2.62}^{+2.48} $ & $ 90.72_{-13.06}^{+11.87} $ & $ 39.42_{-4.17}^{+3.98} $ & 23.6 & 2.28 & \\
092648.0+050124 & 6875 & $ 141.7002 $ & $+5.0234 $ & 16.0 & 26.5 & 0.45 & $ 7.04_{-2.29}^{+4.00} $ & $ 6.57_{-1.35}^{+1.58} $ & $ 21.70_{-4.80}^{+5.49} $ & $ 10.04_{-2.07}^{+2.42} $ & $ 5.77_{-1.16}^{+1.78} $ & $ 9.45_{-2.08}^{+2.13} $ & $ 31.16_{-7.13}^{+7.83} $ & $ 14.47_{-3.31}^{+3.05} $ & 13.3 & 1.84 & \\
092650.5+035755 & 671 & $ 141.7106 $ & $+3.9654 $ & 33.0 & 304.8 & 0.377 & $ 2.86_{-0.69}^{+1.58} $ & $ 4.59_{-0.59}^{+0.62} $ & $ 10.53_{-1.62}^{+1.48} $ & $ 9.81_{-1.27}^{+1.30} $ & $ 2.59_{-0.54}^{+1.18} $ & $ 4.97_{-0.60}^{+0.61} $ & $ 11.29_{-1.61}^{+1.62} $ & $ 10.56_{-1.32}^{+1.33} $ & 11.3 & 1.06 & \\
092735.3+014423 & 266 & $ 141.8974 $ & $+1.7399 $ & 32.9 & 43.2 & 0.149 &  - & $ 0.82_{-0.14}^{+0.15} $ & $ 1.74_{-0.32}^{+0.40} $ & $ 14.03_{-2.36}^{+2.50} $ & $ 1.90_{-0.45}^{+0.73} $ & $ 0.98_{-0.15}^{+0.16} $ & $ 2.06_{-0.37}^{+0.45} $ & $ 16.62_{-2.50}^{+2.79} $ & 12.1 & 2.15 & \\
092737.7+020607 & 1924 & $ 141.9074 $ & $+2.1021 $ & 106.4 & 191.5 & 0.419 &  - & $ 4.82_{-0.51}^{+0.50} $ & $ 11.50_{-1.36}^{+1.32} $ & $ 8.15_{-0.84}^{+0.85} $ & $ 2.82_{-0.65}^{+1.00} $ & $ 7.62_{-0.69}^{+0.65} $ & $ 18.04_{-1.76}^{+1.97} $ & $ 12.88_{-1.14}^{+1.05} $ & 11.8 & 2.04 & \\
092821.2+042149 & 666 & $ 142.0884 $ & $+4.3637 $ & 113.0 & 375.3 & 0.23 & $ 3.41_{-0.94}^{+1.88} $ & $ 3.19_{-0.33}^{+0.35} $ & $ 7.10_{-0.83}^{+0.90} $ & $ 19.97_{-2.03}^{+2.13} $ & $ 2.45_{-0.48}^{+0.77} $ & $ 4.08_{-0.46}^{+0.49} $ & $ 9.09_{-1.16}^{+1.20} $ & $ 25.53_{-2.82}^{+2.99} $ & 15.7 & 2.22 & \\
092915.7-001357 & 1549 & $ 142.3158 $ & $-0.2327 $ & 31.5 & 181.0 & 0.322 & $ 1.83_{-0.33}^{+0.49} $ & $ 2.29_{-0.24}^{+0.26} $ & $ 4.19_{-0.50}^{+0.67} $ & $ 6.62_{-0.71}^{+0.77} $ & $ 1.44_{-0.19}^{+0.34} $ & $ 2.76_{-0.30}^{+0.36} $ & $ 5.06_{-0.62}^{+0.87} $ & $ 8.01_{-0.92}^{+1.04} $ & 9.4 & 1.36 & \\
092921.7+040040 & 609 & $ 142.3408 $ & $+4.0112 $ & 51.8 & 297.4 & 0.50 & $ 3.39_{-0.98}^{+1.88} $ & $ 8.34_{-1.32}^{+1.19} $ & $ 22.82_{-4.00}^{+5.39} $ & $ 9.77_{-1.49}^{+1.39} $ &  - & $ 9.37_{-1.41}^{+1.26} $ & $ 25.63_{-4.57}^{+5.75} $ & $ 10.91_{-1.54}^{+1.52} $ & 11.6 & 1.12 & \\
092955.8-003403 & 1967 & $ 142.4829 $ & $-0.5676 $ & 39.4 & 137.1 & 0.150 & $ 1.76_{-0.24}^{+0.56} $ & $ 0.73_{-0.07}^{+0.07} $ & $ 1.43_{-0.15}^{+0.16} $ & $ 12.14_{-1.08}^{+1.20} $ & $ 1.67_{-0.27}^{+0.41} $ & $ 0.92_{-0.10}^{+0.09} $ & $ 1.78_{-0.19}^{+0.21} $ & $ 15.19_{-1.60}^{+1.58} $ & 11.1 & 2.35 & \\
093003.3+035630 & 288 & $ 142.5140 $ & $+3.9417 $ & 82.1 & 593.1 & 0.327 & $ 2.22_{-0.34}^{+0.62} $ & $ 6.01_{-0.53}^{+0.55} $ & $ 14.40_{-1.59}^{+1.67} $ & $ 18.08_{-1.60}^{+1.66} $ & $ 2.81_{-0.56}^{+1.22} $ & $ 6.48_{-0.51}^{+0.55} $ & $ 15.51_{-1.57}^{+1.64} $ & $ 19.53_{-1.56}^{+1.59} $ & 15.3 & 1.30 & \\
093009.0+040144 & 536 & $ 142.5376 $ & $+4.0289 $ & 76.4 & 376.0 & 0.328 & $ 2.59_{-0.54}^{+1.03} $ & $ 4.61_{-0.45}^{+0.40} $ & $ 10.65_{-1.31}^{+1.48} $ & $ 13.68_{-1.33}^{+1.18} $ & $ 2.66_{-0.67}^{+1.51} $ & $ 5.04_{-0.44}^{+0.45} $ & $ 11.66_{-1.29}^{+1.66} $ & $ 14.93_{-1.31}^{+1.34} $ & 13.0 & 1.30 & \\
093141.2-004717 & 1705 & $ 142.9219 $ & $-0.7883 $ & 13.7 & 30.5 & 0.093 &  - & $ 0.13_{-0.04}^{+0.05} $ & $ 0.21_{-0.06}^{+0.08} $ & $ 5.99_{-1.77}^{+2.26} $ & $ 0.50_{-0.01}^{+0.01} $ & $ 0.14_{-0.04}^{+0.05} $ & $ 0.22_{-0.07}^{+0.08} $ & $ 6.21_{-1.91}^{+2.31} $ & 7.8 & 1.52 & \\
093149.8-020143 & 6494 & $ 142.9576 $ & $-2.0289 $ & 27.5 & 39.9 & 0.13 & $ 1.01_{-0.09}^{+0.09} $ & $ 0.18_{-0.03}^{+0.03} $ & $ 0.29_{-0.04}^{+0.05} $ & $ 3.77_{-0.53}^{+0.58} $ & $ 0.97_{-0.09}^{+0.10} $ & $ 0.23_{-0.05}^{+0.04} $ & $ 0.37_{-0.07}^{+0.07} $ & $ 4.86_{-0.98}^{+0.85} $ & 7.5 & 2.16 & \\
093151.3-002212 & 836 & $ 142.9638 $ & $-0.3701 $ & 54.0 & 307.1 & 0.336 & $ 2.53_{-0.50}^{+0.93} $ & $ 4.00_{-0.38}^{+0.41} $ & $ 8.74_{-0.96}^{+1.05} $ & $ 11.08_{-1.05}^{+1.12} $ & $ 2.36_{-0.43}^{+0.76} $ & $ 4.77_{-0.41}^{+0.46} $ & $ 10.46_{-1.13}^{+1.11} $ & $ 13.19_{-1.16}^{+1.19} $ & 12.1 & 1.52 & \\
093403.5-001422 & 900 & $ 143.5148 $ & $-0.2397 $ & 79.7 & 279.5 & 0.24 & $ 1.87_{-0.25}^{+0.37} $ & $ 2.43_{-0.21}^{+0.24} $ & $ 4.92_{-0.48}^{+0.67} $ & $ 14.30_{-1.27}^{+1.39} $ & $ 1.78_{-0.24}^{+0.47} $ & $ 3.02_{-0.26}^{+0.28} $ & $ 6.07_{-0.56}^{+0.88} $ & $ 17.71_{-1.48}^{+1.66} $ & 13.6 & 2.04 & \\
093431.3-002309 & 1641 & $ 143.6304 $ & $-0.3860 $ & 35.5 & 87.4 & 0.342 & $ 3.02_{-0.75}^{+1.17} $ & $ 3.36_{-0.32}^{+0.36} $ & $ 12.42_{-2.28}^{+1.99} $ & $ 9.86_{-0.88}^{+1.07} $ &  - & $ 4.85_{-0.42}^{+0.50} $ & $ 17.91_{-3.25}^{+2.87} $ & $ 14.26_{-1.25}^{+1.42} $ & 11.8 & 2.18 & \\
093500.7+005417 & 649 & $ 143.7532 $ & $+0.9048 $ & 104.4 & 330.3 & 0.361 & $ 3.27_{-0.78}^{+1.54} $ & $ 6.34_{-0.52}^{+0.56} $ & $ 15.07_{-1.46}^{+1.47} $ & $ 15.19_{-1.26}^{+1.35} $ & $ 2.75_{-0.54}^{+0.91} $ & $ 7.69_{-0.59}^{+0.59} $ & $ 18.16_{-1.56}^{+1.65} $ & $ 18.36_{-1.34}^{+1.37} $ & 14.2 & 1.58 & \\
093513.0+004757 & 213 & $ 143.8046 $ & $+0.7994 $ & 332.8 & 836.9 & 0.356 &  - & $ 12.93_{-1.12}^{+1.11} $ & $ 40.29_{-5.57}^{+6.05} $ & $ 33.92_{-2.91}^{+2.82} $ & $ 5.32_{-1.37}^{+2.24} $ & $ 20.24_{-2.00}^{+1.72} $ & $ 62.75_{-8.42}^{+9.79} $ & $ 52.93_{-5.01}^{+4.55} $ & 27.5 & 3.04 & \\
093520.9+023234 & 82 & $ 143.8374 $ & $+2.5429 $ & 175.9 & 1645.8 & 0.516 & $ 3.39_{-0.54}^{+0.89} $ & $ 27.87_{-1.85}^{+1.74} $ & $ 81.36_{-6.89}^{+10.61} $ & $ 31.01_{-1.96}^{+1.86} $ & $ 4.38_{-0.88}^{+1.59} $ & $ 34.33_{-2.33}^{+2.15} $ & $ 100.27_{-8.49}^{+11.91} $ & $ 38.12_{-2.39}^{+2.37} $ & 23.5 & 1.89 & \\
093522.2+032329 & 1242 & $ 143.8427 $ & $+3.3915 $ & 86.3 & 248.3 & 0.324 & $ 2.75_{-0.62}^{+1.20} $ & $ 3.77_{-0.37}^{+0.34} $ & $ 8.41_{-0.97}^{+0.90} $ & $ 11.35_{-1.10}^{+1.01} $ & $ 2.44_{-0.50}^{+0.97} $ & $ 4.86_{-0.40}^{+0.44} $ & $ 10.85_{-1.14}^{+1.28} $ & $ 14.64_{-1.20}^{+1.26} $ & 12.6 & 1.80 & \\
093531.4+022710 & 2609 & $ 143.8811 $ & $+2.4530 $ & 36.6 & 95.3 & 0.225 &  - & $ 1.04_{-0.13}^{+0.13} $ & $ 1.99_{-0.32}^{+0.45} $ & $ 6.98_{-0.87}^{+0.87} $ & $ 1.49_{-0.62}^{+0.83} $ & $ 1.21_{-0.16}^{+0.17} $ & $ 2.31_{-0.36}^{+0.50} $ & $ 8.12_{-1.06}^{+1.09} $ & 8.4 & 1.50 & \\
093709.7+014143 & 6251 & $ 144.2908 $ & $+1.6954 $ & 27.9 & 50.0 & 0.38 &  - & $ 3.16_{-0.39}^{+0.41} $ & $ 5.88_{-0.83}^{+0.95} $ & $ 6.29_{-0.75}^{+0.78} $ & $ 1.48_{-0.22}^{+0.56} $ & $ 4.67_{-0.54}^{+0.51} $ & $ 8.67_{-1.13}^{+1.31} $ & $ 9.31_{-1.06}^{+0.98} $ & 9.0 & 1.80 & \\
\hline
\end{longtable}

\tablefoot{Column 1: Cluster identification. Column 2: Unique source ID in the main eFEDS source catalog in \citet{Brunner2021}. Columns 3 and 4: RA and Dec in units of degrees. Column 5: Extent likelihood. Column 6: Detection likelihood. RA, Dec, extent likelihood, and detection likelihood are provided by the \texttt{eSASS} source detection algorithm (see Sect.~\ref{sec:data_detec}). Column 7: photometric redshifts provided by MCMF (see Sect.~\ref{sec:optical}). Values with three digits are spectroscopic redshifts. Column 8: Temperature in units of keV within 300~kpc. Column 9: Luminosity in units of $10^{43}~{\rm erg}~{\rm s}^{-1}$ in the soft band (0.5--2~keV) within 300~kpc. Column 10: Same as column 9 for bolometric (0.01--100~keV) luminosity. Column 11: Flux in units of $10^{-14}~{\rm erg}~{\rm s}^{-1}~{\rm cm}^{-2}$ within 300~kpc. Column 12: temperature in units of keV within 500~kpc. Column 13: Luminosity in units of $10^{43}~{\rm erg}~{\rm s}^{-1}$ in the soft band (0.5--2~keV) within 500~kpc. Column 14: same as column 13 for bolometric (0.01--100~keV) luminosity. Column 15: Flux in units of $10^{-14}~{\rm erg}~{\rm s}^{-1}~{\rm cm}^{-2}$ within 500~kpc. Column 16: Maximum S/N in the soft band. Column 17: Radii in units of arcmin corresponding to the maximum S/N in column 16.  }
\end{landscape}

\end{document}